\documentclass[iop,numberedappendix,appendixfloats,tighten]{emulateapj}
\pdfoutput=1

\usepackage{amsmath}
\usepackage{graphicx}
\usepackage{subfigure}
\usepackage{color}
\usepackage{url}
\usepackage[usenames,dvipsnames]{xcolor}
\usepackage{gensymb}

\usepackage{hyperref}
\hypersetup{colorlinks=true,citecolor=blue}

\newcommand{\ANNz}{ANN$z$}
\newcommand{\ttt}{\texttt}
\newcommand{\mrm}{\mathrm}
\newcommand{\ti}{~$\times$~}
\newcommand{\zmed}{z_\mathrm{med}}
\newcommand{\dsqu}{deg$^2$}
\newcommand{\eg}{e.g.\ }
\newcommand{\ie}{i.e.\ }
\newcommand{\WISC}{WISE~$\times$~SCOS}
\newcommand{\phz}{photo-$z$}
\newcommand{\phzs}{photo-$z$'s}
\newcommand{\de}{\mathrm{d}}

\newcommand{\tcb}{\textcolor{blue}}

\shorttitle{WISE\ti{}SuperCOSMOS \phz\ catalog of 20 million galaxies}
\shortauthors{Bilicki, Peacock, Jarrett, Cluver and GAMA}

\submitted{The Astrophysical Journal Supplement Series, 225:5 (24pp), 2016 July}

\begin{document}

\title{WISE~$\times$~SuperCOSMOS photometric redshift catalog:\\20 million galaxies over 3$\pi$ steradians}

\author{Maciej~Bilicki\altaffilmark{A,B,C}, John~A.~Peacock\altaffilmark{D}, Thomas~H.~Jarrett\altaffilmark{A}, Michelle~E.~Cluver\altaffilmark{E},\\Natasha~Maddox\altaffilmark{F}, Michael~J.~I.~Brown\altaffilmark{G}, Edward~N.~Taylor\altaffilmark{H}, Nigel~C.~Hambly\altaffilmark{D}, Aleksandra~Solarz\altaffilmark{I,C},\\Benne~W.~Holwerda\altaffilmark{B}, Ivan~Baldry\altaffilmark{J}, Jon~Loveday\altaffilmark{K}, Amanda Moffett\altaffilmark{L}, Andrew~M.~Hopkins\altaffilmark{M}, Simon~P.~Driver\altaffilmark{L,N}, Mehmet~Alpaslan\altaffilmark{O} and Joss~Bland-Hawthorn\altaffilmark{P}}

\altaffiltext{A}{Department of Astronomy, University of Cape Town, Private Bag X3, Rondebosch, 7701, South Africa\\ E-mail: \email{maciek(at)ast.uct.ac.za}}
\altaffiltext{B}{Leiden Observatory, Leiden University, Niels Bohrweg 2, NL-2333 CA Leiden, the Netherlands}
\altaffiltext{C}{Janusz Gil Institute of Astronomy, University of Zielona G\'{o}ra, ul. Szafrana 2, 65-516, Zielona G\'{o}ra, Poland}
\altaffiltext{D}{Institute for Astronomy, University of Edinburgh, Royal Observatory, Edinburgh EH9 3HJ, United Kingdom}
\altaffiltext{E}{Department of Physics, University of the Western Cape, Robert Sobukwe Road, Bellville, 7530, South Africa}
\altaffiltext{F}{ASTRON, The Netherlands Institute for Radio Astronomy, Postbus 2, 7990 AA Dwingeloo, The Netherlands}
\altaffiltext{G}{School of Physics \& Astronomy, Monash University, Clayton, Victoria 3800, Australia}
\altaffiltext{H}{School of Physics, University of Melbourne, VIC 3010, Australia}
\altaffiltext{I}{National Centre for Nuclear Research, ul. Ho\.{z}a 69, Warsaw, Poland}
\altaffiltext{J}{Astrophysics Research Institute, Liverpool John Moores University, IC2, Liverpool Science Park, 146 Brownlow Hill, Liverpool, L3 5RF, UK}
\altaffiltext{K}{Astronomy Centre, University of Sussex, Falmer, Brighton BN1 9QH, UK }
\altaffiltext{L}{ICRAR, The University of Western Australia, 35 Stirling Highway, Crawley, WA 6009, Australia}
\altaffiltext{M}{Australian Astronomical Observatory, PO Box 915, North Ryde, NSW 1670, Australia}
\altaffiltext{N}{School of Physics and Astronomy, University of St Andrews, North Haugh, St Andrews, KY16 9SS, UK}
\altaffiltext{O}{NASA Ames Research Center, N232, Moffett Field, Mountain View CA 94035, USA}
\altaffiltext{P}{Sydney Institute for Astronomy, School of Physics A28, University of Sydney, NSW 2006, Australia}

\begin{abstract}
We cross-match the two currently largest all-sky photometric catalogs, mid-infrared WISE and SuperCOSMOS scans of UKST/POSS-II photographic plates, to obtain a new galaxy sample that covers $3\pi$ steradians. In order to characterize and purify the extragalactic dataset, we use external GAMA and SDSS spectroscopic information to define quasar and star loci in multicolor space, aiding the removal of contamination from our extended-source catalog. After appropriate data cleaning we obtain a deep wide-angle galaxy sample that is approximately 95\% pure and 90\% complete at high Galactic latitudes. The catalog contains close to 20 million galaxies over almost 70\% of the sky, outside the Zone of Avoidance and other confused regions, with a mean surface density of over 650 sources per square degree. Using multiwavelength information from two optical and two mid-IR photometric bands, we derive photometric redshifts for all the galaxies in the catalog, using the \ANNz\ framework trained on the final GAMA-II spectroscopic data. Our sample has a median redshift of $z_\mrm{med} = 0.2$ but with a broad $\de N/\de z$ reaching up to $z>0.4$. The photometric redshifts have a mean bias of $|\delta z|\sim10^{-3}$, normalized scatter of $\sigma_{z} = 0.033$ and less than 3\% outliers beyond $3\sigma_{z}$.  {Comparison with external datasets shows no significant variation of \phz\ quality with sky position.} Together with the overall statistics, we also provide a more detailed analysis of photometric redshift accuracy as a function of magnitudes and colors. The final catalog is appropriate for `all-sky' 3D cosmology to unprecedented depths, in particular through cross-correlations with other large-area surveys. It should also be useful for source pre-selection and identification in forthcoming surveys such as TAIPAN or WALLABY.
\end{abstract}


\section{Introduction}
\setcounter{footnote}{0}

Direct mapping of the three-dimensional distribution of galaxies in the Universe requires their angular coordinates and redshifts. Dozens of such wide-angle galaxy redshift catalogs now exist, the most notable of which include the Sloan Digital Sky Survey \citep[SDSS,][]{SDSS}, the Two-degree Field Galaxy Redshift Survey \citep[2dFGRS,][]{2dF}, or the Six-degree Field Galaxy Survey \citep[6dFGS,][]{6dF}. 

For some applications, it is an advantage if the survey can cover the majority of the sky: for example,
searches for violation of the Copernican principle in the form of large-scale inhomogeneities or anisotropies \citep{GH12,AS14,Alonso15,YH15} and coherent motions \citep{BCJM11,BDN12,Carrick15}, as well as cross-correlations of galaxy data with external wide-angle datasets. Examples of the latter are studies of the integrated Sachs-Wolfe effect \citep[see][for a review]{Nishizawa15}, of gravitational lensing of the cosmic microwave background (CMB) on the large-scale structure \citep{LC06}, or searches for sources of the extragalactic $\gamma$-ray background \citep[e.g.][]{Xia15}, including constraints on annihilating or decaying dark matter \citep{Cuoco15}. These analyses are limited by cosmic variance, and also frequently much of the signal lies at large angular scales -- both factors that make it desirable to have the largest possible sky coverage.

But there is a practical limit to the number of spectroscopic redshifts that can be measured in a reasonable time. Spectroscopic galaxy catalogs covering the \textit{whole} extragalactic sky, \eg the IRAS Point Source Catalog Redshift Survey \citep[PSCz,][]{PSCz} and the 2MASS Redshift Survey \citep[2MRS,][]{2MRS}, thus tend to be relatively shallow ($z < 0.1$) -- and the same applies to hemispherical samples such as the 6dFGS. This problem can be addressed by using only rare tracers, as with the highly successful BOSS program \citep{BOSS} or planned projects such as the Dark Energy Spectroscopic Instrument (DESI, \citealt{DESI}) or Wide Area VISTA Extra-galactic Survey (WAVES, \citealt{WAVES}) within the 4MOST program; but for many applications it is desirable to have a fully-sampled galaxy density field. For that reason, new wide-field surveys such as the Dark Energy Survey \citep{DES}, Pan-STARRS \citep{Pan-STARRS} or the Kilo-Degree Survey \citep{KiDS} focus on measuring the photometric properties of objects, with only a partial spectroscopic follow-up. In the longer term, the same will apply to forthcoming multi-billion-object facilities including Euclid \citep{Euclid} and the Large Synoptic Survey Telescope \citep{LSST}.  {Lying somewhat in between the spectroscopic and photometric surveys, the currently starting Javalambre-Physics of the Accelerated Universe Astrophysical Survey \citep[J-PAS,][]{J-PAS} is expected to reach sub-percent redshift precision over $\sim8000$ \dsqu, thanks to the usage of $56$ narrow-band filters. Of a similar nature, but aiming to cover 100 \dsqu\ to a greater depth than J-PAS, is the Physics of the Accelerating Universe survey \citep[PAU,][]{PAU}.}

In order for such surveys to yield cosmological information of comparable or even better quality than from traditional spectroscopic samples, one needs to resort to the technique of \textit{photometric redshifts} (\phzs). In the near future, this approach will dominate those cosmological analyses where the benefit from larger volumes outweighs the loss of redshift accuracy. Although some small-scale analyses are not feasible with the coarse accuracy of \phz\ estimation (typically a few \% precision), there are many applications where this level of measurement is more than adequate. This is particularly true when there is an angular signal that changes slowly with redshift, requiring a \textit{tomographic} analysis in broad redshift bins (\eg \citealt{FP10}); but until recently the necessary \phz\ information has only been available for relatively shallow subsamples of all-sky catalogs. 

To improve on this situation, in \citet[][hereafter \tcb{B14}]{2MPZ} we combined three all-sky photometric samples -- optical SuperCOSMOS, near-infrared 2MASS and mid-infrared WISE -- into a multiwavelength dataset. We used various spectroscopic calibration samples to compute photometric redshifts for almost 1 million galaxies over most of the extragalactic sky: the 2MASS Photometric Redshift catalog (2MPZ)\footnote{Available for download from the Wide Field Astronomy Unit, Edinburgh, at \url{http://surveys.roe.ac.uk/ssa/TWOMPZ}}.  The 2MPZ is currently the deepest three-dimensional full-sky galaxy dataset, with a median redshift of $z\simeq0.1$ and a typical uncertainty in photometric redshift of about $12\%$ (scatter $\sigma_{z} = 0.013$). Ideally, these estimates should be superseded by actual spectroscopy -- and recently prospects have emerged for this to happen, thanks to the new hemispherical TAIPAN survey \citep{TAIPAN} in the South, starting in 2016, as well as the recently proposed LoRCA \citep{LORCA} in the North. These efforts, if successful, will provide spectroscopic information for all the 2MASS galaxies which do not have redshifts, although at their planned depths ($r\lesssim 18$ for the former and $K_s<14$ for the latter) they will not replace the need for the catalog presented in the current paper. We note, however, the \textit{SPHEREx} concept by \cite{SPHEREx} to probe much deeper on most of sky.

The depth of 2MPZ is limited by the shallowest of the three photometric surveys combined for its construction,  the 2MASS Extended Source Catalog \citep[XSC,][]{2MASSXSC,XSCz}. However, as was shown in \tcb{B14}, one can go beyond the 2MASS data and obtain a much deeper all-sky \phz\ catalog based on WISE and SuperCOSMOS only. In \tcb{B14} we predicted that such a sample should have a typical redshift error of $\sigma_{z}\simeq0.035$ at a median $z\simeq0.2$ (median relative error of $14\%$). The construction of this catalog is the focus of the present paper, and indeed we confirm and even exceed these expectations on the \phz\ quality. We note that in a related effort \cite{KoSz15} presented a wide-angle sample deeper than the 2MASS XSC, based on WISE and the 2MASS \textit{Point} Source Catalog (PSC). Its depth is, however, still limited by 2MASS: PSC has an order of magnitude smaller surface density than WISE \citep{Jarrett16}. Overall, the \cite{KoSz15} sample includes 2.4 million sources at $\zmed\simeq0.14$ over half of the sky, of which 1/3 are in common with the 2MASS XSC. Here we  map the cosmic web to much higher redshifts than can be accessed with 2MASS, yielding a third shell of presently available all-sky redshift surveys. The first, with exact spectroscopic redshifts at $\zmed=0.03$, is provided by the 2MRS, flux-limited to $K_s\leq11.75$ (Vega) and containing 44,000 galaxies at $|b|>5\degree$ ($|b|>8\degree$ by the Galactic Bulge). The second is the 2MPZ, which includes almost a million 2MASS galaxies at $K_s<13.9$ with precise \phzs\ at $\zmed=0.07$, based on 8-band 2MASS\ti{}WISE\ti{}SuperCOSMOS photometry. This present work concerns 20 million galaxies with $\zmed=0.2$, thus reaching three times deeper than 2MPZ, over $3\pi$ steradians of the sky outside the Galactic Plane.

This paper is organized as follows. In \S\ref{Sec: Contributing catalogs} we provide a detailed description of the catalogs contributing to the sample and their cross-matching. In \S\ref{Sec: x-matches with GAMA} we analyze the properties of the input photometric datasets by pairing them up with GAMA spectroscopic data. \S\ref{Sec: Catalog cleanup} describes the use of external GAMA and SDSS spectroscopic information to remove quasars and stellar blends from the cross-matched catalog. The construction of the angular mask to be applied to the data is also presented there in \S\ref{Sec: Mask}. Next, in \S\ref{Sec: Photometric redshifts} we show how photometric redshifts were obtained for the sample and discuss several tests of their performance; \S\ref{Sec: All-sky photo-z catalog} discusses the properties of the final all-sky catalog. In \S\ref{Sec: Summary} we summarize and list selected possible applications of our dataset.

\vspace{2mm}
\section{Contributing catalogs}
\label{Sec: Contributing catalogs}

The galaxy catalog presented in this paper is a combination of two major photometric surveys of the whole celestial sphere: optical SuperCOSMOS  scans of photographic plates (SCOS for short) and mid-IR WISE. Each of these two datasets includes about 1 billion sources, a large fraction of which are extragalactic. WISE is  deeper than SCOS, but its poorer resolution and lack of morphological information (the latter available from SCOS) prevents the selection of galaxies without the optical criterion of an extended image. Pairing up these two datasets  thus provides a natural means of obtaining a deep wide-angle extragalactic sample, as we proposed already in \tcb{B14}. With appropriate spectroscopic calibration data, the wide wavelength range yields robust photometric redshifts for each of the \WISC\ galaxies.

In this Section we describe the properties of the underlying photometric catalogs and the preselections applied to them. We aim at the highest depth possible for the cross-matched sample, while optimizing its reliability, purity and completeness. By reliability we chiefly refer to the quality of the photometry; purity refers to the percentage of our sources that are indeed galaxies and not stars, high redshift quasars nor blends thereof; completeness is the fraction of all galaxies that are included in the catalog, within adopted magnitude limits. As our focus in the present paper is to derive photometric redshifts for all the galaxies in our sample, which requires multi-wavelength coverage, we select from the two catalogs only those sources that have detections in at least four bands: $W1$ and $W2$  (3.4 and 4.6 $\mu$m) in WISE, $B$ and $R$ in SCOS. The additional bands available from the two surveys, $W3$ and $W4$ (12 and 23 $\mu$m)\footnote{The $W4$ channel effective wavelength was recalibrated from the original 22 $\mu$m by \cite{BJC14}.} from WISE, and $I$ from SCOS, are not used due to their low sensitivity and non-uniform sky coverage.

\begin{figure*}
\centering
\includegraphics[width=\textwidth]{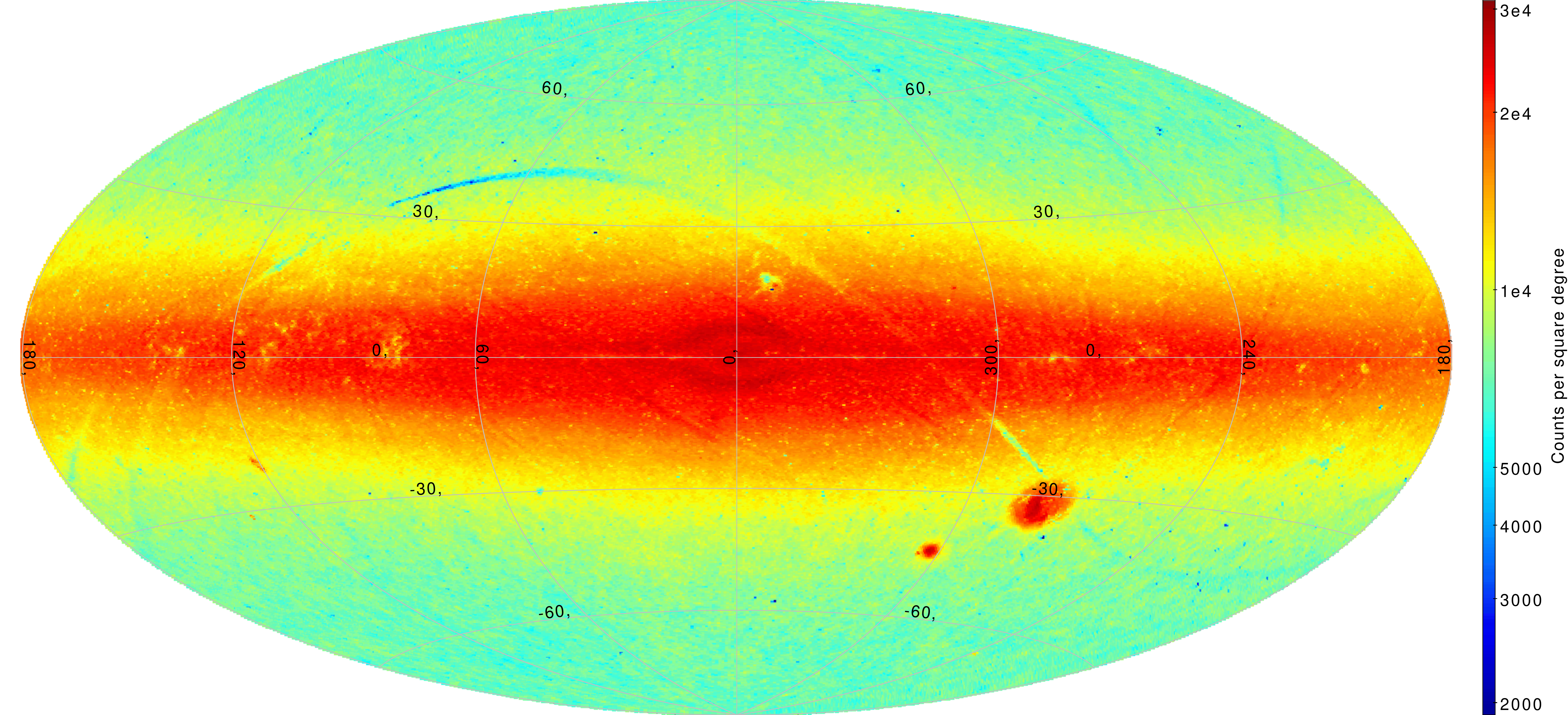}
\caption{\label{Fig: WISE.allsky}WISE all-sky Aitoff map, in Galactic coordinates, of 488 million sources preselected from AllWISE with $W1<17$, before cross-match with SuperCOSMOS and purification. This sample contains both galaxies and stars, and the latter dominate at low latitudes. The missing data in a strip crossing the Galactic plane is due to saturation in $W1$ at the onset of the post-cryogenic phase and can be supplemented by using only data from the cryogenic stage in this region. The color bar shows counts per square degree at each pixel.}
\end{figure*}

This exercise cannot be expected to yield a fully all-sky catalog: both WISE and SCOS suffer at low Galactic latitudes from severe blending of stars with other stars and with galaxies, and high Galactic extinction levels effectively censor the optical bands. In \S\ref{Sec: Catalog cleanup} we will discuss how to minimize such  foreground contamination, and will develop a mask within which the overall catalog has an acceptable completeness and purity. In practice, we find that this can be done over about 70\% of the sky.

\subsection{WISE}
\label{Sec: WISE}

The Wide-field Infrared Survey Explorer (WISE; \citealt{WISE}) is a NASA space-based mission that surveyed the celestial sphere in four infrared bands: 3.4, 4.6, 12 and 22 $\mu$m ($W1$ -- $W4$), with angular resolution of $6.1''$, $6.4''$, $6.5''$ and $12''$, respectively. In our work we use the `AllWISE' full-sky release\footnote{Available for download from NASA/IPAC Infrared Science Archive at \url{http://irsa.ipac.caltech.edu}.} \citep{AllWISE}, which combines data from the cryogenic and post-cryogenic survey phases and provides the most comprehensive picture of the full mid-infrared sky currently available. The AllWISE Source Catalog and Image Atlas have enhanced sensitivity and accuracy compared with earlier WISE data releases, especially in its two shortest bands. This results in a larger effective depth than available from an earlier `All-Sky' release \citep{WISEAllSky}, used \eg in \tcb{B14}.  AllWISE includes over 747 million sources (mostly stars and galaxies) detected with $S/N\geq5$ in at least one band. The $5\sigma$ sensitivities  in the four respective bands are approximately\footnote{\url{http://wise2.ipac.caltech.edu/docs/release/allwise/expsup/sec2\_3a.html}} 0.054, 0.071, 0.73 and 5 mJy, and its 95\% completeness averaged over large areas of unconfused sky is about\footnote{\url{http://wise2.ipac.caltech.edu/docs/release/allwise/expsup/sec2\_4a.html}} $W1<17.1$, $W2<15.7$, $W3<11.5$ and $W4<7.7$ in the Vega system\footnote{Conversions of WISE magnitudes from Vega to AB are provided by \cite{Jarrett11}; for the bands of interest in this paper, $W1$ and $W2$, one needs to add respectively $2.70$ and $3.34$ to the Vega magnitudes to switch to the AB system.}. The depth of coverage does, however, vary over the sky due to the survey strategy, being much higher in the ecliptic poles and the lowest near the ecliptic plane \citep{Jarrett11}; there are also some anomalous stripes resulting from Moon avoidance maneuvers and instrumental issues. 

The WISE photometric pipeline was not optimized for extended sources and the online database does not include a formal extended source catalog. The basic magnitudes (which we use here) are the \ttt{w?mpro} mags, based on PSF profile-fit measurements, where `\ttt{?}' stands for the particular channel number, from 1 to 4. This information is available for all objects, whereas existing attempts to handle extended images are somewhat heterogeneous. For instance, the \ttt{w?gmag}s, which are measured in elliptical apertures derived from associated 2MASS XSC sources, are  available only for the 483,000 largest WISE galaxies.  Circular aperture magnitudes are in fact provided for practically all sources, namely the \ttt{w?mag\_n}, where $n=1,2,...,8$; these were obtained from the coadded Atlas images in a series of different fixed radii. But the angular sizes of the sources have not been determined; in addition, this photometry does not account for source ellipticities, is prone to contamination from nearby objects, and is not compensated for saturated or missing pixels in the images. 

In any case, as we eliminate all the  bright ($W1<13.8$) sources from our cross-matched catalog (see \S\ref{Sec: Catalog cleanup}), we are thus left with galaxies typically smaller than the WISE resolution threshold, which are well-described by PSF magnitudes{, although we note that their fluxes might be underestimated by WISE}. This is supported by independent analyses showing that the eventual WISE XSC will include mostly 2MASS XSC galaxies and be limited to $W1\lesssim14$ \citep{Cluver14,Jarrett16}. In any case, any residual biases in photometry for resolved sources, which may influence for instance source colors, will not be propagated to the photometric redshifts derived via the neural network framework employed here, as such systematics are automatically accounted for in the empirical training procedure.

Initially, we selected AllWISE sources with signal-to-noise ratios larger than 2 in its two shortest bands. This selection, meaning that we use detections in the two bands and not upper limits (the latter having $S/N<2$ in WISE),  is practically equivalent to selecting objects with $\mathtt{w1snr}\geq5$, as those with low $S/N$ in $W1$ but high in $W2$ are extremely rare. Having cleaned the sample of obvious artifacts (\ttt{cc\_flags[1,2]$=$`DPHO'}) and saturated sources ($\mathtt{w?sat}>0.1$), we ended up with more than 603 million AllWISE objects over the whole sky. In order to optimize all-sky uniformity, we applied a global magnitude cut of $W1<17$. This removes $\sim20\%$ of AllWISE (mostly around the ecliptic poles, where the WISE depth is greatest), leaving 488 million objects (pictured in Fig.\ \ref{Fig: WISE.allsky}). From this image, it is apparent that low Galactic latitudes are entirely dominated by stars and blends thereof; as we will show below, stellar contamination remains significant even at high latitudes (see also \citealt{Jarrett11,Jarrett16}). A minimal Galactic restriction to $|b|>10\degree$ lowers the total to 340 million sources (see Table \ref{Table surveys} for a summary), but we will show that the final masking needs to be more severe than this.

Note that  some sources observed during the early three-band cryo survey phase are not captured by the above selection, as they have missing $W1$ magnitude uncertainties and are listed as upper limits in the database. This is discussed in detail in the AllWISE Explanatory Supplement\footnote{\url{http://wise2.ipac.caltech.edu/docs/release/allwise/expsup/sec2\_2.html\#w1sat}} and applies mostly to two strips within ecliptic longitudes of $44.7\degree < \lambda < 54.8\degree$ or $230.9\degree < \lambda < 238.7\degree$ (visible in Fig.\ \ref{Fig: WISE.allsky}). This will be rectified in our final galaxy sample cross-matched with SuperCOSMOS by adding data from the earlier WISE data release, `All-Sky' \citep{WISEAllSky}. Some other issues are caused by variable coverage due to Moon avoidance maneuvers, which results in several under- or oversampled stripes crossing the Ecliptic\footnote{\url{http://wise2.ipac.caltech.edu/docs/release/allwise/expsup/sec4\_2.html\#lowcoverage}}. 

Galactic extinction corrections are very small in the WISE bands, over an order of magnitude smaller than in the optical, which does not mean they are totally negligible. Following \cite{Indebetouw05} and \cite{SF11}, we use $A_{W1}/E(B-V)=0.169$ and $A_{W2}/E(B-V)=0.130$ as coefficients to be applied to the original \cite{SFD} maps; these values in part implement a general recalibration of the original $E(B-V)$ values, which need to be lowered by 14\% \citep{SF11}.

\begin{deluxetable*}{lccc}
\tabletypesize{\footnotesize}
\tablewidth{0pt}
\tablecolumns{4} 
\tablecaption{\label{Table surveys}Statistics of the parent photometric catalogs and the final WISE\ti{}SuperCOSMOS cross-match used in this paper.}
\tablehead{ 
\colhead{ catalog } & 
\colhead{ flux limit(s) } & 
\colhead{ sky cut } & 
\colhead{ \# of sources }
}
\startdata
{WISE} & { none } & { none } & { $604 \times 10^6$ } \\
{(preselected in $W1$ and $W2$)} & { none } & { $|b|>10^\circ$ } & { $457 \times 10^6$ } \\
{  } & { $W1<17$ } & { none } & { $488 \times 10^6$ } \\
{  } & { $W1<17$ } & { $|b|>10^\circ$ } & { $343 \times 10^6$ } \\
\hline \\
{SuperCOSMOS XSC} & { none } & { none } & { $ 288 \times 10^6$ } \\
{(preselected in $B$ and $R$)} & { none } & { $|b|>10^\circ$ } & { $ 158 \times 10^6$ } \\
{  } & { $B<21$ \& $R<19.5$ } & { none } & { $ 208 \times 10^6$ } \\
{  } & { $B<21$ \& $R<19.5$ } & { $|b|>10^\circ$ } & { $ 85.1 \times 10^6$ } \\
\hline \\
{WISE\ti{}SuperCOSMOS XSC}  & { none } & { none } & { $ 109 \times 10^6$ } \\
{ } & { none } & { $|b|>10^\circ$ } & { $ 78.3  \times 10^6$ } \\
{ } & {  $W1<17$ \& $B<21$ \& $R<19.5$ } & { none } & { $ 77.9  \times 10^6$ }\\
{ } & {  $W1<17$ \& $B<21$ \& $R<19.5$ } & { $|b|>10^\circ$ } & { $ 47.7  \times 10^6$ }\\
{\qquad after star \& quasar cleanup } & {  $13.8<W1<17$ \& $B<21$ \& $R<19.5$ } & { $|b|>10^\circ$ + Bulge, masked } & { $ 18.8 \times 10^6$} \\
\hline \\
{galaxies in WISE, not in SCOS XSC} & { $W1<17$ } & { $|b|>10^\circ$ } & { $\sim100 \times 10^6$}
\enddata
\end{deluxetable*}


\subsection{SuperCOSMOS}
\label{Sec: SuperCOSMOS}

The SuperCOSMOS Sky Survey \citep[SCOS,][]{SCOS3,SCOS2,SCOS1} was a program of automated scanning and digitizing of sky atlas photographic plates in three bands ($B,R,I$), the source material having been obtained in the last decades of the twentieth century with the United Kingdom Schmidt Telescope (UKST) in the South and the Palomar Observatory Sky Survey-II (POSS-II) in the North. The data are stored in the SuperCOSMOS Science Archive\footnote{Available for download from \url{http://surveys.roe.ac.uk/ssa/}.}, with multicolor information provided for 1.9 billion sources, in the form of integrated quasi-total and point-source photometry (where available). The derived resolved-source data were accurately calibrated using SDSS photometry when possible, with the calibration extended over the full sky by matching plate overlaps and by using the average color between the optical and 2MASS $J$ bands as a constraint to prevent large-scale drifts in zero point \citep{FP10,Peacock16}. The typical resolution of SuperCOSMOS images is $\sim2''$ \citep{SCOS2} and the photometric depth $R\simeq19.5$, $B\simeq 21$ in a pseudo-AB system, in which SCOS and SDSS coincide for objects with the color of the primary SDSS standards; detailed color equations are given in \cite{Peacock16}. The third band available from the catalog, $I$, offers shallower coverage and will not be used here.

For the present work, we are interested only in resolved images. SCOS supplies a classification flag for every image in each of the three bands, as well as a combined one, \ttt{meanClass}. These are equal to 1 if a source is non-stellar, 2 if it is consistent with unresolved, 3 if unclassifiable and 4 if likely to be noise; the two latter classes constitute a negligible fraction of all the sources ($\ll 1\%$) in any given plate.  {The image classification is based on image morphology via a two-stage process \citep[][and references therein]{SCOS2}. The first stage uses image surface  brightness, size and shape to identify isolated, point-like images with good reliability and high completeness. The second stage takes this first-pass selection and analyses the 1D radial profile of those unresolved images as a function of plate position and source brightness. Finally every image is assigned a profile statistic~$\eta$, the probability distribution of which has zero mean and unit variance, to quantify the point-likeness. This continuously distributed statistic is used to define discrete classification codes when cut at fixed thresholds: sharper--than--point-like images with $\eta < -3$ are assigned class=4 (noise); point-like images with $-3 < \eta < 2.5$ are assigned class=2 (stellar); and resolved images having $\eta > 2.5$ are assigned class=1 (non-stellar). Where data from two or more plates are available for the same image, the individual profile statistics are averaged to form a single zero mean, unit variance statistic via $\overline{\eta}=\Sigma\eta/\surd{N}$ for $N$ plates with a discrete merged classification code \ttt{meanClass} assigned using the same ranges as above for the individual class codes.} For brighter images with good detections in all bands, this increases the precision of classification; but for faint objects lacking good $I$-band data, this overall classification may be less reliable than the $B$ or $R$ plates individually. But the data we use here have cuts that eliminate the faintest objects, so we have chosen to use the \ttt{meanClass} parameter in all cases. We have verified by comparison with SDSS test regions that this choice leads to better star-galaxy separation than using the individual $B$ and $R$ classes. 

\begin{figure*}
\centering
\includegraphics[width=\textwidth]{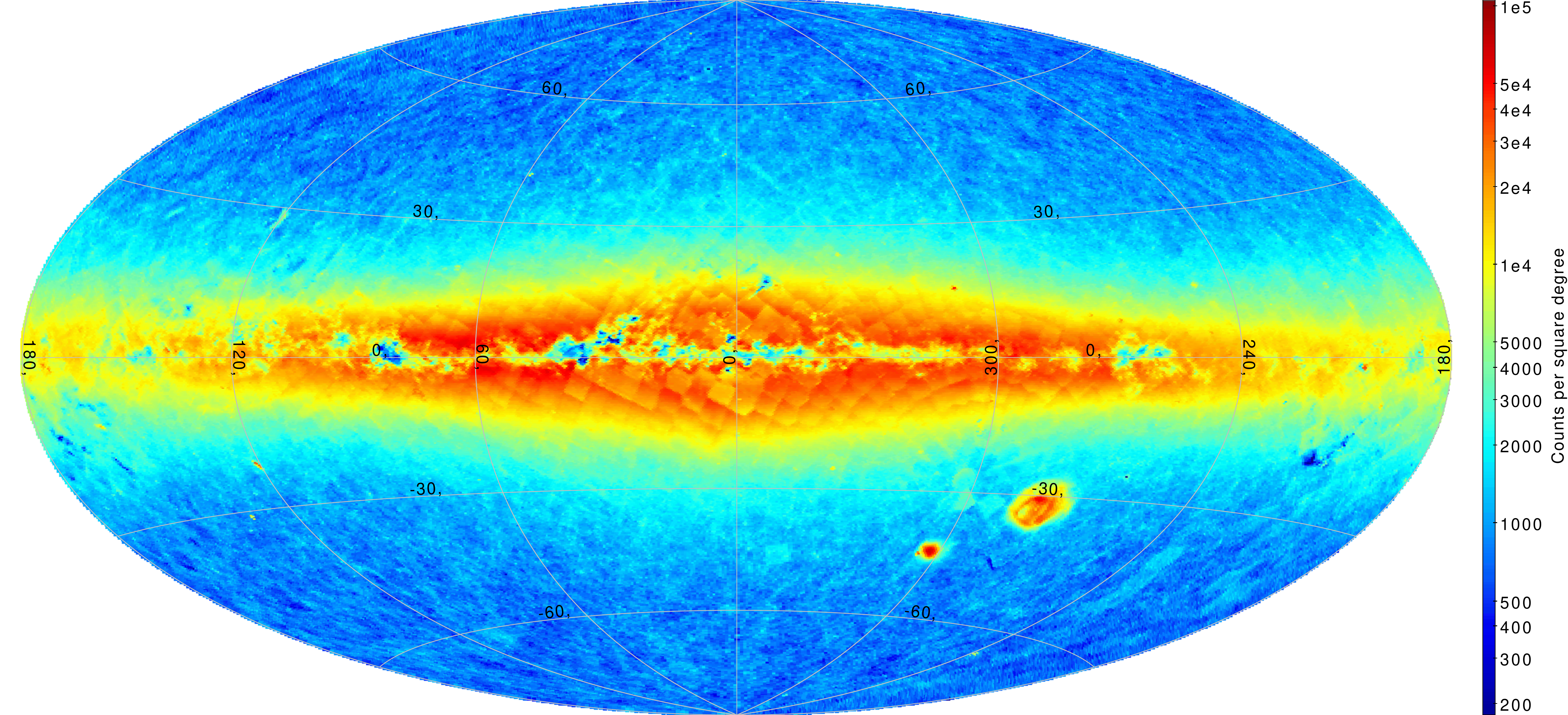}
\caption{\label{Fig: SCOS.allsky}SuperCOSMOS all-sky Aitoff map, in Galactic coordinates, of 208 million extended sources preselected with $B<21$ and $R<19.5$, before cross-match with WISE and purification. The spurious overdensities in the Galactic Plane and at the Magellanic Clouds arise due to star blending. The color bar shows counts per square degree at each pixel.}
\end{figure*}

Note that the classification flags also affect the photometric calibration procedure \citep{SCOS2}: separate calibrations were applied for stars and galaxies. This is mainly because of the limited dynamic range of SCOS when compared to, for example, some of the much slower modified PDS scanning machines or the highly optimised `flying spot' APM system (\citealt{SCOS1} and references therein). SCOS employed a linear CCD in the imaging system and a strip of emulsion was therefore illuminated to quickly scan lanes of $\sim1$~cm width. When scanning over denser spots in an otherwise less dense emulsion the core density measured was limited by light from the entire illuminated strip diffracting in the imaging lens. This was not the case for extended objects since the amount of light subject to diffraction was significantly reduced. For stars the diffraction limit of the measurement process occurred at much lower densities than any emulsion saturation in the photographic emulsions themselves. Hence the calibration curve of instrumental magnitude versus externally measured magnitude bifurcated into separate star and galaxy loci only a magnitude or so above the plate limit despite both the point and extended images being well exposed on the log--linear part of the photographic response curve. In any case, the galaxy calibration was performed at a later date \citep{Peacock16}, following the wider availability of SDSS photometry.

Because of a slight difference between the passbands of the UKST and POSS-II, there is in effect a small color-dependent offset in the SCOS magnitudes between the North and the South (here meaning above and below $\delta_{1950}=2.5\degree$). As discussed in \tcb{B14}, direct corrections were designed by comparison with SDSS to compensate for this effect. The following appropriate formulae (revised over \tcb{B14}) aim to correct the Southern $B$ \& $R$ data ($\delta_{1950}<2.5\degree$) to be consistent with the North\footnote{Unfortunately, the  corresponding equations in \tcb{B14} [eqs.\  (1--2) therein] are incorrect, owing to an inadvertent swapping of North and South. A revised version of the 2MPZ catalog will be issued that incorporates this correction.}:
\begin{equation}\label{Eq: Bcal}
B_{\rm S}^{\rm cal} = B + 0.03 (B - R)^2 - 0.005 (B - R)\; ,
\end{equation}
\begin{equation}\label{Eq: Rcal}
R_{\rm S}^{\rm cal} = R + 0.03 (B - R)^2 - 0.06 (B - R) + 0.015 \; .
\end{equation}
However, even these corrections may not fully guarantee N$-$S uniformity:
within our fiducial flux limits, the mean high-latitude surface density in the North is up to 4\% larger than in the South; it is hard to be sure whether this is a remaining very small calibration offset or genuine cosmic variance.  {On the other hand, these offsets do not induce significant additional  scatter to the corrected magnitudes. For typical galaxy colors, $B-R\sim1$, by error propagation in Eqs.\ \eqref{Eq: Bcal}--\eqref{Eq: Rcal} we see that the random error in $B_{\rm S}^{\rm cal}$ is increased by less than 6\% with respect to the original values, while for $R_{\rm S}^{\rm cal}$ there is a fortuitous cancellation and the error is not changed at all. For a general discussion of SCOS magnitude errors, see \cite{Peacock16}.}

We have also revised the extinction corrections used in 2MPZ: a series of papers using SDSS \citep{Schlafly10,SF11} and Pan-STARRS data  \citep{Schlafly14} show that the original \cite{SFD} maps overestimate the $E(B-V)$ values by roughly 14\% and that one should use the \cite{Fitzpatrick} reddening coefficients rather than the \cite{CCM89} ones. Based on the revised extinction coefficients for the SDSS $g$ and $r$ bands \citep{SF11}, the new corrections for the $B$ and $R$ SCOS bands are, respectively, $A_B/E(B-V)=3.44$ and $A_R/E(B-V)=2.23$ \citep{Peacock16}, for the full sky\footnote{Note that in \tcb{B14} we incorrectly provided different extinction corrections for the two hemispheres; as the magnitudes had already been calibrated N--S, one should use a single coefficient (N) in a given band for the full sky.}. These numbers already incorporate the rescaling of the $E(B-V)$ values by \cite{SF11}: they should thus be applied to the original \cite{SFD} $E(B-V)$ to obtain band-dependent extinction corrections for a given galaxy in magnitudes. In what follows, all the quoted SCOS magnitudes will refer to hemisphere-calibrated and extinction-corrected values, in the AB-like system. 

\begin{figure*}
\centering
\includegraphics[width=\textwidth]{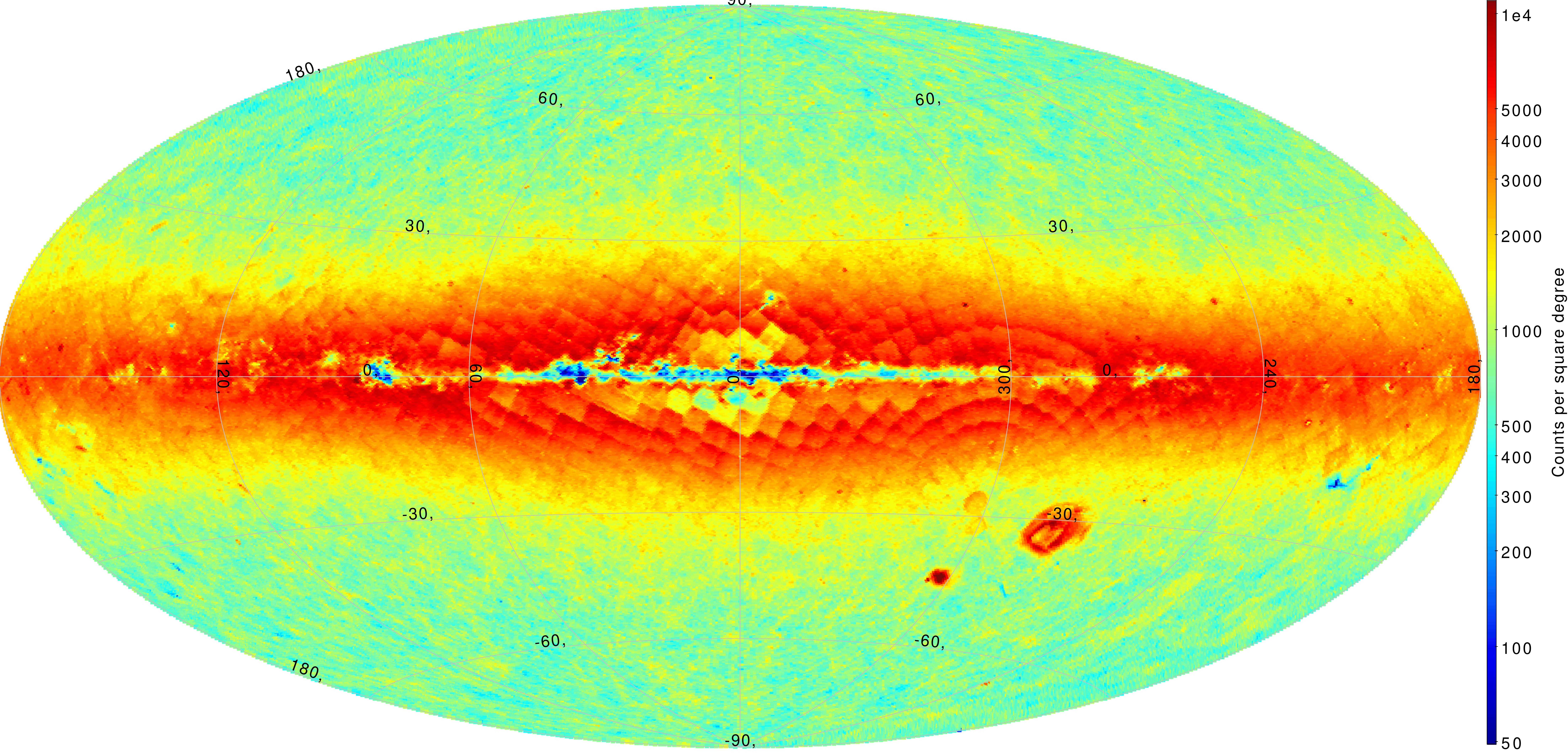}
\caption{\label{Fig: WIxSC.noclean}WISE\ti{}SuperCOSMOS cross-matched catalog of extended sources, before purification of stars and masking, in an all-sky Aitoff map in Galactic coordinates. The map contains 78 million objects flux-limited to $B<21$ \& $R<19.5$ \& $W1<17$. Low latitudes and Magellanic Clouds are dominated by star blends mimicking extended sources. The color bar shows counts per square degree at each pixel.}
\end{figure*}

For the purposes of the present work, our requirements for SCOS preselection were that the sources be properly detected with aperture photometry in $B$ and $R$ bands: \ttt{gCorMagB} and \ttt{gCorMagR2} not null in the database, quality flags \ttt{qualB} and $\mathtt{qualR2}<2048$ (no strong warnings nor severe defects: \citealt{SCOS2}). In addition, as described above, we used the sources with SCOS morphological classification flag $\mathtt{meanClass}=1$. This selection greatly enhances the \textit{purity} of our final cross-matched sample, by eliminating most of the stars from unconfused regions, as well as many quasars (see further discussion on these issues in \S\ref{Sec: Catalog cleanup} below). On the other hand, it only slightly reduces the \textit{completeness} of the catalog, removing less than 1\% of galaxies, which we estimated based on GAMA and SDSS galaxies cross-matched to our data.  As with WISE, also from SCOS we will not be using low-latitude sources in the present work (almost 50\% of SCOS `extended' sources are in the $|b|<10\degree$ strip -- mostly blends of stars). On the other hand, we have supplemented our catalog over what is publicly available by adding sources originally omitted from the SCOS catalogs due to areas excluded around stepwedges, which affected mostly plate corners (564,000 objects in our case).

The above selections in SCOS resulted in the `SCOS extended source catalog' (XSC), with about 158 million sources at $|b|>10\degree$. Owing to remaining low-latitude stellar blends, only a part of these sources are actually extragalactic. Simple cross-matching of this catalog with AllWISE would give a highly contaminated sample, therefore extra effort was needed to derive the best possible purity and completeness criteria for our eventual catalog. This is discussed in \S\ref{Sec: Catalog cleanup}.

As far as reliability is concerned, the main limitation here and for the cross-matched catalog is the depth of the SCOS data. We decided to adopt $B<21$ and $R<19.5$ as the optical limits, motivated by our analysis of galaxy counts from direct comparison with very deep SDSS photometric data (\citealt{SDSS.DR10}; see also \citealt{Peacock16}). Applying these magnitude cuts to the $|b|>10\degree$ sample removes almost 50\% of the SCOS XSC there{, leaving 85 million sources}. Had we included the Galactic Plane data, the flux-limited sample would count almost 208 million objects (see Table \ref{Table surveys}). Their distribution is shown in Fig.\ \ref{Fig: SCOS.allsky}; in addition to the Galactic Plane, the Magellanic Clouds are also clearly dominated by spurious overdensities from star blends. The plate pattern is noticeable at low latitudes because the degree of blending varies with plate quality. Note also much wider dynamic range of the counts than in the case of WISE.

\subsection{WISE\ti{}SuperCOSMOS cross-match}
\label{Sec: WISE x SCOS x-match}

In the following, all the cross-matches will be performed within a radius of $2''$, unless otherwise specified\footnote{Catalog cross-matching was done using the TOPCAT/STILTS software \citep{TOPCAT,STILTS} available for download from \url{http://www.star.bristol.ac.uk/\~mbt/}.}. In the case of the \WISC\ cross-match, this radius is motivated by the large beam of the former ($\sim6''$ in the $W1$ band: \citealt{WISE}) and the angular resolution of the latter ($\sim2''$). The mean matching radius for the resolved \WISC\ sources that pair up is $0.54''\pm0.42''$, and less than 14\% of the cross-matched sources are separated by over $1''$. It is important to note that both surveys offer comparable, sub-arcsecond astrometric accuracy: $\lesssim0.15''$ for WISE \citep{WISE} and $\lesssim0.3''$ for SCOS \citep{SCOS3}. It is then highly unlikely for a source identified in the two catalogs and detected in the four bands used here to be spurious. 

As already mentioned, all the WISE-based magnitudes are in the Vega system, while the SCOS ones are AB-like. We will keep this convention also for source colors derived from the two catalogs. From this point on, all magnitudes are corrected for extinction as described earlier.

After selecting the AllWISE and SCOS objects as discussed above, the resulting cross-match at $|b|>10\degree$ gave us over 78 million sources if no flux limits were applied, of which almost 48 million were within (extinction-corrected) magnitude cuts of $W1<17$ \& $B<21$ \& $R<19.5$ (Table \ref{Table surveys}). These numbers include sources that were added to the sample from the earlier WISE release ('All-Sky') to remove the incompleteness in AllWISE data visible as undersampled strips in Fig.\ \ref{Fig: WISE.allsky} and discussed in \S\ref{Sec: WISE}, as well as the SCOS objects lost through stepwedge exclusion. Fig.\ \ref{Fig: WIxSC.noclean} shows the sky distribution of this flux-limited sample. One expects the angular distribution of extended (extragalactic) sources to be relatively uniform on the sphere, whereas here clearly the foreground Milky Way dominates the counts at low latitudes, as well as the Magellanic Clouds do at their respective positions. Although the contamination from stellar blends is much reduced with respect to the two parent catalogs considered individually, less than half of these sources are actually extragalactic, despite them being classified by SCOS as \textit{extended}. 

In order to purify this sample, in \S\ref{Sec: Catalog cleanup} we present the color cuts aimed at removing some problematic quasars (\S\ref{Sec: QSO removal}) and the remaining stars (\S\ref{Sec: Star removal}). We also describe the mask that needs to be applied to the data in order to remove regions where the stellar and other contaminations cannot be corrected (\S\ref{Sec: Mask}). However, first in \S\ref{Sec: x-matches with GAMA} we analyze the properties of the photometric catalogs used here by pairing them up with the GAMA spectroscopic sample. Table \ref{Table surveys} summarizes the surveys contributing to our sample for different flux and sky cuts, including the cross-match after removal of stars and quasars as described later in \S\ref{Sec: Catalog cleanup}.

\section{Properties of the input photometric catalogs: cross-match with GAMA}
\label{Sec: x-matches with GAMA}

In order to explore the properties of our input catalogs, we cross-matched them with the Galaxy And Mass Assembly (GAMA) data covering three equatorial fields. GAMA \citep{Driver09} is an ongoing multiwavelength spectroscopic survey of the low-redshift Universe:  its input catalog (including star and quasar removal) is discussed in \cite{Baldry10}; the tiling strategy is described in \cite{Robotham10}; while the spectroscopic pipeline is explained in \cite{Hopkins13}. \cite{Baldry14} present a fully automatic redshift code  (AUTOZ) developed to homogenize the redshift measurements and improve their reliability, and \cite{Liske15} discuss the accuracy of these new measurements in context. The dataset we use here, taken from GAMA-II (\ttt{TilingCat v43}, not publicly released yet), covers three GAMA Equatorial Regions, G09, G12 and G15, centered on 9h, 12h and 14.5h in right ascension, respectively. Each of these fields spans over $5\degree \times 12\degree$, which gives 180 \dsqu\ in total. This sample is preselected in the SDSS Petrosian $r$ magnitude, and within the limit of $r_\mrm{Petro} \leq 19.8$ its galaxy redshift completeness is 98.4\% \citep{Liske15}. This makes the catalog ideal for our purposes, as in the fields it covers it is deeper and more complete than our core flux-limited \WISC\ sample, and at the same time free from stellar and quasar contamination by construction. GAMA is also unique in comparison to other surveys as it offers a plethora of ancillary data and parameters derived by the team, and of particular usefulness for our purposes are some of the intrinsic properties of galaxies as presented by \cite{Taylor11} and more recently by \cite{Cluver14}. The latter paper focused in particular on sources common to GAMA and WISE  {in the equatorial fields.} 

\begin{figure}
\includegraphics[width=0.5\textwidth]{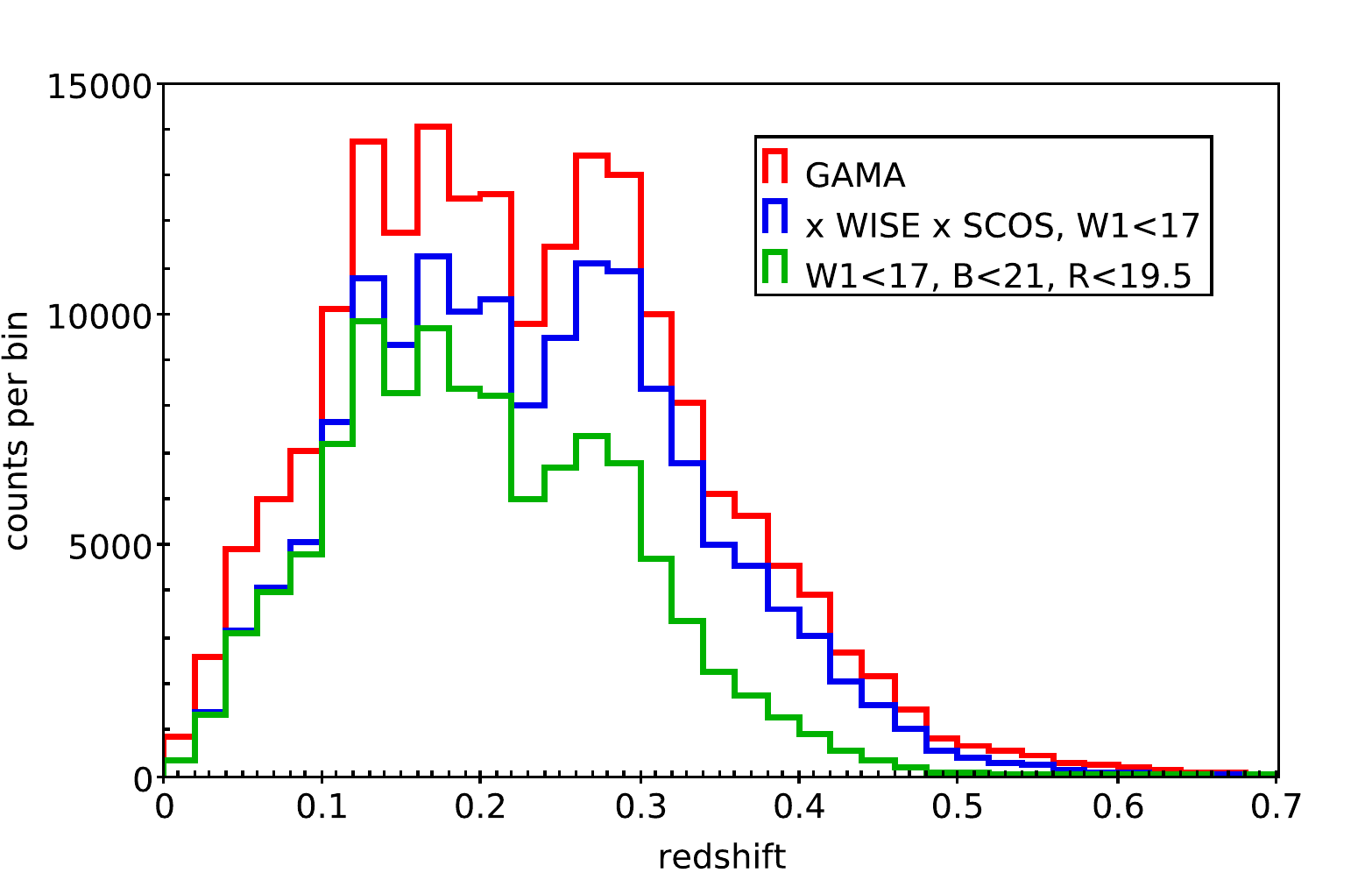} 
\caption{\label{Fig: GAMA x-match dNdz}Redshift distributions of GAMA (red line) and of its cross-matches with the WISE\ti{}SuperCOSMOS extended source catalog. Two flux limits for the cross-matches are shown: WISE-based only (blue line) and WISE+optical (green line).}
\end{figure}

\begin{deluxetable*}{lccc}
\tabletypesize{\footnotesize}
\tablewidth{0pt}
\tablecolumns{4} 
\tablecaption{\label{Table GAMA fields}Properties of photometric surveys in the GAMA equatorial fields and of their cross-matches with GAMA.}
\tablehead{ 
\colhead{ sample } & 
\colhead{ flux limit(s) } & 
\colhead{ \# of sources } &
\colhead{ $z_\mrm{med}$ }  
}
\startdata
{GAMA-II} & { none \tablenotemark{a} } & { $193,500$ \tablenotemark{b} } & { $0.23$ } \\
{  } & { $r\leq19.8$ } & { $183,000$ \tablenotemark{b} } & { $0.22$ } \\
\hline \\
{WISE \tablenotemark{c}  } & { none } & { $2,000,000$ } & {N/A} \\
{WISE\ti{}GAMA} & { none } & { $167,000$ } & { $0.23$ } \\
{GAMA but not WISE  } & { none } & { $26,500$ } & { $0.17$ } \\
\hline \\
{SCOS XSC \tablenotemark{c} } & { none } & { $484,000$ } & {N/A} \\
{ } & {  $B<21$ \& $R<19.5$ } & { $183,000$ } & {N/A}\\
{SCOS\ti{}GAMA} & {  $r\leq19.8$ (GAMA) } & { $167,000$ } & { $0.21$ } \\
{ } & { $B<21$ \& $R<19.5$ } & { $117,000$ } & { $0.19$ } \\
{GAMA but not SCOS} & {  $r\leq19.8$ (GAMA) } & { $16,000$ } & { $0.26$ } \\
\hline \\
{WISE\ti{}SCOS XSC \tablenotemark{c} } & { $W1<17$ } & { $294,000$ } & {N/A} \\
{ } & {  $W1<17$ \& $B<21$ \& $R<19.5$ } & { $151,000$ } & {N/A}\\
{WISE\ti{}SCOS\ti{}GAMA } & { $W1<17$ } & { $153,000$ } & { $0.22$ } \\
{  } & {  $W1<17$ \& $B<21$ \& $R<19.5$ } & { $109,000$ } & { $0.19$ }
\enddata
\tablenotetext{a}{Most of the sources are within the flux limit of $r\leq19.8$.}
\tablenotetext{b}{Preselected with $z>0.002$ and $\mathtt{NQ} \geq 3$.}
\tablenotetext{c}{In the GAMA equatorial fields.}

\end{deluxetable*}

The GAMA-II sample we use includes almost 203,000 sources with redshift measurements (some fainter than $r=19.8$). Of these, we have preselected confirmed galaxies ($z>0.002$) and with reliable redshifts (quality $\mathtt{NQ} \geq 3$). This gave us over 193,500 sources with $\zmed=0.22$; their redshift distribution is presented in Fig.\ \ref{Fig: GAMA x-match dNdz} (red line).  {This plot displays a dip at $z\simeq0.23$; this feature is observed in all the three equatorial fields at roughly the same redshift.} We interpret this as a coincidence in cosmic variance, as the three areas are too widely separated to trace the same large-scale structures.   {In fact, it is a projection effect mostly due to filaments and walls present in the 3 fields at $z\sim0.2$ and $z\sim0.26$, as can be seen in cone plots of \cite{Eardley15}, where environmental classification is also provided.} In addition, this pattern is not observed in the Southern GAMA fields (G02 and G23),
 {for which the spectroscopy was processed in the same way as for the equatorial ones,} so it cannot reflect an error in the redshift determination  {(cf.\ footnote \#10 in \citealt{Liske15}). The two additional fields available from GAMA-II are significantly less complete than the equatorial ones \citep{Liske15} and they will not be used in this part of the present work; we will however employ them for photometric redshift quality tests discussed in \S\ref{Sec: External redshifts}.}

A detailed analysis of WISE sources common with GAMA was presented in \cite{Cluver14}. Two of the three equatorial fields were studied there, and the WISE data originated from the earlier, `All-Sky' release \citep{WISEAllSky}. \cite{Cluver14} analyzed mid-infrared properties of GAMA galaxies, paying particular attention to characterizing and measuring resolved WISE sources. Many other issues were explored therein, in particular the empirical relations between optically determined stellar mass and the $W1$ and $W2$ measurements (using the synthetic stellar population models of \citealt{Taylor11}).

In the present work we use the updated AllWISE release together with the complete information in the three GAMA equatorial fields. Out of over 2 million AllWISE sources (of any kind) in these areas, our cross-match with GAMA gives almost 167,000 objects, which constitutes 86\% of the GAMA galaxy sample (see Table \ref{Table GAMA fields} for these and other details). This is a similar percentage to the one reported by \cite{Cluver14}, where a larger matching radius ($3''$) was used. The GAMA sources with no AllWISE counterparts are mostly faint and at lower redshifts ($\zmed=0.23, 0.17$ respectively for the matches and non-matches), \ie they are more local low-luminosity   galaxies. Some of the non-matches arise due to WISE blending GAMA galaxies at smaller angular separations than the beam of the former \citep{Jarrett16}.

The source density of AllWISE is some 10 times that of GAMA, and objects that are in AllWISE and not in GAMA belong to two general classes: either mostly bright, having colors consistent with stellar ones (\eg $W1-W2\lesssim 0$) -- stars filtered out by GAMA preselection -- or those at the the faint end ($W1>16$), where galaxies dominate over stars \citep{Jarrett11,Jarrett16}, with colors typical for an extragalactic population. Some are also quasars, which were eliminated from GAMA via morphological and color preselections \citep{Baldry10}. All this leads to the conclusion that a significant fraction of the unmatched AllWISE sources will be galaxies too faint for GAMA, and that $\zmed$ of the former should be significantly larger than that of the latter. This is further supported by the results of \cite{Jarrett16} where it is shown that the WISE\ti{}GAMA cross-match becomes incomplete for WISE galaxies fainter than $W1=15$ ($0.3$~mJy).

Next, we paired up the GAMA galaxy sample with the SCOS XSC. Here we have used only the $r_\mrm{Petro}\leq 19.8$ GAMA galaxies (183,000 with $z>0.002$ and $\mathtt{NQ} \geq 3$) to have a complete and unbiased sample. Not applying any flux limit on SCOS gave 9\% of GAMA without SCOS counterparts. The unmatched GAMA sources were mostly at high redshifts, with $\zmed=0.26$ (see Table \ref{Table GAMA fields}), in contrast to the AllWISE case -- confirming that the GAMA data are deeper than SCOS.   The SCOS magnitudes for the fainter GAMA galaxies have a substantial random error, hence to capture most of the true $r \leq 19.8$ GAMA objects one would need to go to SCOS $R \la 21$, beyond its reliability limit \citep{Peacock16}. Flux-limiting the SCOS sample to our fiducial values of $R<19.5$ \& $B<21$ resulted in  64\% of GAMA galaxies found also in the photographic data, with $\zmed=0.19$ for the matched sources, and over 90\% of unmatched GAMA galaxies having $r_\mrm{Petro}>19.2$.

Finally, we analyzed the \WISC\ cross-match in the three equatorial GAMA fields, focusing on the sources of interest for the present work, namely those resolved by SCOS. Out of 484,000 SCOS $\mathtt{meanClass} = 1$ sources in these areas, roughly 294,000 (61\%) had counterparts in AllWISE $W1<17$ if no magnitude cuts were applied to SCOS data. If we preselect SCOS as $R<19.5$, $B<21$, we end up with over 150,000 \WISC\ XSC objects, which is 83\% of the flux-limited extended SCOS sources. Of these two \WISC\ samples (with no SCOS magnitude limit and the flux-limited one), respectively 51\% and 71\% have GAMA counterparts. If no SCOS flux limit is applied, the median redshift of the WISE\ti{}SCOS\ti{}GAMA sample is $\zmed=0.22$ and decreases to $\zmed=0.19$ if only the $R<19.5, B<21$ sources are used; see Fig.\ \ref{Fig: GAMA x-match dNdz} for relevant redshift distributions and Table \ref{Table GAMA fields} for a summary. The sources present in the flux-limited \WISC\ resolved sample and not identified among GAMA galaxies are mostly bright and have colors (especially $W1-W2$) consistent with Milky Way stars, which illustrates the already mentioned fact that SCOS morphological classification is prone to misidentifying stellar blends as extended sources. 

The analysis of this Section has confirmed that the present GAMA data are appropriate for photometric redshift training of the wide-angle (`all-sky') \WISC\ catalog that we aim to produce. On the other hand, as is visible in Fig.\ \ref{Fig: GAMA x-match dNdz}, we cannot hope to reach beyond $z \simeq 0.45$ with our present sample due to the depth of the SCOS data; but WISE alone with no optical limit reaches up to $z\sim 1$ as shown in \cite{Jarrett16}. We plan to explore the latter property in future work.

\section{Purifying the WISE\ti{}SuperCOSMOS galaxy catalog}
\label{Sec: Catalog cleanup}

Despite preselecting the sources as extended in SCOS, our catalog will be contaminated with blended stellar images that masquerade as galaxies: this problem also affects WISE, and becomes more pronounced as we approach the Galactic plane. In addition, a number of high-$z$ quasars projected on more local galaxies will be present in the all-sky dataset{, thus contaminating the colors of the galaxies}. In this Section we propose relatively simple cuts to clean our data of this quasar and stellar contamination. To these one should also add an angular mask based on for instance Galactic extinction and star density, as well as encompassing such objects as the Magellanic Clouds, other large nearby galaxies or very bright stars. We discuss such a mask in \S\ref{Sec: Mask}. A separate work \citep{Krakowski16} will be devoted to another, machine-learning-based attempt at all-sky galaxy selection from the \WISC\ catalog.

As already mentioned, GAMA includes practically no stars or quasars, so it cannot be used as a calibration set to identify them in our \WISC\ sample. We have thus employed the Sloan Digital Sky Survey (SDSS, \citealt{SDSS.III}) spectroscopic data from Data Release 12 (DR12, \citealt{SDSS.DR12}) for the purpose of star and quasar cleanup. At the moment, SDSS is the most appropriate deep and wide-angle dataset that contains stars, galaxies and quasars comprehensively identified based on their spectral properties \citep{Bolton12}. SDSS assigns a class to spectroscopic sources at the same time as deriving their redshift (velocity)\footnote{\url{http://www.sdss3.org/dr12/algorithms/redshifts.php}}, which ensures far better reliability of this procedure over the photometric-only (morphological) classification. For these reasons, properly cleaned SDSS spectroscopic data form the best calibration sample for star-galaxy-quasar identification in wide-angle $z \lesssim 0.5$  photometric catalogs, such as ours. The trade-off is limited and variable depth of the spectroscopic sample, which is not as uniformly selected in SDSS as the photometric data.

The full SDSS DR12 spectroscopic catalog, which encompasses earlier releases (properly recalibrated where necessary), contains almost 3.9 million sources, of which 61\% are classified as galaxies, 16\% as quasars and the remaining 23\% as stars. Not all of these objects have, however, sufficient classification and redshift quality for our purposes. To maintain reliability, we have cleaned this sample of $\mathtt{zWarning}\neq0$ sources (problematic redshifts), as well as of those without a redshift error estimate ($\Delta z<0$) or with low-accuracy redshifts ($\Delta z / z > 0.01$). This gave as over 2.6M sources listed in SDSS DR12 as extragalactic (galaxies+quasars), including both the `Legacy' \citep{SDSS.DR7} and `BOSS' \citep{BOSS} samples, plus 750,000 stars.

\subsection{Quasar removal}
\label{Sec: QSO removal}

Our core dataset of extended objects is expected to contain a number of AGN and quasars. These will occasionally be outliers in the size distribution, much rarer than stars, as well as blends. Low-luminosity, relatively low-redshift, morphologically extended AGN, which will dominate the quasar population in our sample, are acceptable as long as their redshifts can be reliably reproduced photometrically. However, blends of a high-redshift quasar with a foreground star, which will mimic extended sources and have peculiar colors, as well as quasar-galaxy projections which also can  have compromised colors, will be problematic for the \phz\ procedure. Such blends lying at high redshifts should preferably be removed from the catalog before the photometric redshift estimation, because their presence may contaminate the derived galaxy sample which is expected to reach up to $z \sim 0.5$. In what follows, we will often use the terms `AGN' and `quasar' interchangeably.

Most of the quasars from the \WISC\ sample had been eliminated through the morphological preselection of resolved SCOS sources, as well as through the flux limits in the optical and infrared bands: high-redshift quasars are typically fainter than low-redshift galaxies in terms of their apparent magnitudes. There are, however, some quasars bright enough to be captured in the sample, while still classified as extended: about 30\% of the Sloan quasars/AGN (\ie \ttt{CLASS}=\ttt{QSO} in SDSS) identified also in our flux-limited sample have SCOS $\mathtt{meanClass} = 1$ (30,000 sources). These are mostly at redshifts of $z<0.6$, but some reach up to $z>3.5$. The latter are blends of  background point-like quasars with a low redshift foreground galaxy or with a foreground Galactic star, and might be problematic for photometric redshift  estimation, irrespective of the method used to obtain the \phzs. AGN experiencing significant dust obscuration, such as type 2 AGN, where the accretion disk and the broad line region are completely obscured, have colors similar to galaxies. In broad-band photometry, quasars at $z \gtrsim 2.3$ can be mistaken for low-redshift galaxies as the Lyman alpha spectral break can mimic the 4000~{\AA} break. Additionally, at $z\sim2.7$ and $\sim3.5$, the optical colors of broad-line, unreddened quasars and Galactic stars are the same \citep{Richards06}.

\begin{figure}
\centering
\includegraphics[width=0.49\textwidth]{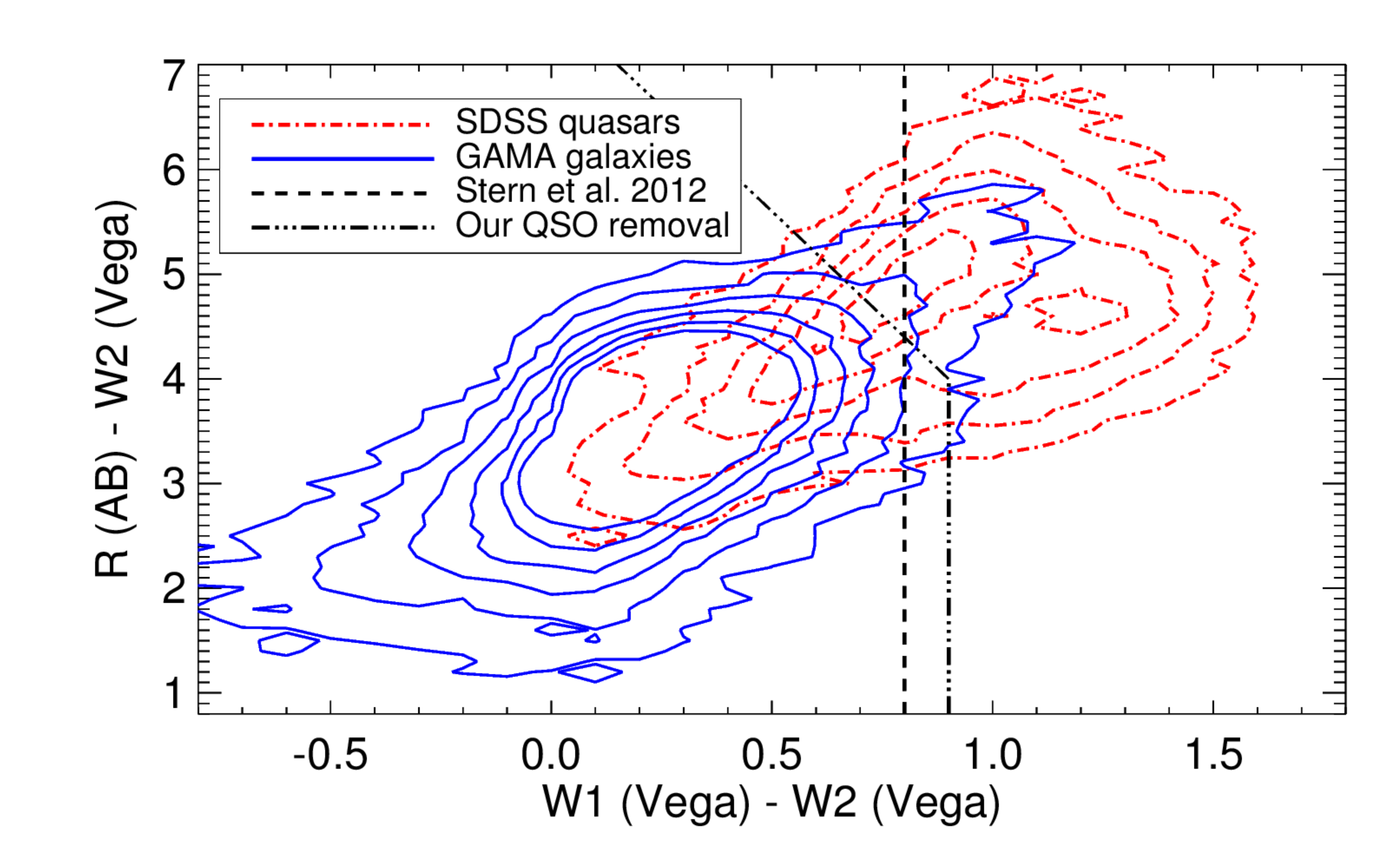} 
\caption{\label{Fig: QSO color cuts} Color-color plot ($W1-W2$ vs. $R-W2$) for GAMA galaxies (blue  {solid} contours) and SDSS quasars (red  {dot-dashed}) classified as extended in \WISC, together with the cuts that are used to remove quasars. The quasars in this plot are either low-redshift AGN or blends of a high-$z$ quasar with a star or galaxy. The contours are linearly spaced.}
\end{figure} 

In order to remove as many as possible of the quasars remaining in our sample, we have analyzed their multi-color properties, based on the SDSS spectroscopic data cross-matched with our catalog. \cite{Stern12} proposed $W1-W2>0.8$ as an efficient AGN finder in WISE, which was subsequently used in several other studies to select quasars \citep[e.g.][]{DYSA14,DiPompeo14}. By looking additionally at GAMA sources (which are practically free of quasar contamination), we have slightly revised this cut to preserve completeness of the galaxy sample, and added a second criterion using also the optical $R$ band (see Fig.\ \ref{Fig: QSO color cuts}). Our color cuts to remove quasars are:
\begin{equation}
\label{Eq: QSO cut}
R-W2>7.6-4(W1-W2) \quad \mrm{or} \quad W1-W2>0.9\;,
\end{equation}
where the WISE magnitudes are Vega and the SCOS one AB-like. These criteria remove 71\% of SDSS quasars present in our extended-source sample, while affecting less than 1\% of GAMA galaxies. Practically all the quasars with $0.5<z<2.1$ are eliminated through this cut; those that remain are mostly at low redshifts (peaking at $z\sim0.2$), with some at $2.1<z<3.5$. An alternative way of selecting quasars by using only WISE information through a comparison of $W1-W2$ and $W2-W3$ colors \citep{Jarrett11,Mateos12} cannot be applied here because of too low a detection rate of our sources in the $W3$ band.

The cut defined in Eq.\ \eqref{Eq: QSO cut} removed almost 300,000 quasar candidates from our flux-limited, $|b|>10\degree$ photometric sample of \WISC\ extended sources. Rescaling from the SDSS-based numbers, we thus estimate that there are about 115,000 quasars remaining in the all-sky catalog, which is about 0.6\% of the total number of galaxies, so our \phzs\ derived in \S\ref{Sec: Photometric redshifts} will be only minimally affected by the high-$z$ quasars that were not filtered out.
 
\subsection{Star removal}
\label{Sec: Star removal}

We also paired the SDSS DR12 stars with reliable spectra ($\mathtt{zWarning}=0$ and $0<\Delta z<0.001$) against our core sample, and used the result to derive typical stellar colors for star removal. Thanks to the morphological information from SCOS ($\mathtt{meanClass} = 1$ only sources) many of the stars had already been eliminated and only 8\% of those common to Sloan spectroscopic and \WISC\ are present in our sample. These `stars' in the catalog of extended sources are expected to be blends, which might be the reason why their separations from their SDSS counterparts are usually larger than in the case of extragalactic sources: $0.31''\pm0.26''$ for SDSS galaxies, $0.30''\pm0.29''$ for quasars, but $0.40''\pm0.24''$ for stars. To avoid mismatches, when deriving the color cuts for star removal we have used only the stars paired up within $1''$ with our photometric catalog. Note that poorer matching accuracy for the stars might  also be partly due to proper motions between the epochs of SCOS photographic material and these of the SDSS (see \citealt{MG13} for a related discussion).

\begin{figure*}
\centering
\includegraphics[width=\textwidth]{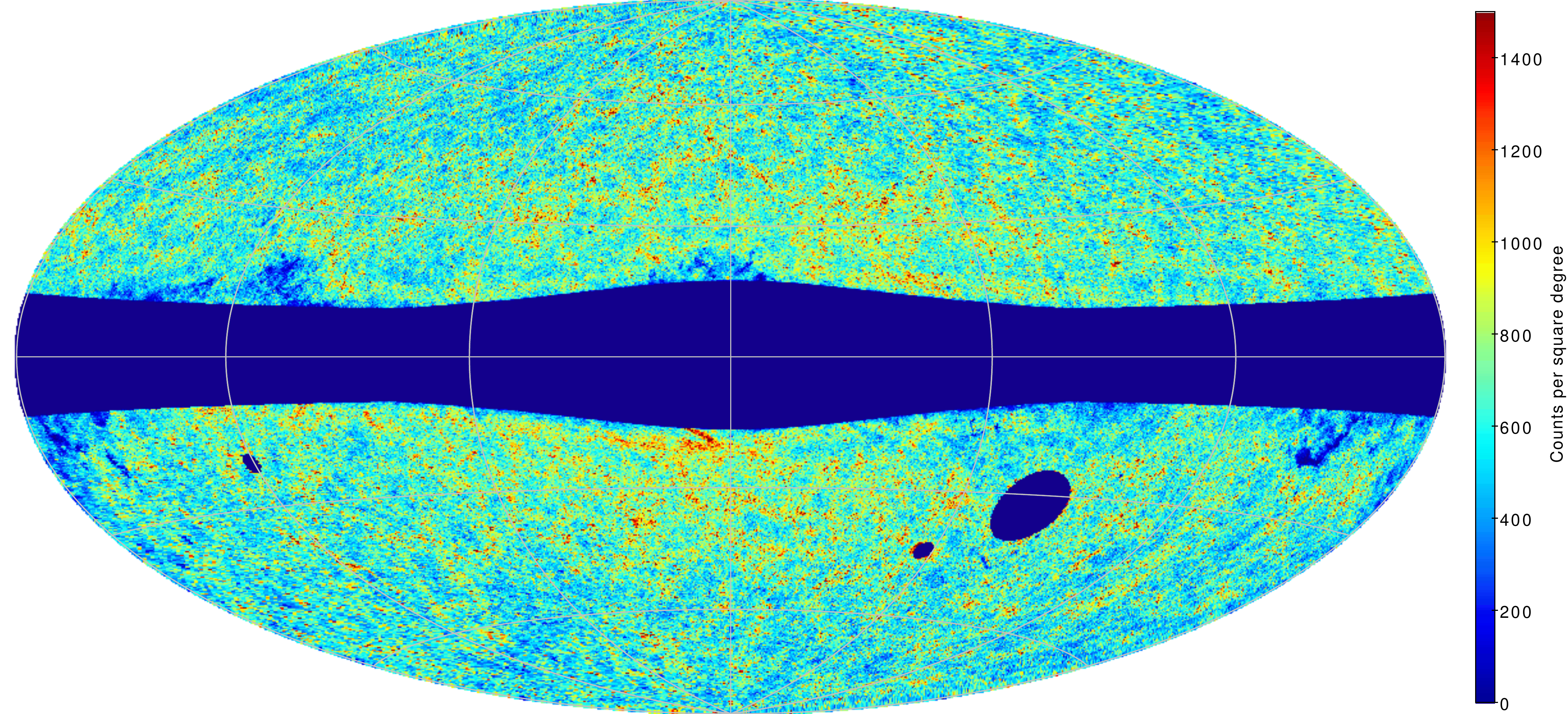} 
\caption{\label{Fig: WIxSC.allsky.NOTmasked}WISE\ti{}SuperCOSMOS galaxy catalog, after star and quasar cleanup and manual cutouts of the Galaxy, Magellanic Clouds and M31, but before final masking, in an all-sky Aitoff map in Galactic coordinates. The map contains 21.5 million sources flux-limited to $B<21$ \& $R<19.5$ \& $13.8<W1<17$.}
\end{figure*}

To remove stellar contamination, we have examined the source distribution presented in Fig.\ \ref{Fig: WIxSC.noclean}, which clearly shows spurious overdensities (caused by stellar blends) at low Galactic latitudes and at the Magellanic Clouds. We began by rejecting by hand regions in the Galactic Plane and Bulge where the contamination was too severe to contemplate reliable correction (an enhancement in surface density by a factor $\sim 10$): we applied a latitude cut depending on the distance from the Galactic Center: it goes smoothly from $|b|<17\degree$ at $\ell=0\degree$ to $|b|<10\degree$ at $\ell\sim80\degree$ or $\ell\sim280\degree$. Detailed equations are provided in Appendix \ref{App: position-dependent cuts}.  This cut removed almost 6 million sources from the flux-limited sample at $|b|>10\degree$. To this we added circular cutouts around the most prominent nearby galaxies, namely the Magellanic Clouds and M31.

We have investigated what other cuts need to be taken to purify the sample further. This is traditionally done in multi-color space, and we explored different combinations of the available bands{, based on the cross-match with SDSS spectroscopy}. Stars are much more difficult to remove than quasars from our catalog without seriously compromising the completeness of the galaxy sample; this is due to blends of stars with other stars and with galaxies, especially at low redshift, where galaxies from our dataset often have colors similar to stellar ones. In particular, the SCOS optical bands were found not to be useful for star identification. We were left with the option to use only WISE colors for star-galaxy separation, as had been discussed in earlier studies \citep{Jarrett11,Goto12,Yan13,FSS15}. An advantage of applying infrared-only cuts to our sample is less sensitivity to variations in plate zero points, or in extinction corrections and their errors. 

The colors usually considered for WISE source identification are $W1-W2$ and $W2-W3$, and especially the former is particularly useful for this task \citep{Jarrett11}. We have found that using $W2-W3$ does not add much information, mostly due to the low level of signal-to-noise in the $W3$ band, and a similar effect is observed in the automatic galaxy identification of  \cite{Krakowski16}. In a related effort, \cite{FSS15} treated as stellar anything with $W1-W2<0$ or ($W1<10.5$ and $W2-W3<1.5$ and $W1-W2<0.4$). These conditions applied to WISE leave, however, a certain degree of contamination, dependent on the distance from the Galactic Center (GC) -- see Fig.\ 1 in \cite{FSS15}. The same is found in our \WISC\ catalog, namely a fixed $W1-W2$ color cut would give purity levels largely varying over the sky; this is also expected because we are using extinction-corrected magnitudes, hence effective stellar colors will be correlated with the $E(B-V)$ map.  

In order to account for star contamination changing with Galactic coordinates, we have examined source density and the $W1-W2$ color as a function of distance from the GC, and have found that the stellar locus shifts as the GC is approached, which we interpret as a reflection of older stellar populations being located towards the Bulge. This lead us to design a position-dependent color cut, the details of which are provided in Appendix \ref{App: position-dependent cuts}. In brief, at high latitudes we remove sources with $W1-W2<0$, while closer to the GC this cut is gradually shifted towards  $W1-W2<0.12$. This adaptive star removal, together with the sky cuts discussed earlier, eliminated over half  of the sample, mostly from low Galactic latitudes as expected (90\% of removed sources are within $|b|<34\degree$). This approach slightly degrades the completeness of the final galaxy sample: almost 6\% of \WISC\ti{}GAMA galaxies are removed with this cut. On the other hand, a completeness level of $\sim90\%$ in the final sample is preserved for $|b|\gtrsim 15\degree$, which gives almost $3\pi$ sr of the extragalactic sky comprehensively sampled with the catalog. A more detailed discussion of completeness and purity of the galaxy dataset is provided in \S\ref{Sec: Completeness and purity}.

\begin{figure*}
\centering
\includegraphics[width=0.48\textwidth]{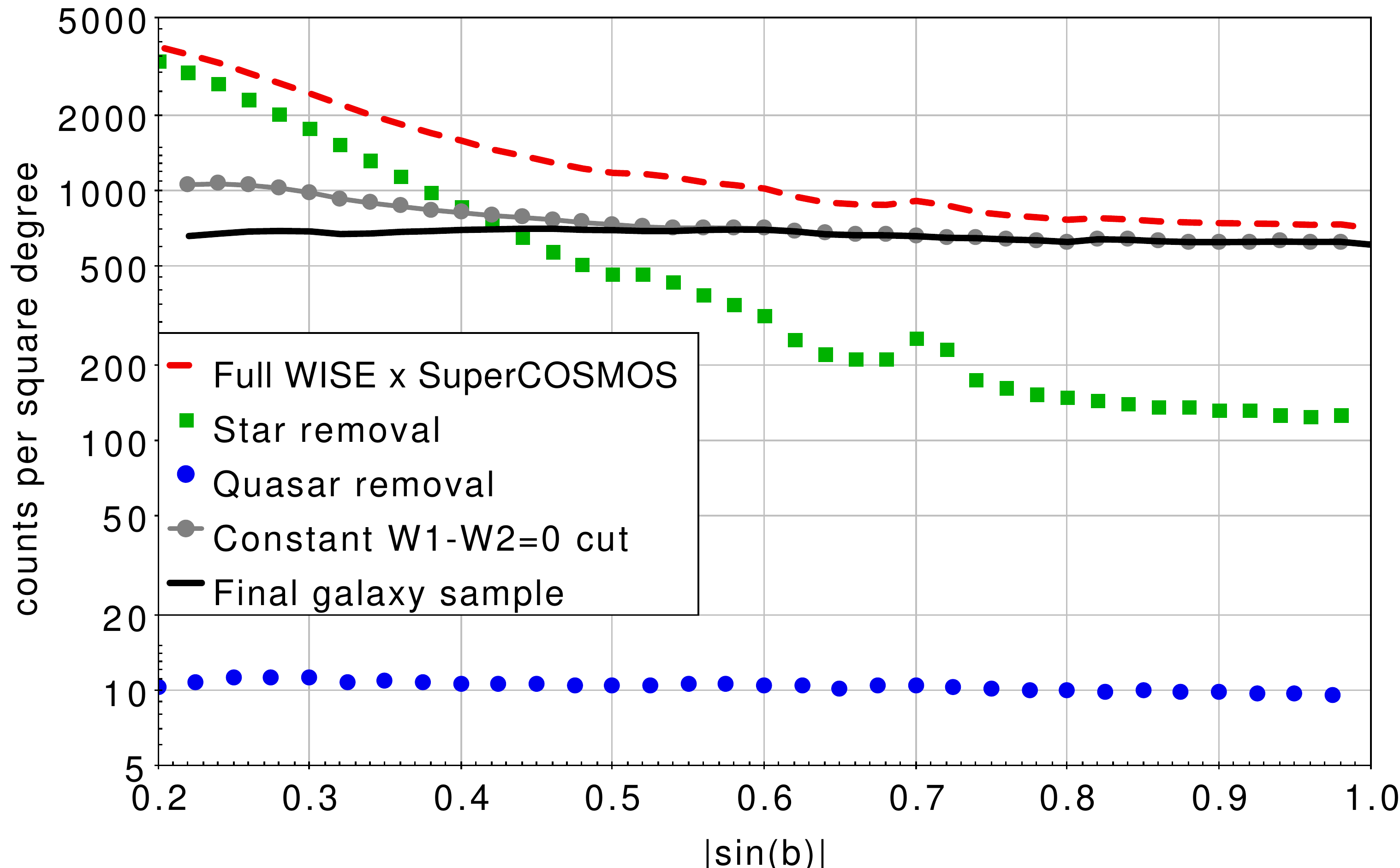} 
\includegraphics[width=0.48\textwidth]{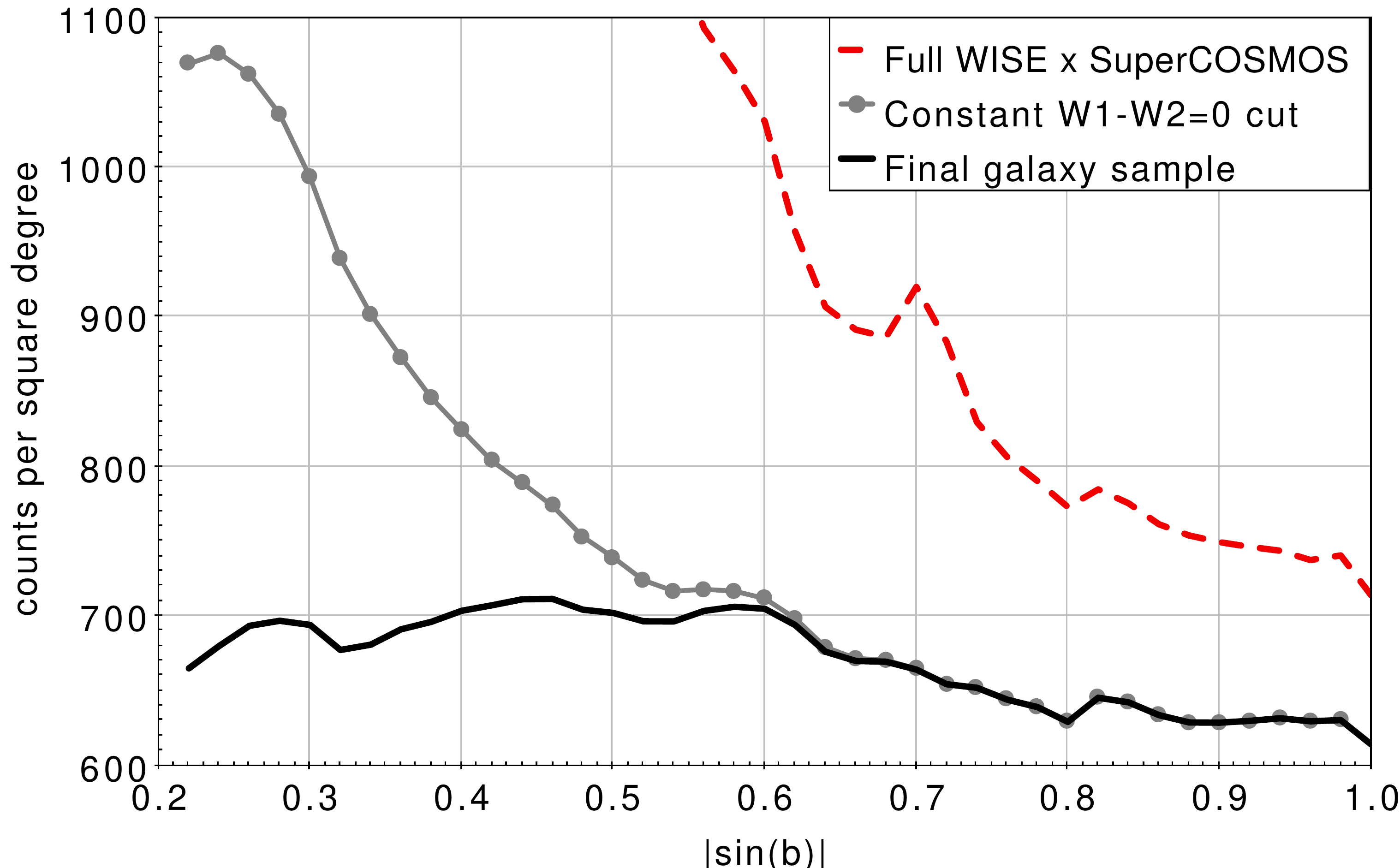} 
\caption{\label{Fig: histograms cleanup}Source counts per square degree as a function of the sine of Galactic latitude in the cross-matched WISE\ti{}SuperCOSMOS extended source catalog: full sample (red  {dashed}), sources removed with our star (green  {squares}) and quasar (blue  {dots}) cleanup, and the final sample (black  {solid}). For comparison, we also show the counts for a sample with a constant $W1-W2$ color cut applied (grey  {solid-dotted}).  {Right-hand panel shows a zoom in on the two latter curves, in linear scaling.}}
\end{figure*}

In addition, we removed the bright end of our sample ($W1<13.8$), for two main reasons. First, the galaxies that have counterparts in the 2MASS XSC $K_s<13.9$ already have precise photometric redshifts derived in the 2MPZ (\tcb{B14}). At low redshifts the typical galaxy color is $K_s-W1\simeq0$, so most of these 2MASS sources are removed by applying this bright end cut in WISE. Secondly, most of the bright WISE sources which are \textit{not} present in 2MASS XSC are stars, as they dominate $W1$ number counts there \citep{Jarrett11} and are concentrated towards the Galactic plane. There were over 5 million objects with $W1<13.8$ in the cross-matched catalog before the cleanup, of which 90\% lay at $|b|<50\degree$.

Fig.\ \ref{Fig: WIxSC.allsky.NOTmasked} shows the all-sky distribution of our sources after the purification and manual cutouts but before final masking, which is addressed in the following Section. In Fig.\ \ref{Fig: histograms cleanup} we show source counts per square degree, as a function of the sine of Galactic latitude $b$, for the cross-matched sample: before and after the star and quasar cleanup, as well as for the sources removed with our cuts. A uniformly distributed (extragalactic) sample should have roughly constant counts in this scaling, which is indeed  {approximately true} for the final dataset, as well as for the quasars removed. The bump at $\left| \sin b \right| \simeq0.7$ in the removed sources is the LMC. For comparison we also show the case of a constant $W1-W2>0$ cut as in \cite{FSS15}. Stellar contamination becomes then prominent already from $|\sin b|=0.5$ ($|b|=30\degree$), \ie for half of the sky {, and the surface density of the sources close to the Galactic Plane is almost twice as large as in the Caps, as is visualised in the right panel of Fig.\ \ref{Fig: histograms cleanup}}.

\subsection{Final mask}
\label{Sec: Mask}

The above cuts helped improve the fidelity of the catalog, reducing the numbers of non-galaxy entries resulting from stellar blends and other problems. Nevertheless, a casual inspection of the sky distribution reveals clear imperfections, especially at low Galactic latitudes: Fig.\ \ref{Fig: WIxSC.allsky.NOTmasked} exhibits some spurious source overdensities and lack of extragalactic data behind Galactic molecular clouds such as Orion, Taurus/Perseus and Ophiuchus. We thus need to develop a mask that excludes significantly affected regions. This is a common task, but not a trivial one: the human eye is highly adept at spotting artifacts of this sort, and it takes some effort to design an objective automated process that performs as well. As a starting point, one can identify pixels where the surface density is discrepant, using the fact that the galaxy surface density very nearly obeys a log-normal distribution \citep{Hubble34} and clipping pixels in the tails of this distribution. This approach can be made more effective if we perform it at a variety of resolutions: large-scale regions where the density is systematically slightly in error can be found more sensitively by using coarse pixels where the pixel-to-pixel variance is reduced. We therefore constructed a HEALPix\footnote{\url{http://healpix.sourceforge.net/}} \citep{HEALPIX} map of the galaxy counts, initially at $N_{\rm side}=256$, identified discrepant pixels, and then repeated the process degrading the resolution by successive factors of 2. The final mask is then the accumulation of flagged sky areas at all resolutions. But this process requires an unsatisfactory compromise: in order to remove all apparent artifacts, the clipping threshold has to be set at a rather high probability ($p(\delta)\sim 0.001$), with the unacceptable result that the extreme regions of real cosmic structures are also removed. 

\begin{figure}
\centering
\includegraphics[width=0.49\textwidth]{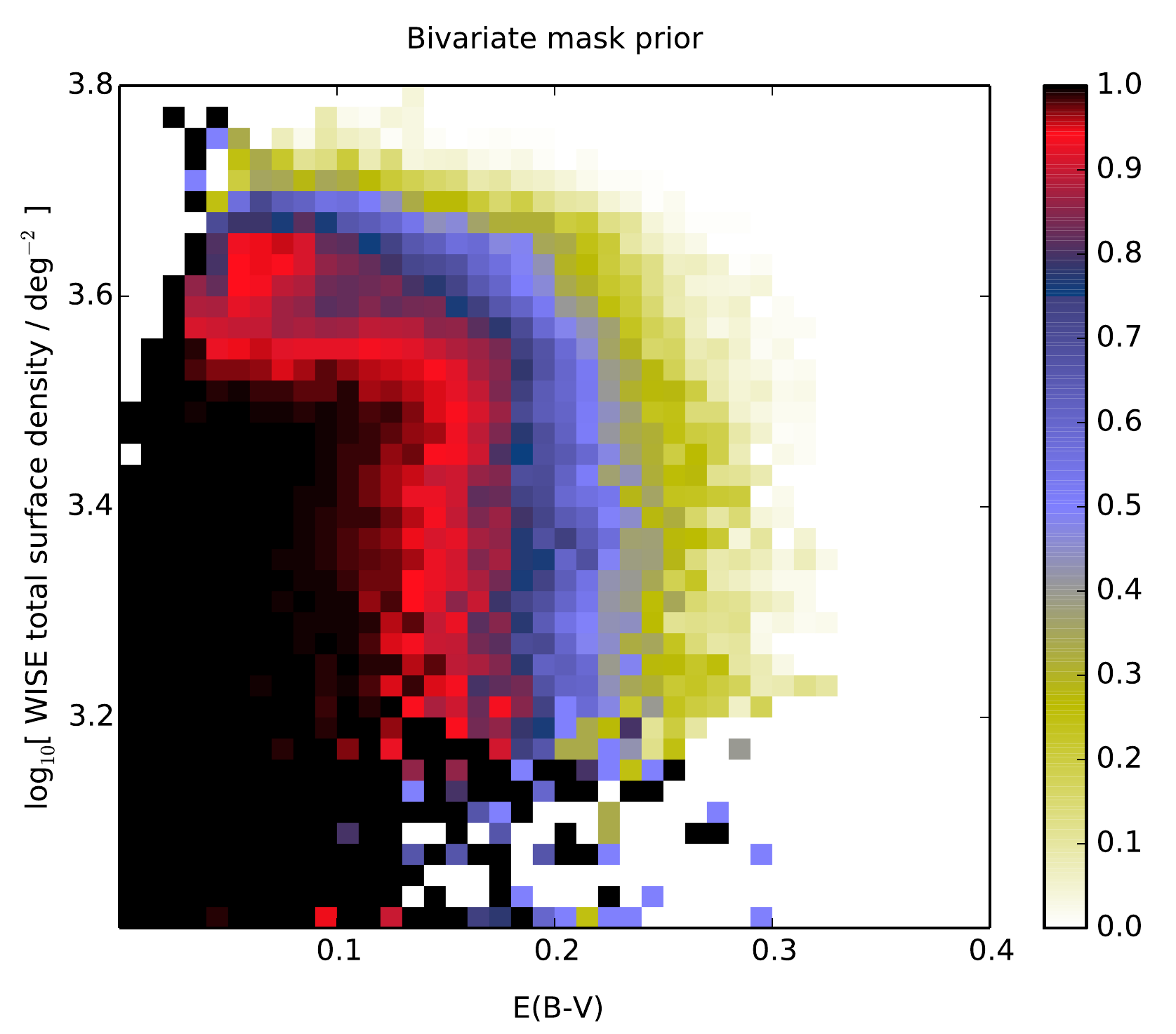}
\caption{\label{Fig: prior}Initial prior for the mask, based on clipping regions of abnormal galaxy density. The probability of a pixel being accepted is shown as a function of extinction and of total WISE surface density at $W1<17$ as a proxy for stellar density.}
\end{figure}
\begin{figure}
\centering
\includegraphics[width=0.49\textwidth]{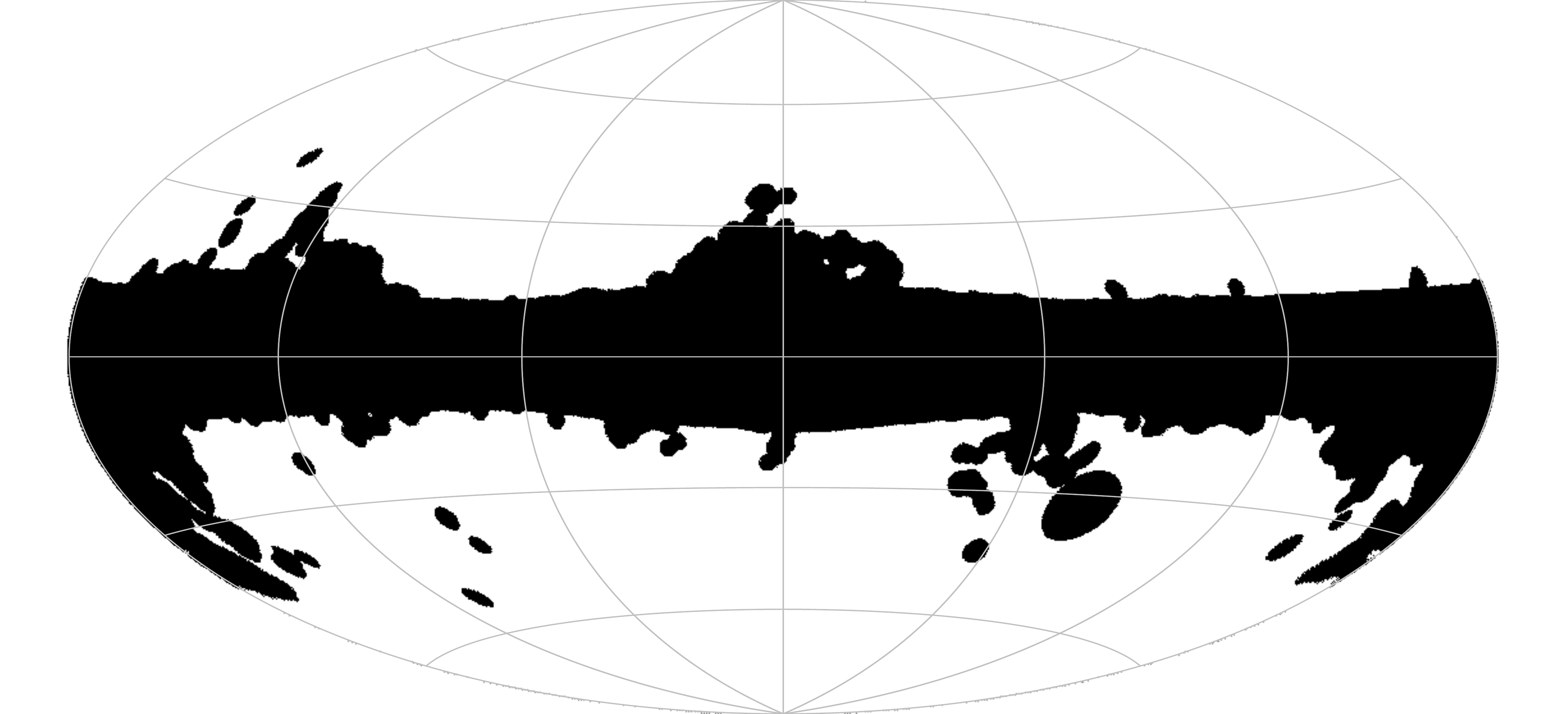}
\caption{\label{Fig: WIxSC.mask}Final mask applied to the WISE\ti{}SuperCOSMOS galaxy catalog, presented here in Galactic coordinates with $\ell=0$, $b=0$ in the center. Black areas are masked and over 68\% of sky is retained for further analysis.}
\end{figure}

\begin{figure*}
\centering
\includegraphics[width=\textwidth]{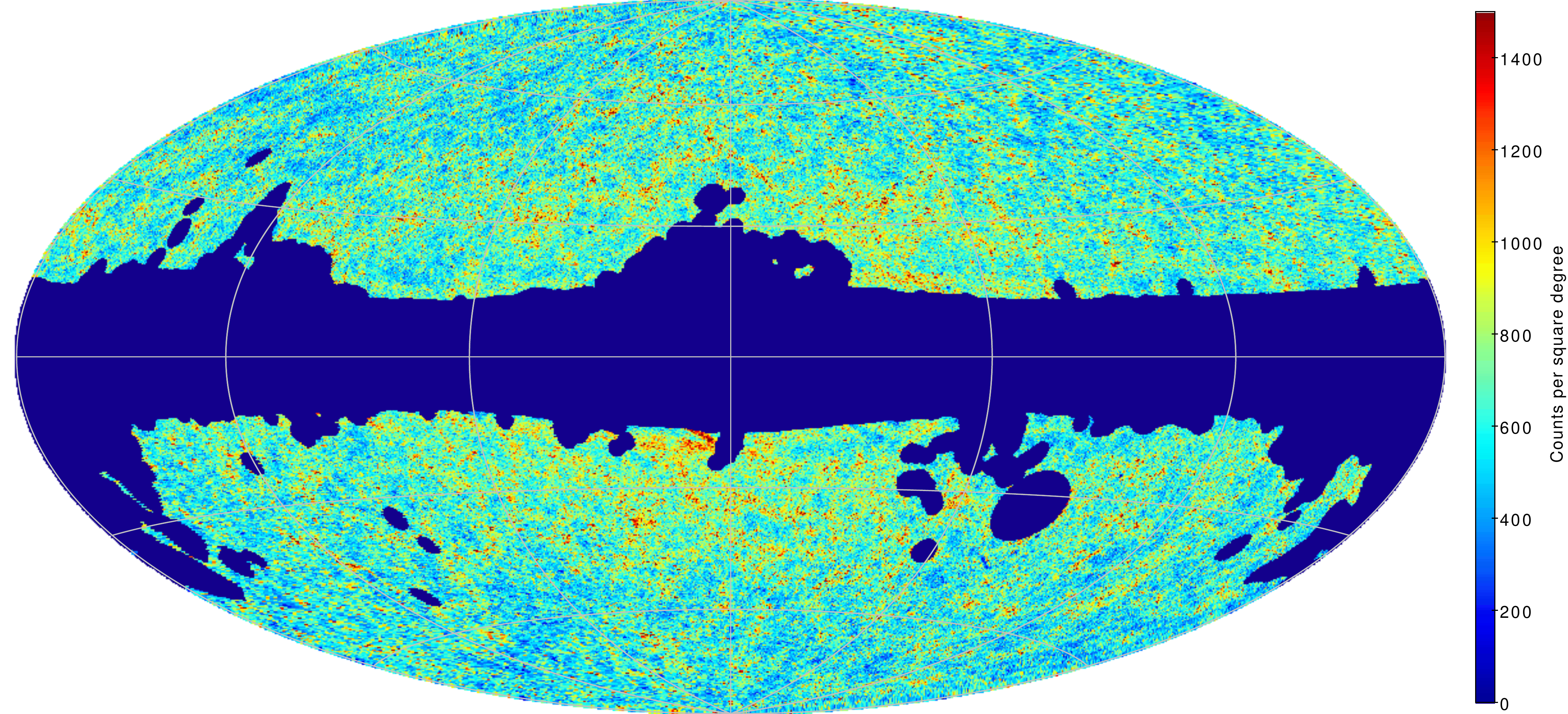} 
\caption{\label{Fig: WIxSC.allsky.masked}WISE\ti{}SuperCOSMOS galaxy catalog, after star and quasar cleanup and masking, in an all-sky Aitoff map in Galactic coordinates. The unmasked region (68\% of the sky) contains 18.7 million sources flux-limited to extinction-corrected limits of $B<21$ \& $R<19.5$ \& $13.8<W1<17$.}
\end{figure*}

To deal with this problem, we take the Bayesian approach of bringing in prior information. Most of the problems are associated with the Milky Way, so we can make a good guess in advance about whether a given region should be masked. We therefore consider two indicators of potential problems: extinction and stellar density, measured via $E(B-V)$ and the empirical total WISE density, $\Sigma$ (to $W1<17$). The latter additionally brings in information on WISE coverage issues that cause spurious over- and underdensities in source distribution. We use the first estimate of the mask derived from clipping to estimate a prior probability that a given pixel is masked as a function of these variables; this is shown in Fig.\ \ref{Fig: prior}. From this, it is clear that regions at $E(B-V)>0.25$ should be completely clipped. We can now repeat the clipping analysis, but considering a full posterior probability that a given pixel is clean: 
\begin{equation}
p_c=f_{\rm prior} \times p(\delta)\;,
\end{equation}
where $f_{\rm prior}$ is the fraction of pixels accepted at that Galactic location. It is now possible to clip with a more discerning threshold, $p_c\sim 10^{-5}$, which removes negligible amounts of real large-scale structure, while still remaining sensitive to anomalies at low latitudes. As a further precaution, we apply `guilt by association', and mask all pixels within 1 degree of a masked pixel. Finally, this process can be iterated, updating the prior when a revised mask has been generated. The final mask is shown in Fig.\ \ref{Fig: WIxSC.mask}, and it removes 32\% of the sky, leaving a satisfactorily clean galaxy sample over 28,000 deg$^2$. Applying the mask to the data gives us the final catalog of almost 18.7 million sources, illustrated in Fig.\ \ref{Fig: WIxSC.allsky.masked}. The mean surface density of the sources is about 670 deg$^{-2}$, which is more than 20-fold increase over 2MASS. We note, however, that for cosmological applications it might be more appropriate to repeat the above masking procedure on data first preselected in \phz\ or other (\eg magnitude) bins.

\subsection{Completeness and purity of the final catalog}
\label{Sec: Completeness and purity}

Having applied all the cuts and the mask aiming at optimizing the reliability of the \WISC\ galaxy catalog, we now quantify its levels of completeness and purity. This was done using external data that will be treated as the `truth', ignoring any imperfections in them. Since our catalog was created by requiring detection in three independently surveyed wavebands, we assume that all our objects are genuine astronomical sources. We then need to measure the purity of the catalogue (\ie the fraction of our objects that are actually galaxies rather than stars) and its completeness (the fraction of all true galaxies that are included). 

Purity is relatively easy to assess via cross-matching with SDSS. We selected a 1-degree-wide strip centered at $\delta = 30\degree$ with magnitude limits much fainter than those of \WISC, which yielded over 130,000 matched sources at $|b|>12\degree$. From this, we found that at high latitudes approximately 95\% of our sources are indeed galaxies. For latitudes of $|b|>60\degree$ the stellar contamination of our catalog does not exceed 6\%, and remains less than 10\% down to almost $|b|=40\degree$. One could further improve the purity at the expense of completeness; for instance, using a color cut of $W1-W2>0.2$ instead of the fiducial one used in \S\ref{Sec: Star removal}, would allow for a catalog with a stellar contamination of $\lesssim 3\%$ down to $|b|\sim40\degree$. Of course this would at the same time lead to a significant drop in completeness, as about 30\% of galaxies have $W1-W2\leq0.2$ (typically early types).

Completeness is slightly more complicated to assess on account of magnitude errors. Even if our catalog was perfect in all other respects, it will miss many galaxies that are really just brighter than our magnitude limits, and include many that are in reality just slightly fainter. Therefore, the assessment of completeness involves two questions that go beyond noisy magnitudes: (1) what fraction of true galaxies are incorrectly classified by SCOS as stars? (2) what further fraction of galaxies are lost as a result of the colour cuts aimed at purifying the sample? The first question can be addressed by looking at a pairing of SDSS data with both the SCOS galaxy \textit{and} star catalogs (the latter classified as $\mathtt{meanClass}=2$). We started with the same SDSS strip centered at $\delta=30\degree$ as above,  synthesised SCOS magnitudes \citep{Peacock16} and cut the sample to $B_\mrm{syn}<21$ \& $R_\mrm{syn}<19.5$, looking at the relative numbers that paired with SCOS galaxies and stars. The conclusion is that the overall misclassification incompleteness is about 15\%, with some dependence on Galactic latitude. Averaged completeness exceeds 90\% for $|b|>40\degree$ and equals 88\% for half of the sky ($|b|>30\degree$). Unsurprisingly, the limited image quality on Schmidt plates leads to compact galaxies being classified as stars. According to \cite{Baldry10}, the corresponding figure for SDSS is about 2\% (GAMA does very much better because it uses near-IR colors in addition to image width). 

Finally, we can use GAMA to measure the effect of the additional color cuts applied in \S\ref{Sec: QSO removal} and \S\ref{Sec: Star removal}. The depth and completeness of GAMA itself are large enough that we can assume that practically each \WISC\ galaxy should be present also in GAMA. We will thus treat the cross match of the two catalogs \textit{before} star and quasar removal as the reference. As already mentioned, the quasar cutout criterion (Eq.\ \ref{Eq: QSO cut}) affects less than 1\% of GAMA galaxies; the most important for the final completeness will be then the color cut used to discard stellar contamination. Using our prescription for star removal (\S\ref{Sec: Star removal}) we lose about $6\%$ of galaxies at $|b|\geq30\degree$, increasing to 10\%  for the lowest latitudes of $b\sim22$ observed by GAMA.  We can thus safely assume that, including the 1\% drop in completeness due to quasar cutout, our catalog is about 93\% complete for half of the sky ($|b|>30\degree$) and more than 90\% complete over at least $2.4\pi$ steradians (in addition to the classification incompleteness). 

This completeness analysis is ultimately limited by the fact that our catalog is a cross-match of two independent samples of different characteristics. Our dataset will thus miss some sources that could not be detected by one of the parent surveys, or by both. For instance, WISE is not sensitive to low surface brightness galaxies, while SCOS is biased against the dusty ones, which WISE does detect very well. Quantification of these effects is beyond the scope of this paper and in particular it would require using much deeper reference catalogs of otherwise very similar preselections as those employed here.

\section{Photometric redshifts}
\label{Sec: Photometric redshifts}

This Section presents the derivation and analysis of photometric redshifts for our full galaxy sample. To compute the \phzs, we used the \ANNz\ package\footnote{Available from \url{http://www.homepages.ucl.ac.uk/\~ucapola/annz.html}.} developed by \cite{ANNz}, an artificial neural network algorithm which estimates photometric redshifts based on a training sample with photometric quantities and spectroscopic redshifts (see also \citealt{FLS03}). The \ANNz\ \phz\ estimator has been shown to be one of the most accurate methods \citep[e.g.][]{ABLR11,DESSanchez}, so long as a sufficiently large and representative spectroscopic sample is available for redshift calibration(the latter being generally true for empirical \phz\ methods). In our case, such a sample is provided by the deep and complete GAMA dataset. We also experimented with another photometric redshift code, GAz \citep{GAz}\footnote{\url{https://github.com/rbrthogan/GAz}}, which gave results similar to \ANNz{}, albeit slightly poorer\footnote{A new version of \ANNz, dubbed \ANNz{}2 \citep{ANNz2} was released when the present work was in an advanced stage, and we postpone its possible application to future work.}.

As was the case for 2MPZ (\tcb{B14}), we have not used any template-fitting \phz\ estimation methods. One cannot efficiently implement them when incorporating SCOS photometry, mostly because the photographic filter transmission curves are not known to sufficient accuracy. We note however that another independent technique of redshift estimation could in principle be employed for the catalog presented here, namely the `clustering redshifts' \citep[e.g.][]{Newman08,Menard13}, applied recently to the SDSS \citep{Rahman15,RahmanSDSS} and 2MASS \citep{Rahman2MASS} samples.

The most optimal neural network architecture for the \ANNz\ code is not known \textit{a priori} and depends on such parameters as the number of photometric bands used and the size of the training sample. For each of the tests we have always tried a number of different architectures,  limiting ourselves to no more than two intermediate layers (adding more layers does not increase \phz\ accuracy). We used `committees' of at least 6 networks and in most cases the most accurate \phzs\ were obtained by applying network architectures with 1 or 2 intermediate layers, 10 to 35 nodes in each.

Before computing photometric redshifts for our full sample, we performed extensive tests of their properties using GAMA redshifts as training and test samples. Such an approach gives the most comprehensive results, as GAMA is a highly complete, flux-limited sample deeper than our catalog (\S\ref{Fig: GAMA x-match dNdz}) and offers much auxiliary information that allowed us to examine \phz\ performance as a function of both observed (apparent) and intrinsic properties of the sources. We also experimented with adding the SDSS DR12 spectroscopic dataset to the GAMA calibration sample, but it is too non-uniformly selected at $r>17.77$ (beyond the Main Sample) to be applicable for photometric redshift training. 

Empirical photometric redshift estimators, such as the \ANNz{}, generally provide better results when more photometric parameters are used in the \phz\ derivation. It would  thus be desirable to add more bands to the four basic ones employed to preselect WISE and SCOS data ($B,R,W1,W2$), but presently this is not possible for our full catalog. In \tcb{B14} we showed that the SCOS $I$ band does not significantly change the \phz\ accuracy for the GAMA-based sample, mainly because this band is shallower than $B$ and $R$. The situation is no better with WISE, where two additional bands are in principle available: mid-IR $W3$ and $W4$ centered respectively on 12 and 23 $\mu$m. However, these bands were of much lower sensitivity than $W1$ and $W2$: only 30\% of our sources have  $S/N>2$ in $W3$ and fewer still  are detected in $W4$ (compare also \citealt{Cluver14}). The $I$, $W3$ and $W4$ bands will thus not be used in the derivation of all-sky photometric redshifts, to preserve uniformity of the catalog. 

\subsection{Tests and calibration on GAMA}
\label{Sec: tests on GAMA}

\begin{deluxetable*}{lcccccccccc}
\tabletypesize{\footnotesize}
\tablewidth{0pt}
 \tablecolumns{11} 
\tablecaption{\label{Table photo-z}\small Statistics for the photometric redshift estimation generated for a test sample of WISE\ti{}SuperCOSMOS sources {in GAMA equatorial fields. Two cases are shown: (i) no flux limits applied to the photometric sample; (ii) fiducial flux limits as in the final dataset. In the former case, we also show statistics for the particular GAMA fields. }}

\tablehead{
\colhead{\# of} & 
 \multicolumn{2}{c}  {mean $\langle z \rangle$} & 
 \multicolumn{2}{c} { median $\overline{z}$} &
 \colhead{ 1$\sigma$ scatter\tablenotemark{d}} &
 \colhead{ scaled } &
  \colhead{ norm. } &
 \colhead{ mean bias\tablenotemark{g} } & 
 \colhead{ median } &
 \colhead { \% of } \\
\colhead{sources\tablenotemark{a}} &
\colhead{spec\tablenotemark{b}} & \colhead{phot\tablenotemark{c}} &
\colhead{spec\tablenotemark{b}} & \colhead{phot\tablenotemark{c}} &
\colhead{  $\sigma_{\delta z/(1+z)}$} &
\colhead{ MAD\tablenotemark{e} } &
\colhead{ SMAD\tablenotemark{f} } &
\colhead{$\langle \delta z \rangle$} &
\colhead{ error\tablenotemark{h} } &
\colhead{ outliers\tablenotemark{i} }
}

\startdata

\sidehead{\bf Trained and tested on GAMA}
\sidehead{\underline{no flux limits}}

127,703 & 0.231 & 0.231 & 0.223  & 0.233 & 0.045 & 0.044 & \textbf{0.036}  & \textbf{-7.3{e}-5}  & \textbf{14.1\%} & \textbf{2.8\%} \\

\sidehead{separate fields}
 (G09) 37,810 & 0.242 & 0.240 & 0.242 & 0.244 & 0.044 & 0.045 & {0.037} & {-2.0{e}-3} & {14.0\%} & {2.4\%} \\
 (G12) 48,974 & 0.229 & 0.230 & 0.221 & 0.231 & 0.045 & 0.044 & {0.036} & {6.8{e}-4} & {14.5\%} & {2.7\%} \\
 (G15) 40,919 & 0.223 & 0.224 & 0.211 & 0.224 & 0.044 & 0.043 & {0.036} & {7.6{e}-4} & {14.6\%} &  {2.9\%} \\

\sidehead{\underline{flux limited $B<21$ \& $R<19.5$ \& $W1<17$}}

86,516 & 0.200 & 0.200 & 0.191  & 0.202 & 0.041 & 0.040 & \textbf{0.033}  & \textbf{-3.7{e}-4}  & \textbf{14.7\%} & \textbf{2.8\%} \\

\enddata
\tablenotetext{a}{In the test set.}
\tablenotetext{b}{Input (spectroscopic) redshift sample.}
\tablenotetext{c}{Output (photometric) redshift sample.}
\tablenotetext{d}{Normalized 1$\sigma$ scatter between the spectroscopic and photometric redshifts, $\sigma_{\delta z/(1+z)}$; unclipped.}
\tablenotetext{e}{Scaled median absolute deviation, $\mathrm{SMAD}(\delta z) = 1.48 \times \mathrm{med}(|\delta z-\mathrm{med}(\delta z)|)$.}
\tablenotetext{f}{Scaled median absolute deviation of the normalized bias, $\mathrm{SMAD}(\delta z/(1+z_\mrm{sp}))$.}
\tablenotetext{g}{Mean bias of $z_{\rm phot}$: $\langle \delta z \rangle = \langle z_{\rm phot}-z_{\rm spec} \rangle$; unclipped.}
\tablenotetext{h}{Median of the relative  error, $\mathrm{med}(|\delta z|/z_\mrm{sp})$; unclipped.}
\tablenotetext{i}{Percentage of outliers for which $|(z_\mrm{ph}-z_\mrm{sp})/(1+z_\mrm{sp})| > 3\, \mathrm{SMAD}(\delta z/(1+z_\mrm{sp}))$.\\}
\end{deluxetable*}


In \tcb{B14} we presented the potential of applying GAMA as a \phz\ training sample for \WISC\ all-sky data. Using the shallower GAMA DR2 public release (complete to $r<19.0$ in two of the equatorial fields and to $r<19.4$ in the third one, \citealt{Liske15}) together with the WISE `All-Sky' and SCOS extended source data, we obtained an accuracy of $\sigma_{z} \simeq 0.035$ at a median redshift of $z\sim0.18$. Here we extend that study using a complete flux-limited $r\leq 19.8$ GAMA-II sample, together with deeper AllWISE data, corrected SCOS color calibrations, and Galactic extinction coefficients revised according to \cite{SF11}.

We start with the \WISC\ti{}GAMA sample as described in \S\ref{Sec: x-matches with GAMA}. We tested \phz\ performance for various cuts applied to this dataset, such as flux limits being the same as in the all-sky sample discussed in \S\ref{Sec: WISE x SCOS x-match}, star and quasar color cuts (\S\ref{Sec: Catalog cleanup}), etc. We have found that \phz\ statistics are practically independent of whether the cuts are first applied both on the training and test sample, or only on the latter once \ANNz\ had been trained on the full sample; in other words, it is enough to train the neural networks on the most general training set possible and apply the required cuts on the test set only.  The only cut applied \textit{a priori} to all the GAMA samples discussed in this Section is $\delta_{1950}<2.5\degree$ to avoid residual passband mismatch between SCOS `North' and `South'. More discussion on this issue will be provided in \S\ref{Sec: All-sky photo-z catalog}.

The \WISC\ti{}GAMA sample of  142,000 sources was divided randomly in proportion 1:9 into training and test sets, and a validation set was additionally separated out from the former (20\% of the full training set). Summary statistics for the \phz\ tests are provided in Table \ref{Table photo-z}. Note that the outlier fraction is defined here relative to the scatter of each of the particular test sets, so the decrease in the scatter may lead to a slight increase in the outlier rate.  {The table includes also the statistics computed separately for the three GAMA equatorial patches (in the case of no flux limits in \WISC), showing that the variations in the \phz\ accuracy between these fields are not significant.}  Comparing the general results with table 2 of \tcb{B14}, one can see that the current \phz\ performance is very similar to the one achieved there with GAMA DR2, taking into account the increased depth of the present sample. It is also worth noting that our uncertainty of $\sigma_{z}\simeq0.03$ for the flux-limited case is comparable to the results of \cite{Christo12}, where GAMA spectroscopy was applied to train a large SDSS $r<19.4$ sample using 5-band $ugriz$ photometry. Last but not least, this accuracy is very close to the prediction for future surveys such as Euclid or LSST \citep{Ascaso15}.

\begin{figure}[t]
\centering
\includegraphics[width=0.4\textwidth]{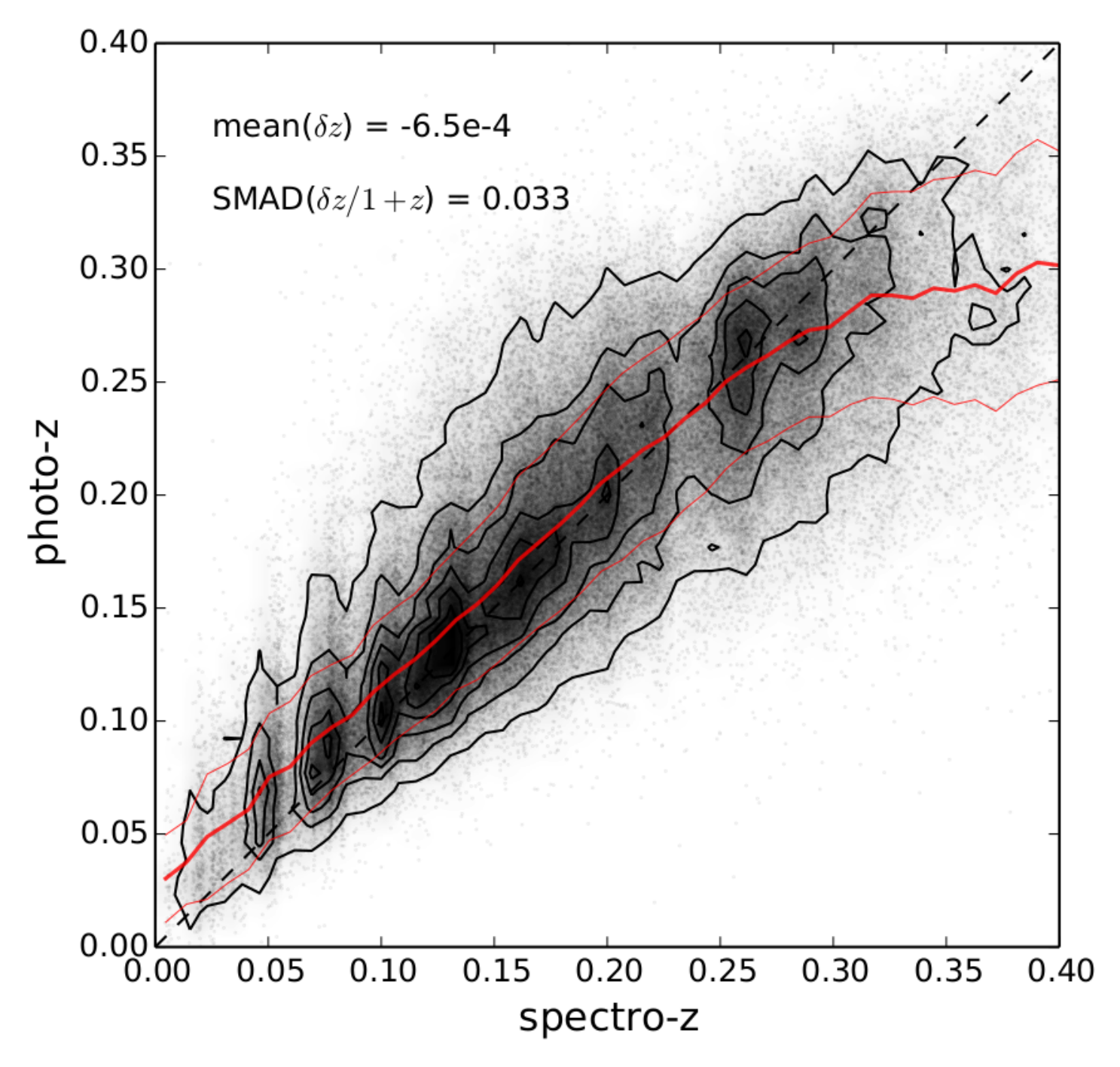} 
\caption{\label{Fig: zsp-zph GAMA}Comparison of GAMA spectroscopic redshifts with the photometric ones derived from WISE\ti{}SuperCOSMOS photometry, for the $B<21$ \& $R<19.5$ \& $W1<17$ flux-limited sample  of \WISC\ti{}GAMA sources. Red lines show the running median \phz\ and its scatter (SMAD). The vertical striping of the density plot results from galaxy overdensities being radially diluted  in the photometric redshift space.}
\end{figure} 
\begin{figure}[t]
\centering
\includegraphics[width=0.48\textwidth]{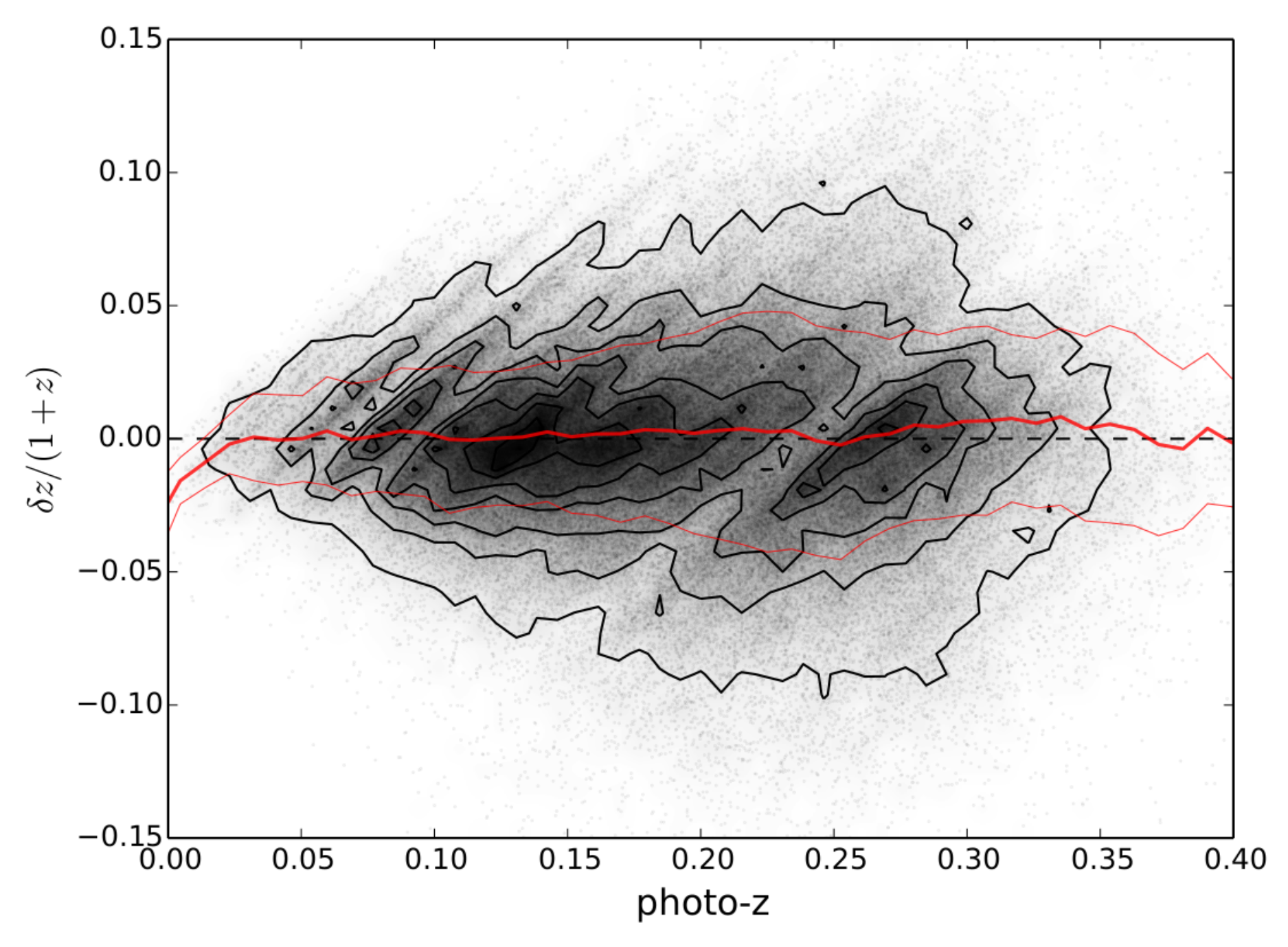} 
\caption{\label{Fig: zph-bias GAMA}Photometric redshift accuracy as a function of the \phz\ derived from WISE\ti{}SuperCOSMOS photometry, for the $B<21$ \& $R<19.5$ \& $W1<17$ flux-limited sample of \WISC\ti{}GAMA sources. Red lines illustrate the running median and scatter (SMAD) of $\delta z/(1+z)$.}
\end{figure} 
\begin{figure}
\centering
\includegraphics[width=0.48\textwidth]{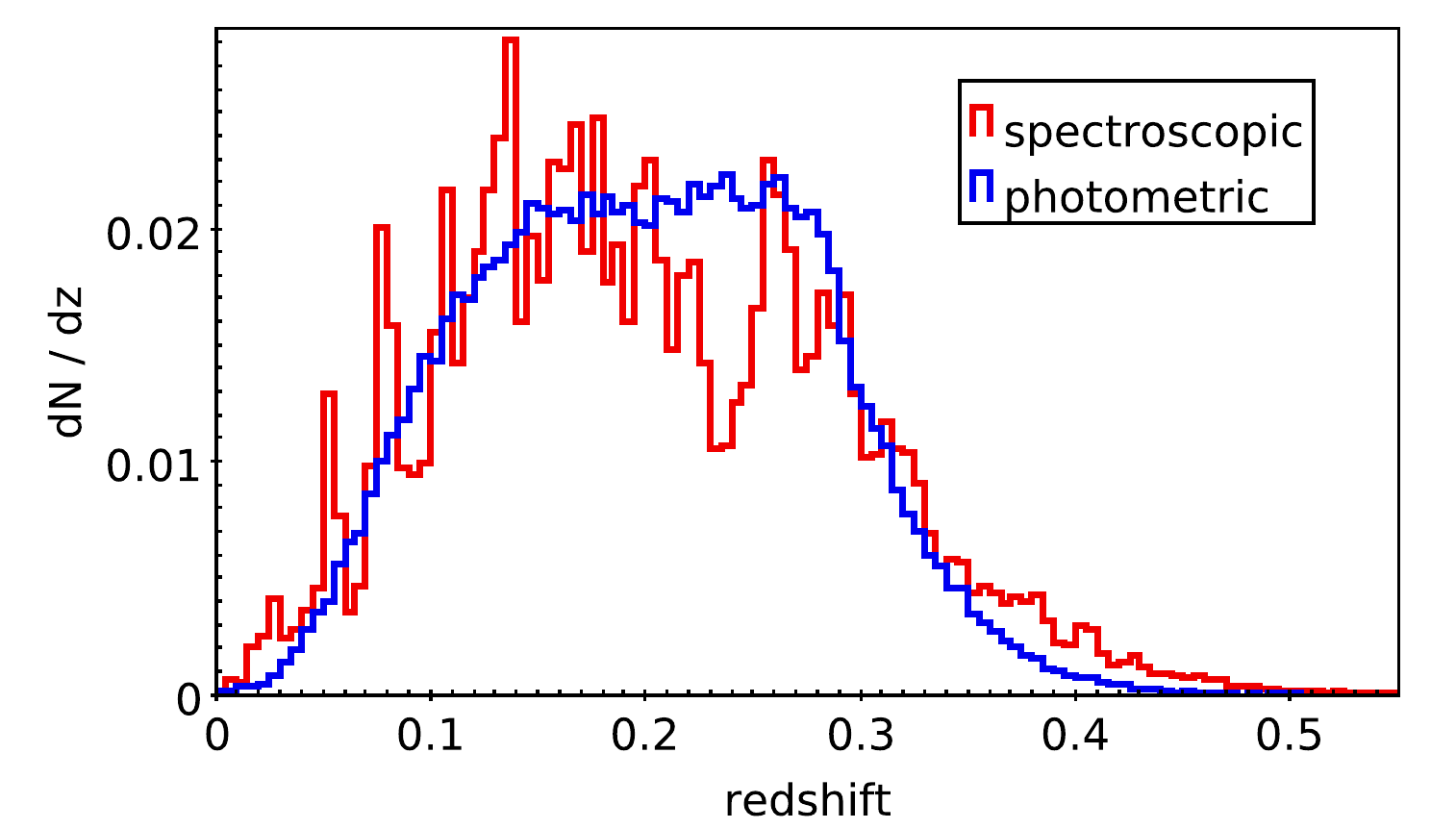} 
\caption{\label{Fig: dNdz GAMA}Comparison of the GAMA\ti{}WISE\ti{}SuperCOSMOS spectroscopic redshift distribution (red) with the photometric one derived from WISE\ti{}SuperCOSMOS photometry (blue), for the $B<21$, $R<19.5$, $W1<17$ flux-limited sample.\\}
\end{figure} 

Figures \ref{Fig: zsp-zph GAMA}--\ref{Fig: dNdz GAMA} illustrate the general performance of our photometric redshifts trained and tested on GAMA, for the flux-limited \WISC\ sample. A comparison of $z_\mrm{spec}$ with  $z_\mrm{phot}$ (Fig.\ \ref{Fig: zsp-zph GAMA}) as well as of $z_\mrm{phot}$  with the difference between them (Fig.\ \ref{Fig: zph-bias GAMA}) confirms the expected property of the \phzs\ being unbiased in the true $z_\mrm{spec}$ at a given $z_\mrm{phot}$ (\citealt{Driver11}; \tcb{B14}; \citealt{ANNz2}); a non-flat $N(z)$ must then lead, however, to the redshifts being photometrically overestimated at the low end and underestimated at high $z$.

The redshift distribution (Fig.\ \ref{Fig: dNdz GAMA}) shows similar features to those in figure 13 of \cite{Driver11}, where \phzs\ were derived using \ANNz\ for an earlier version of GAMA. The $\de N/\de z_\mrm{phot}$ diagram is narrower than the spec-$z$ one and the former is unable to reproduce sharp features in the latter, such as the dip at $z\sim0.23$ and several peaks related to clusters and walls. The quality of our photometric redshifts is however impressive given that we used only two optical bands. The latter limitation cannot be currently overcome when constructing nearly all-sky \phz\ samples with presently available data, covering much more of the sky at $z\sim0.2$ than available from \eg SDSS only \citep{DAbrusco07,Oyaizu08,Brescia14,Beck16}. 

\subsubsection{Dependence on apparent properties}
\label{Sec: apparent properties}

\begin{figure*}
\centering
\subfigure[]
{
\includegraphics[width=0.4\textwidth]{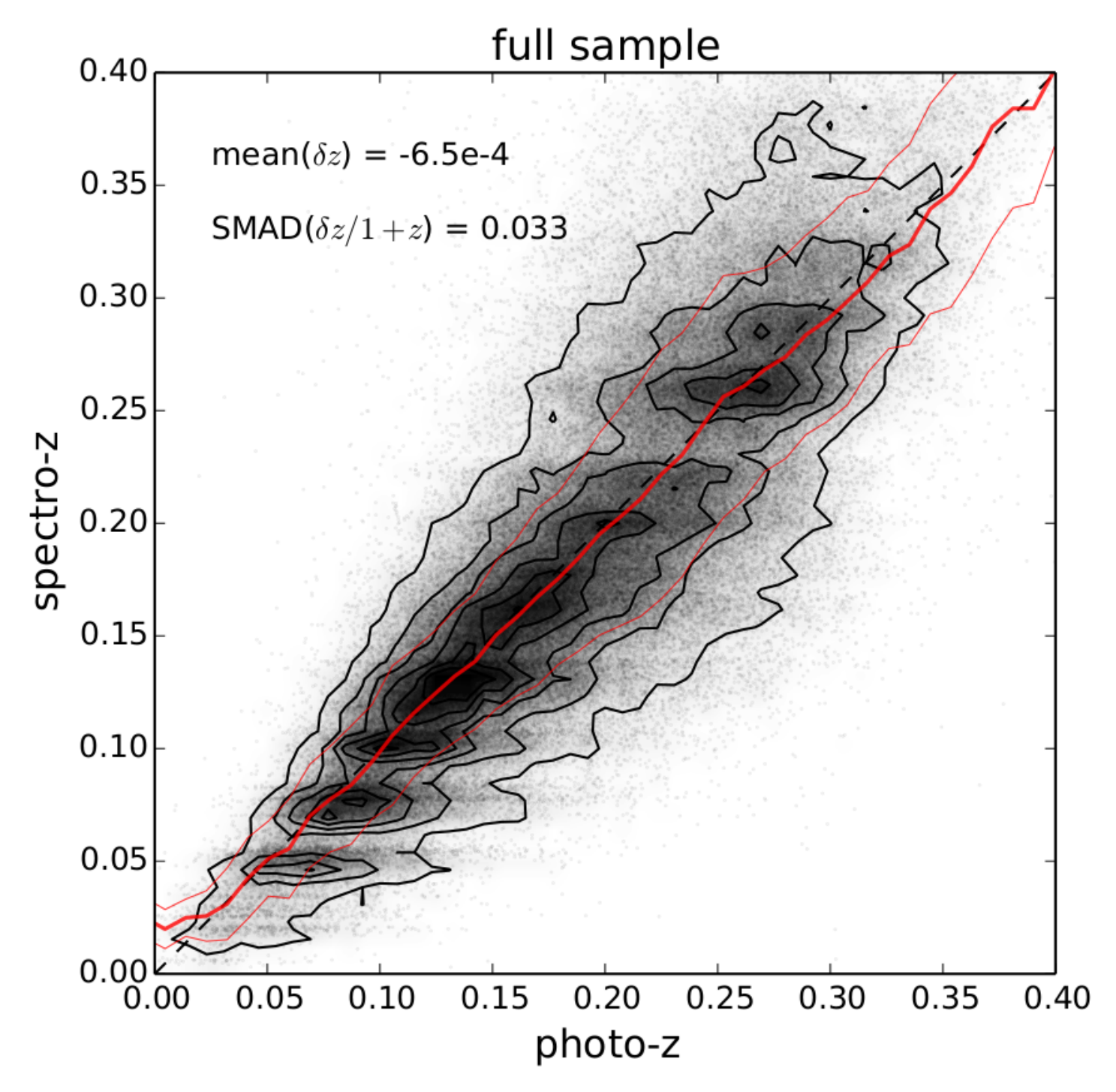} 
}
\subfigure[]
{
\includegraphics[width=0.4\textwidth]{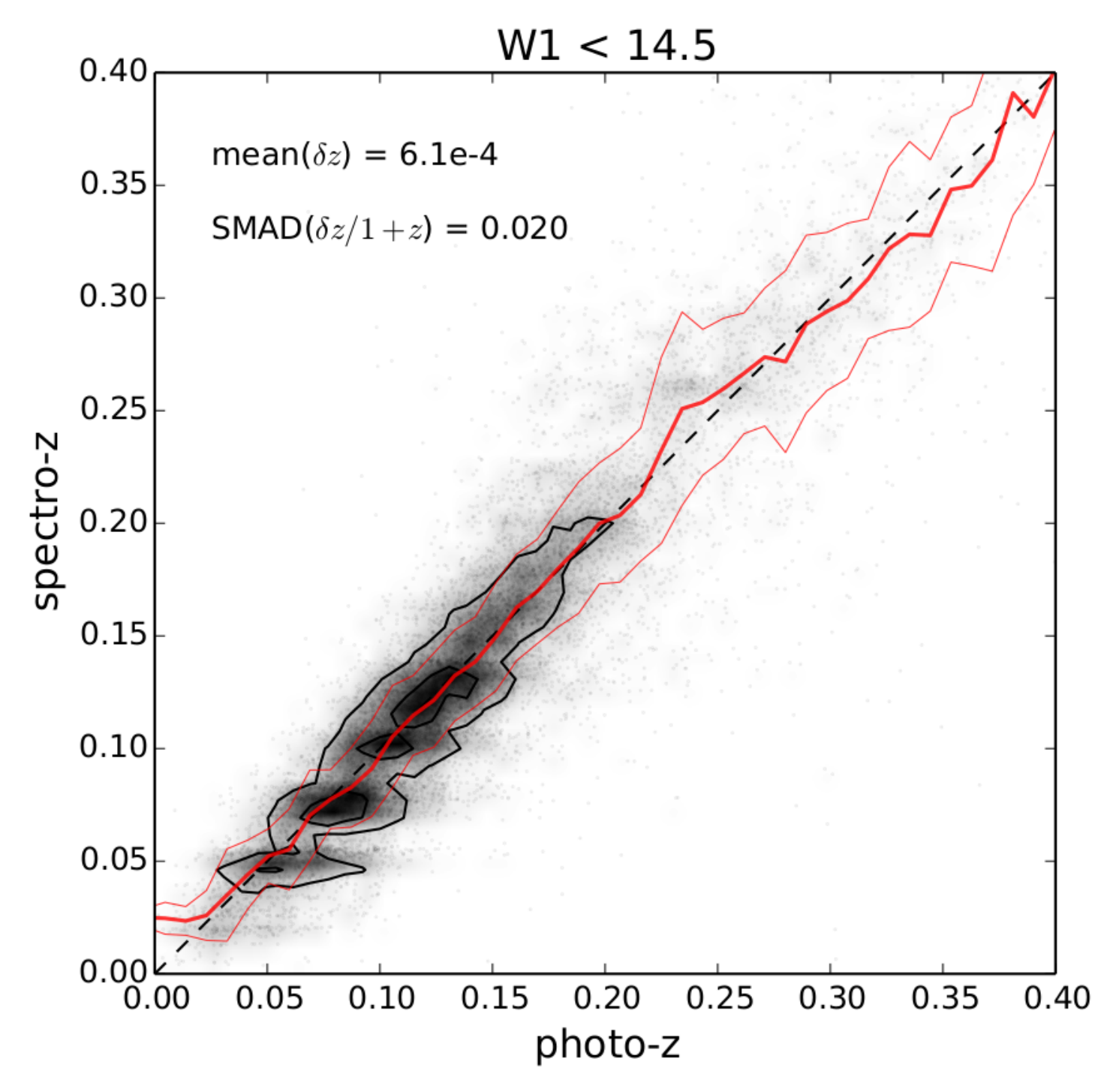} 
}
\subfigure[]
{
\includegraphics[width=0.4\textwidth]{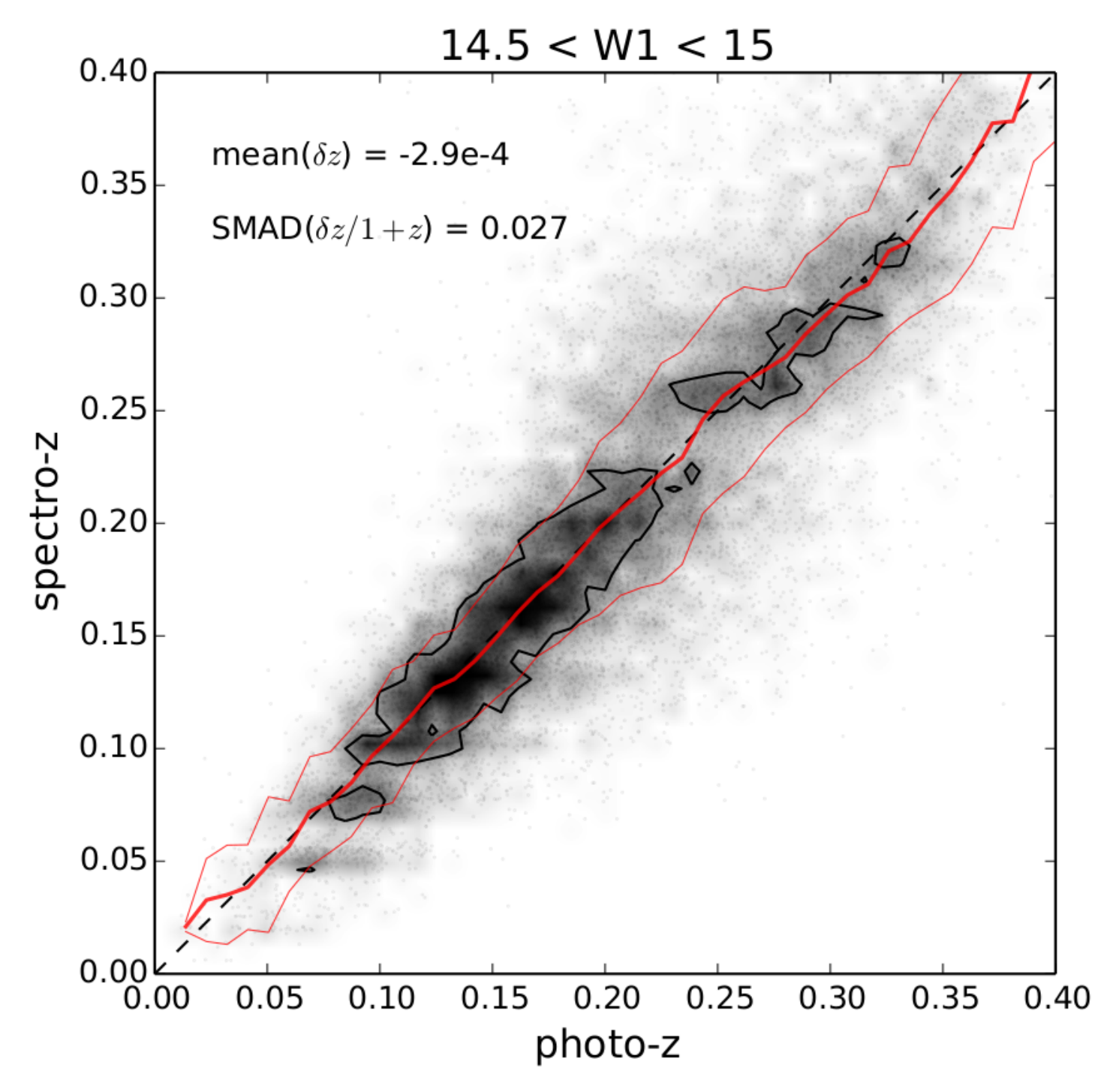} 
}
\subfigure[]
{
\includegraphics[width=0.4\textwidth]{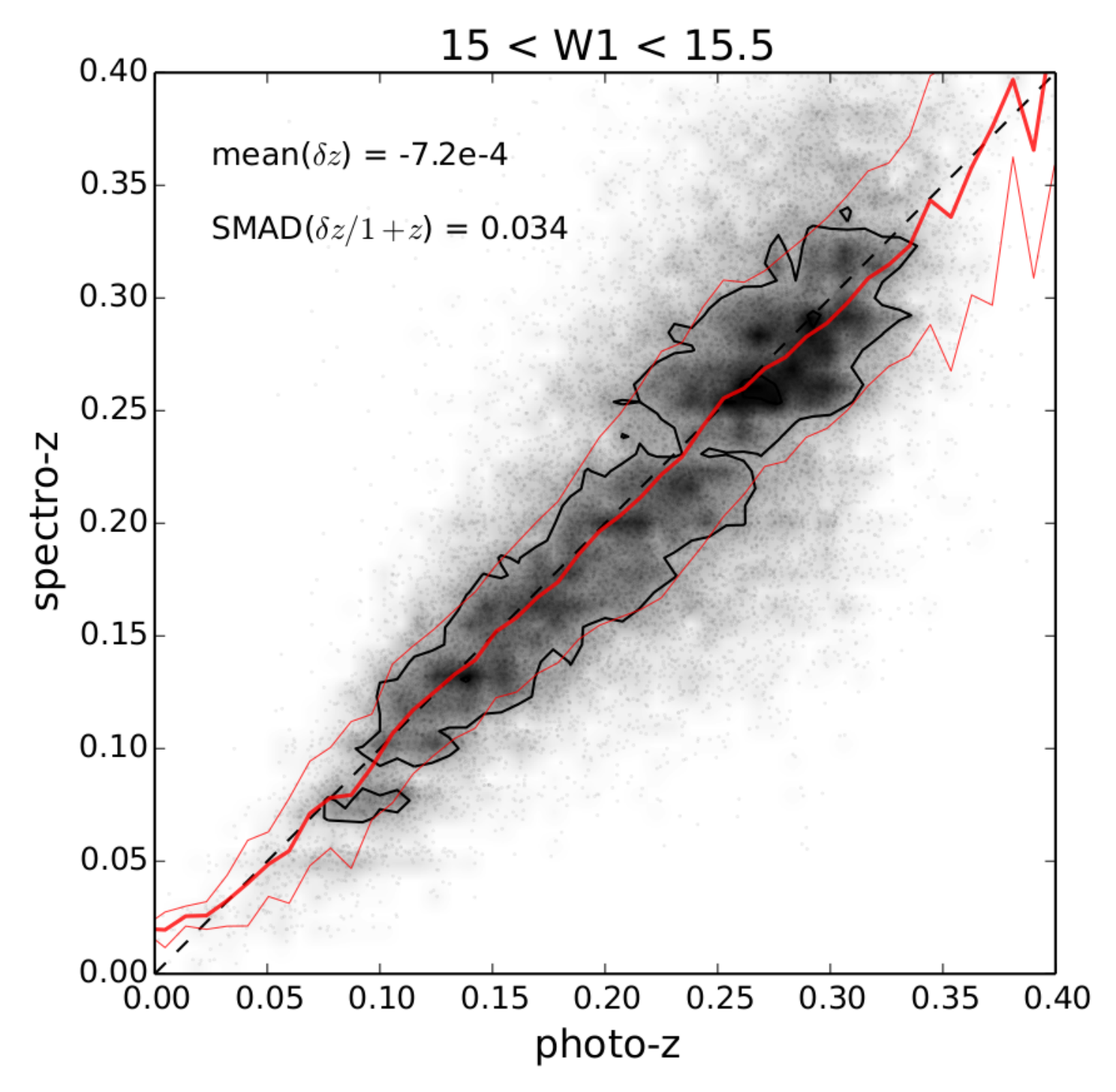} 
}
\subfigure[]
{
\includegraphics[width=0.4\textwidth]{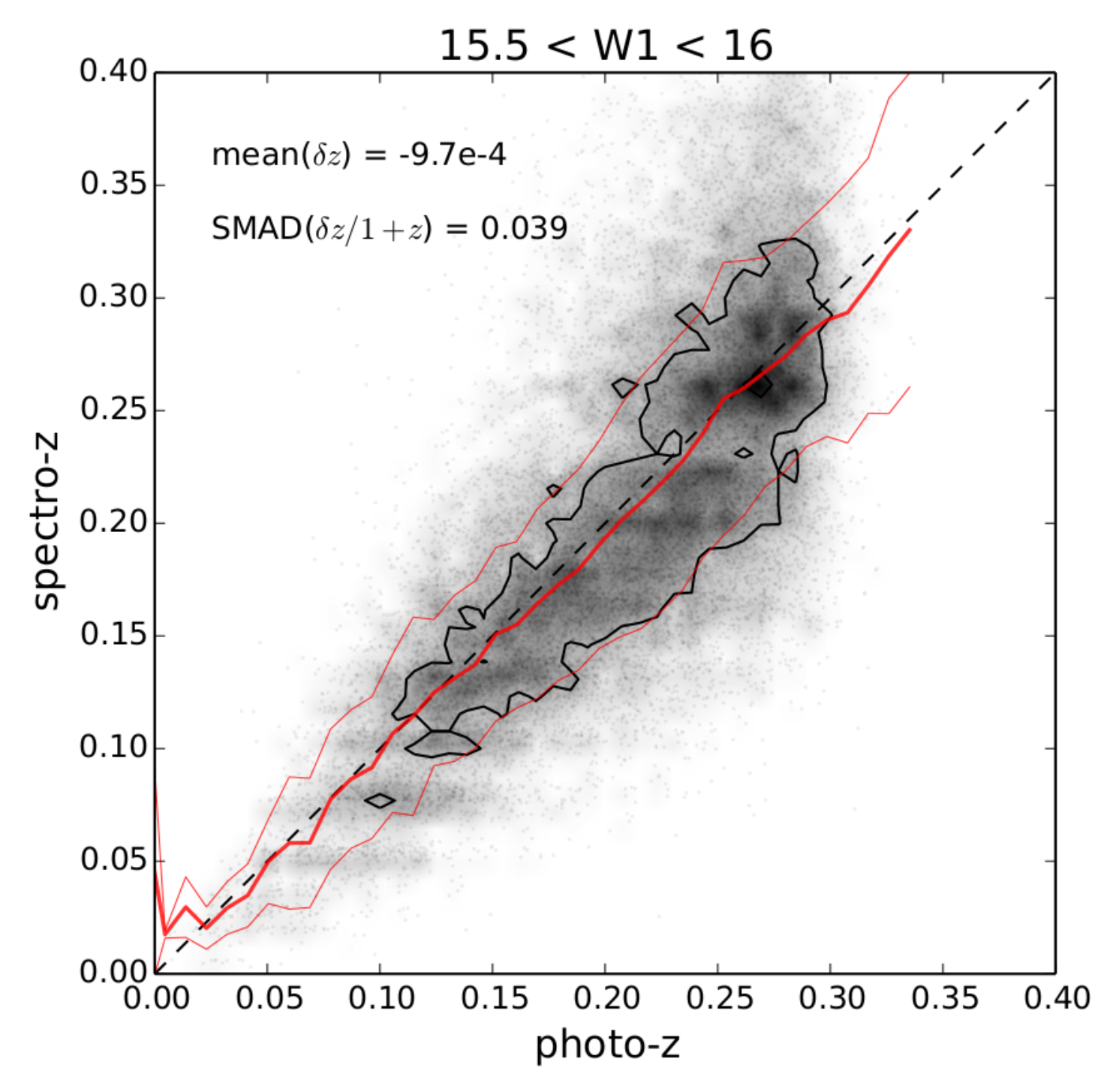} 
}
\subfigure[]
{
\includegraphics[width=0.4\textwidth]{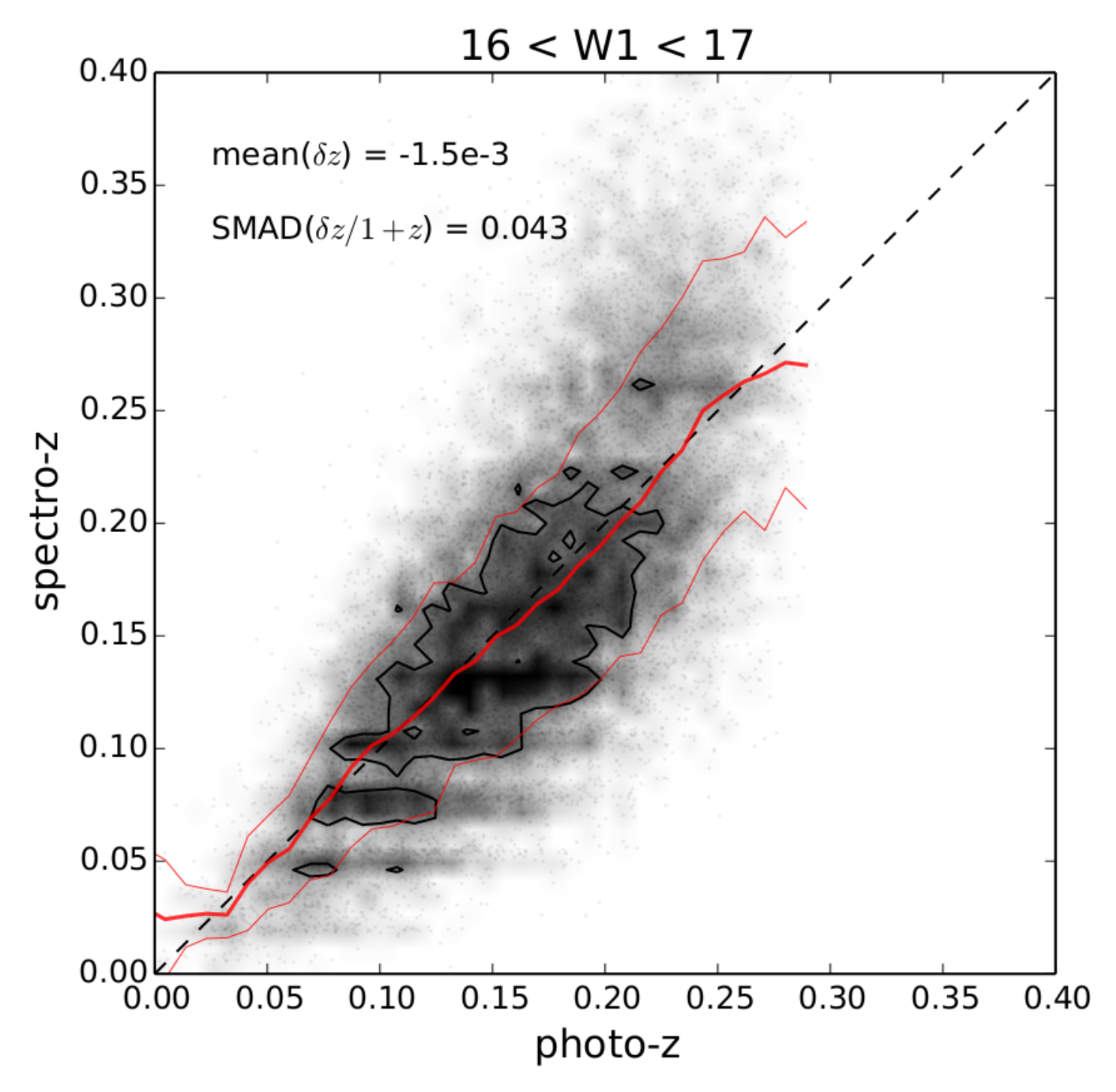} 
}
\caption{\label{Fig: zph-zsp for W1}Dependence of photometric redshift quality on the apparent $W1$ magnitude in the WISE\ti{}SuperCOSMOS sample calibrated on GAMA spectroscopic redshifts. Panel (a) shows the full sample, and the subsequent panels (b)--(f) present data binned in $W1$ intervals. Red lines illustrate the running median and scatter (SMAD) of $z_\mrm{spec}$.}
\end{figure*}

We have examined photometric redshift performance as a function of various observed and intrinsic properties of the galaxies. As far as the former are concerned, we have found no alarming patterns in the \phz\ accuracy as a function of apparent magnitudes in the four bands used in the procedure, other than general deterioration in photometric redshift quality  as the sources become fainter, consistent with expectations. Note, however, that as our catalog is produced by combining optical and infrared preselections, in some magnitude bins the fainter sources are not necessarily more distant. This, together with dependencies of photometric redshifts on varying magnitude, is illustrated for the $W1$ band in Fig.\ \ref{Fig: zph-zsp for W1}. Here we inverted the axes with respect to Fig.\ \ref{Fig: zsp-zph GAMA} to emphasize possible systematics for samples selected in \phz\ and magnitude bins, as will be practical for the full-sky sample where spectroscopic redshifts are not available. Some issues are evident, such as the lack of $z_\mrm{phot}>0.3$ galaxies for $W1>15.5$. 

This analysis of photometric redshift properties can be made more detailed by binning the data further. In particular, in addition to $W1$ intervals, we have also divided the test set into bins of the observed $B-R$ color (in $\Delta(B-R)=0.5$ mag), as well as of $z_\mrm{spec}$ and $z_\mrm{phot}$ (in bins of $\Delta z=0.1$). This gives two 3-dimensional `tables', each cell of which contains \phz\ statistics as in Table \ref{Table photo-z}. The extracts of these two tables are provided in Table \ref{Table extracts 3D} in the Appendix for a particular bin of $z_\mrm{spec}$ ($z_\mrm{phot}$) and of $B-R$. Full electronic versions of these tables can be made available on request.

Variations of photometric redshift quality are also observed  with WISE $W1$ and $W2$ signal-to-noise levels, which are already strongly correlated with the flux. In particular, there is a noticeable decrease in \phz\ accuracy for sources with $\mathtt{w2snr}<5$, as compared with those with $\mathtt{w2snr}>5$: the former have an order of magnitude larger mean bias $\langle\delta z\rangle$ than the latter and a considerably larger scatter in $\delta z$. Interestingly, in the cross-matched \WISC\ti{}GAMA sample the low-$\mathtt{w2snr}$ sources are on average located at \textit{smaller} distances than the high-$\mathtt{w2snr}$ ones: the $\mathtt{w2snr}<5$ galaxies have $\zmed\simeq0.13$ (spectroscopic) and practically never reach beyond $z=0.4$. The low-$\mathtt{w2snr}$ sources, however, constitute a small fraction (less than 5\%) of our galaxy catalog, and are mostly localized in several strips crossing the ecliptic, resulting from Moon avoidance maneuvers. 

\subsubsection{Dependence on intrinsic properties}

As an additional verification, we have examined photometric redshift performance as a function of source intrinsic properties, such as absolute magnitudes, rest-frame colors and stellar masses. Such parameters are not available for the full-sky sample as they require spectroscopic redshifts, but these tests are useful to search for possible issues in the \phz\ dataset. The test were done `blindly', \ie the parameters had not been used in the \phz\ procedure, and they were extracted \textit{a posteriori} from two GAMA data management units (DMUs), namely \ttt{StellarMasses v16} \citep{Taylor11} and \ttt{WISE-GAMA v01}  \citep{Cluver14}. The first one provides optical and near-infrared absolute magnitudes and rest-frame colors, as well as galaxy stellar masses and several other ancillary parameters, while the second one offers WISE-derived mid-infrared photometry, including isophotal magnitudes for resolved sources. \cite{Cluver14} also discuss the derivation of the absolute luminosities and stellar masses of WISE\ti{}GAMA galaxies, which we use here.

In practically all the bands available for the analysis from these GAMA DMUs (from $u$ up to $W3$), the photometric redshifts as a function of absolute magnitude are typically underestimated for bright galaxies, and overestimated for faint ones. This is expected and related to the previously mentioned property that the \phzs\ are not unbiased in the true redshift at a given $z_\mrm{spec}$, but are unbiased at $z_\mrm{phot}$.  A similar dependence is found for stellar masses, which again is not unexpected because of the correlation between galaxy's absolute luminosity and stellar mass (being the tightest in near-infrared bands). Interestingly, our photometric redshifts are relatively unbiased as a function of rest-frame colors, such as $u-r$ or $g-i$. 

\subsection{Final all-sky catalog}
\label{Sec: All-sky photo-z catalog}

\begin{figure}
\centering
\includegraphics[width=0.45\textwidth]{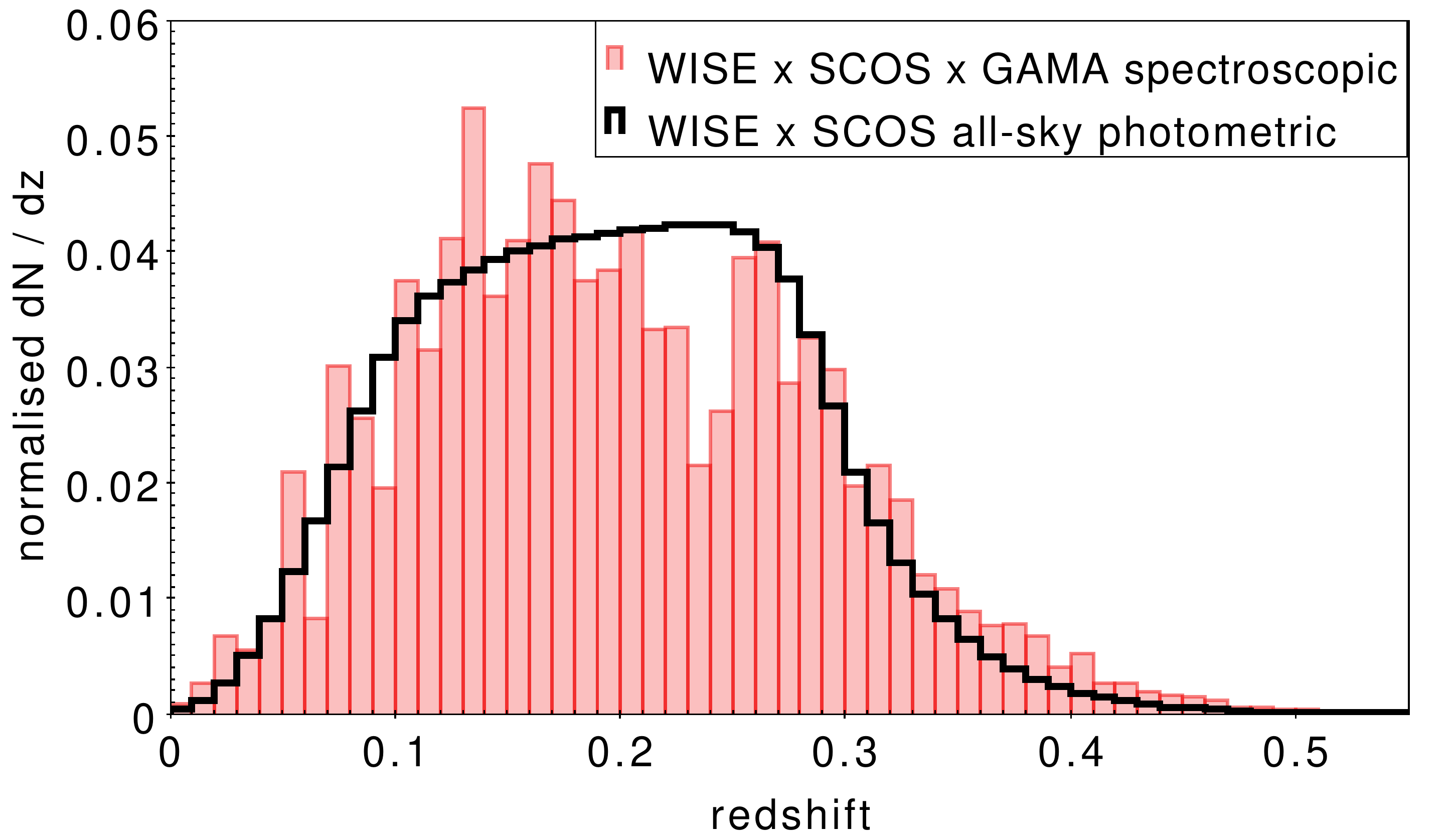} 
\caption{\label{Fig: GAMA+WIxSC.dNdz}Normalized redshift distributions of the \WISC\ti{}GAMA spectroscopic training set (red bars) and of the final all-sky WISE\ti{}SuperCOSMOS photometric sample (black line).}
\end{figure}

\begin{figure}
\centering
\includegraphics[width=0.45\textwidth]{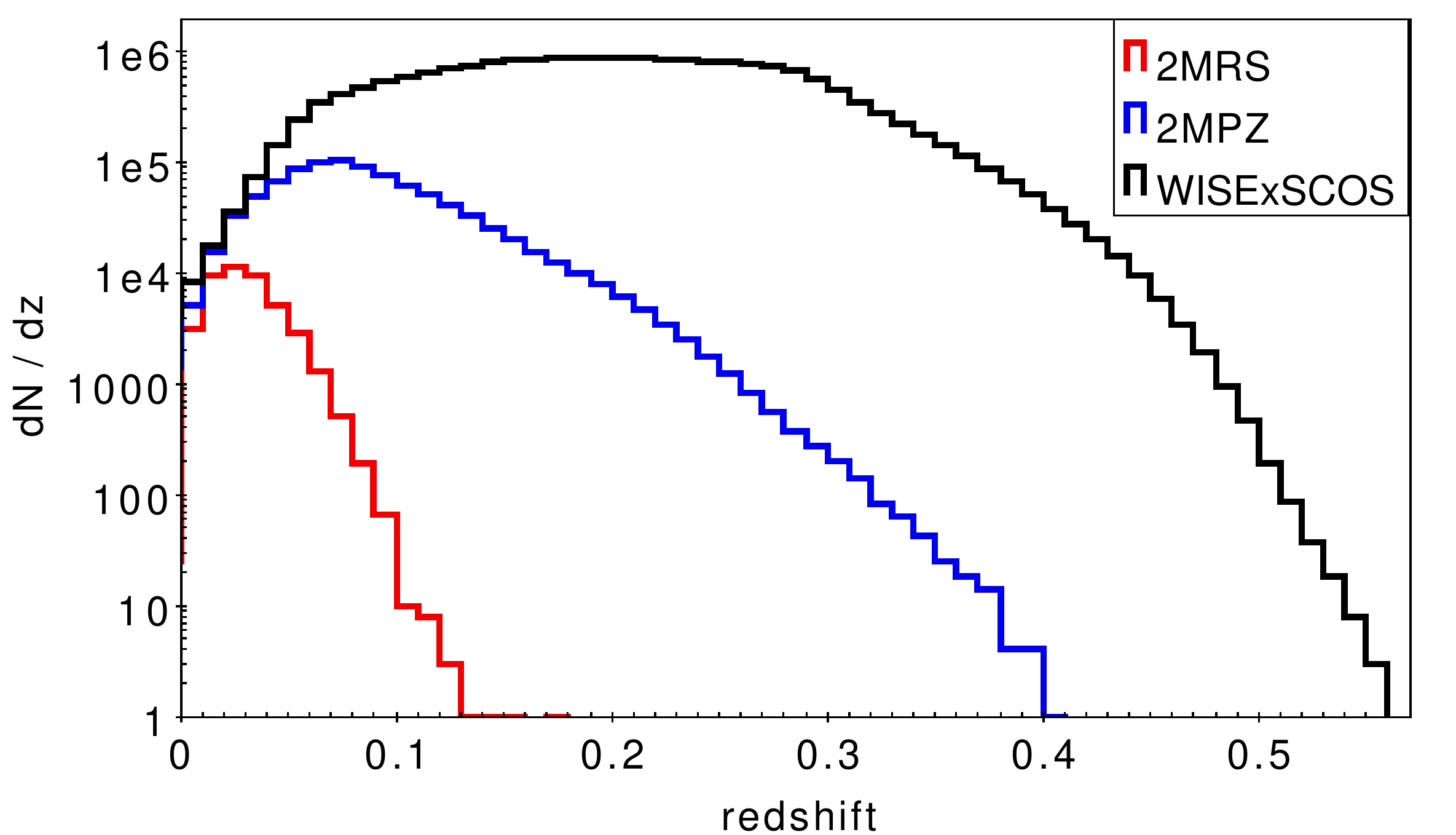}
\caption{\label{Fig: 3surveys.dNdz}Redshift distributions for three major all-sky surveys: 2MRS spectroscopic (red), 2MPZ photometric (blue) and \WISC\ photometric (black).}
\end{figure}

After performing all the tests discussed above, we trained the \ANNz\ algorithm on the full GAMA-South sample (142,000 sources, of which 98,000 fall within the flux limit of the core \WISC\ catalog) and applied the resulting networks to the cleaned dataset described in \S\ref{Sec: Catalog cleanup}. Fig.\ \ref{Fig: GAMA+WIxSC.dNdz} compares normalized redshift distributions of the \WISC\ti{}GAMA spectroscopic input (red bars) with the all-sky \WISC\ photometric output (black line). In Fig.\ \ref{Fig: 3surveys.dNdz} we present the absolute $\de N / \de z$, in logarithmic scaling, for three `all-sky' datasets: 2MASS Redshift Survey (2MRS, \citealt{2MRS}), 2MASS Photometric Redshift catalog (2MPZ, \tcb{B14}) and the \WISC\ \phz\ sample. This illustrates clearly the great improvement in information that \WISC\ brings at $z>0.1$ as compared with 2MASS. 

As described in \S\ref{Sec: SuperCOSMOS}, and discussed earlier in \cite{FP10} and \tcb{B14}, the SCOS passbands between the North (`N', $\delta_{1950}>2.5\degree$) and the South (`S') were slightly different; if not accounted for, this can lead to a bias between the photometric redshifts in the two `hemispheres'. The first step towards making N and S consistent was to calibrate the two parts using the Eqs.\ \eqref{Eq: Bcal}-\eqref{Eq: Rcal} from \S\ref{Sec: SuperCOSMOS}. In \tcb{B14} a possible residual \phz\ bias due to imperfect calibration was avoided by training the neural networks separately for N and S, which was possible thanks to comprehensive training sets in both parts of the sky. At the depths of the present sample, using GAMA data for \phz\ derivation in each hemisphere is not practical as the majority of the dataset is below $\delta=2.5\degree$: only 9\% of the GAMA sample is in the North. Such a training set of $\sim10,000$ sources is too small for proper calibration of the \phzs\ for $\sim10^7$ \WISC\ objects.

\begin{figure}
\centering
\includegraphics[width=0.45\textwidth]{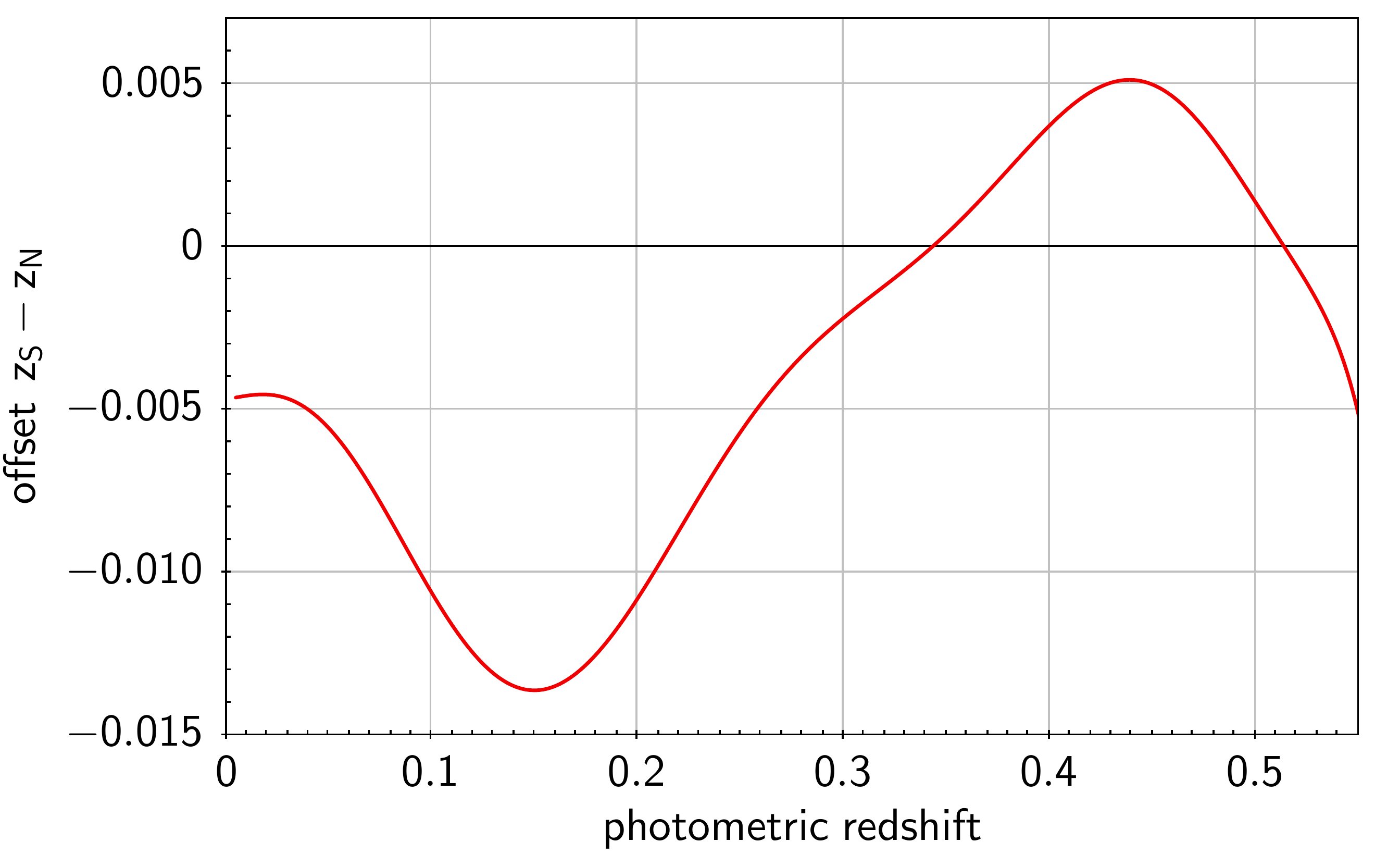} 
\caption{\label{Fig: N-S offset}Average offset between the photometric redshifts in the North and in the South, plotted as a function of the redshift.  {For each galaxy located in the North ($\delta_{1950}>2.5\degree$), this correction is added to the \phzs\ derived with \ANNz.}\\}
\end{figure}

In the absence of a large Northern spectroscopic dataset of GAMA's depth, there is no direct solution to this problem. We therefore rely in the first instance on the color corrections by which we attempt to place SCOS photometry in both hemispheres on a uniform basis. However, there exist remaining inconsistencies in the \phzs\ between N and S, which we address in the same way as in \cite{FP10}: examine the probability distribution of \phzs\ in the two `hemispheres' and make an adjustment to the redshift scale so that these distributions are consistent. The  {result of the} procedure is illustrated in Fig.\ \ref{Fig: N-S offset}, where the derived offset $z_S - z_N$ is plotted as a function of $z_N$. In all cases {, such a correction is} a small fraction of the \phz\ precision, but it does seem to be real.  {Therefore, for the sources at $\delta_{1950} > 2.5\degree$, we added this offset to the individual \phzs\ derived with \ANNz.} 

An additional test of photometric redshift quality comes from extending the purity analysis of \S\ref{Sec: Completeness and purity}. Examining \phz\ distributions of the sources identified as galaxies and as stars in the cross-match with the photometric SDSS data, we found that the stars contaminating the full sky sample are assigned \phzs\ peaking at $z\sim0.05$ rather than zero; since the training sample used to derive the \phzs\ contains no stars, the neural networks will then assign them redshifts of galaxies being the closest in the parameter space. But interestingly, some stars are assigned very high redshifts: at $z_\mrm{phot}>0.4$, over 40\% of the \WISC\ \phz\ sample is in fact stellar contamination. However the absolute source numbers are very small at these redshifts: only 144,000 of the sources have such \phzs\ (less than 1\% of the full sample). In general, we conclude that the final sample can be purified further over what was discussed in \S\ref{Sec: Catalog cleanup} by removing the lowest- and highest-\phz\ bins. The contamination is always smaller than 20\% for a cutout of $0.085<z_\mrm{phot}<0.345$, which is 90\% of the full sample; the purity improves further if the lowest Galactic latitudes are discarded. We would like to emphasize that for cosmological studies benefiting from `tomographic' slicing in redshift bins, the $z<0.1$ range is better sampled by 2MPZ (\tcb{B14}) rather than the present catalog owing to much higher completeness, purity and \phz\ accuracy of the former.

\begin{figure*}
\centering
\includegraphics[width=0.75\textwidth]{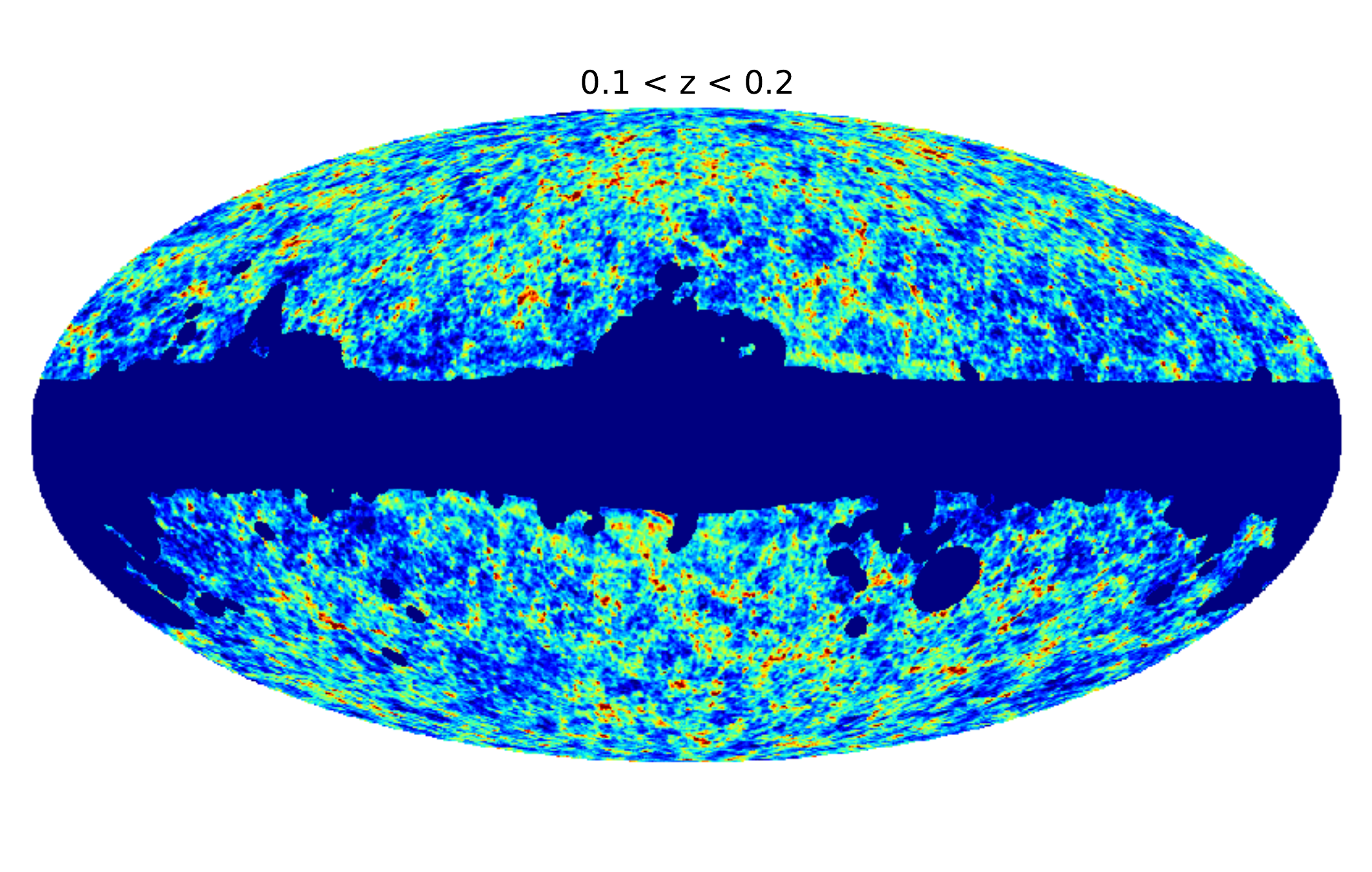}\vglue -3em
\includegraphics[width=0.75\textwidth]{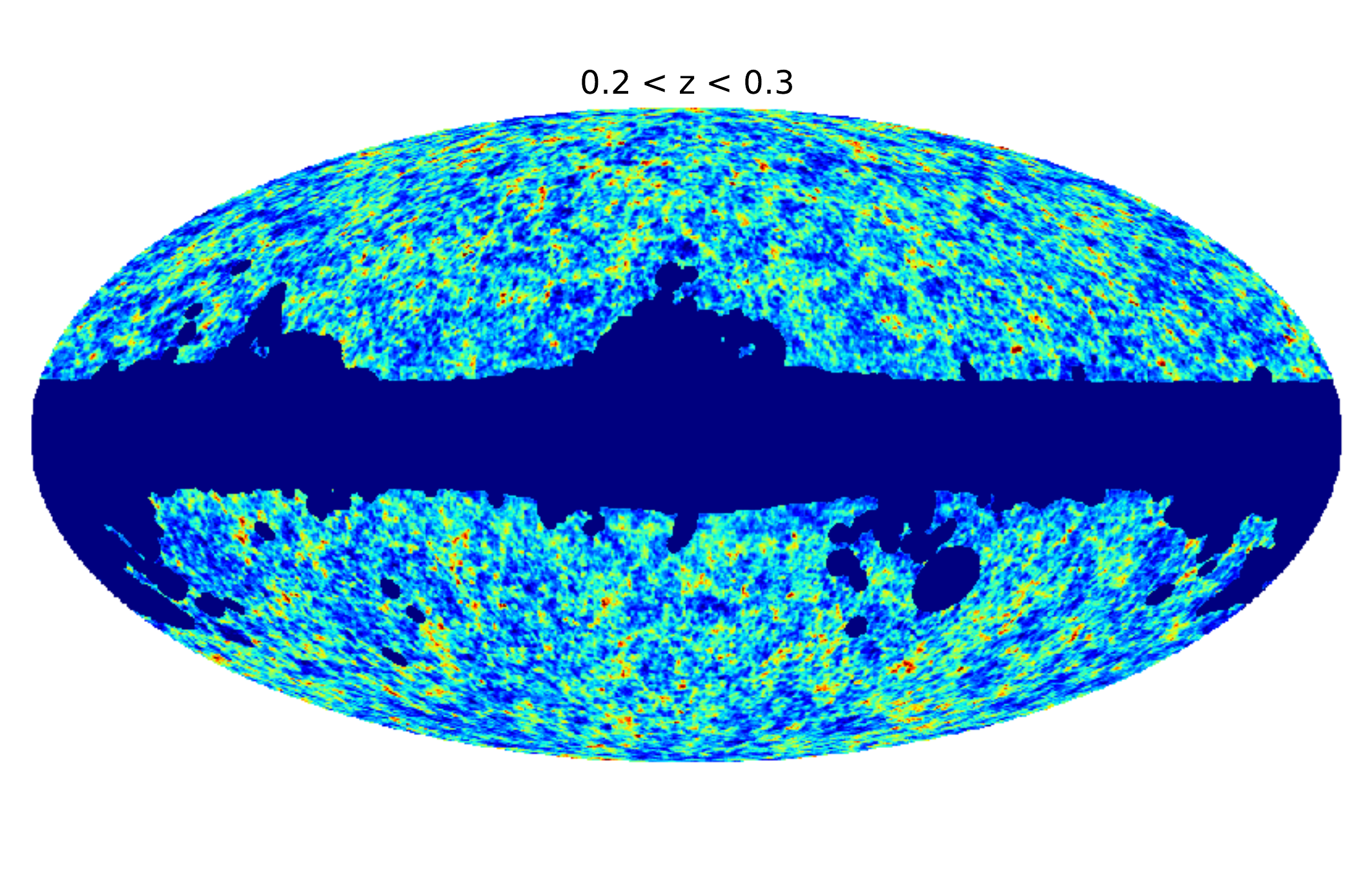}\vglue -3em
\includegraphics[width=0.75\textwidth]{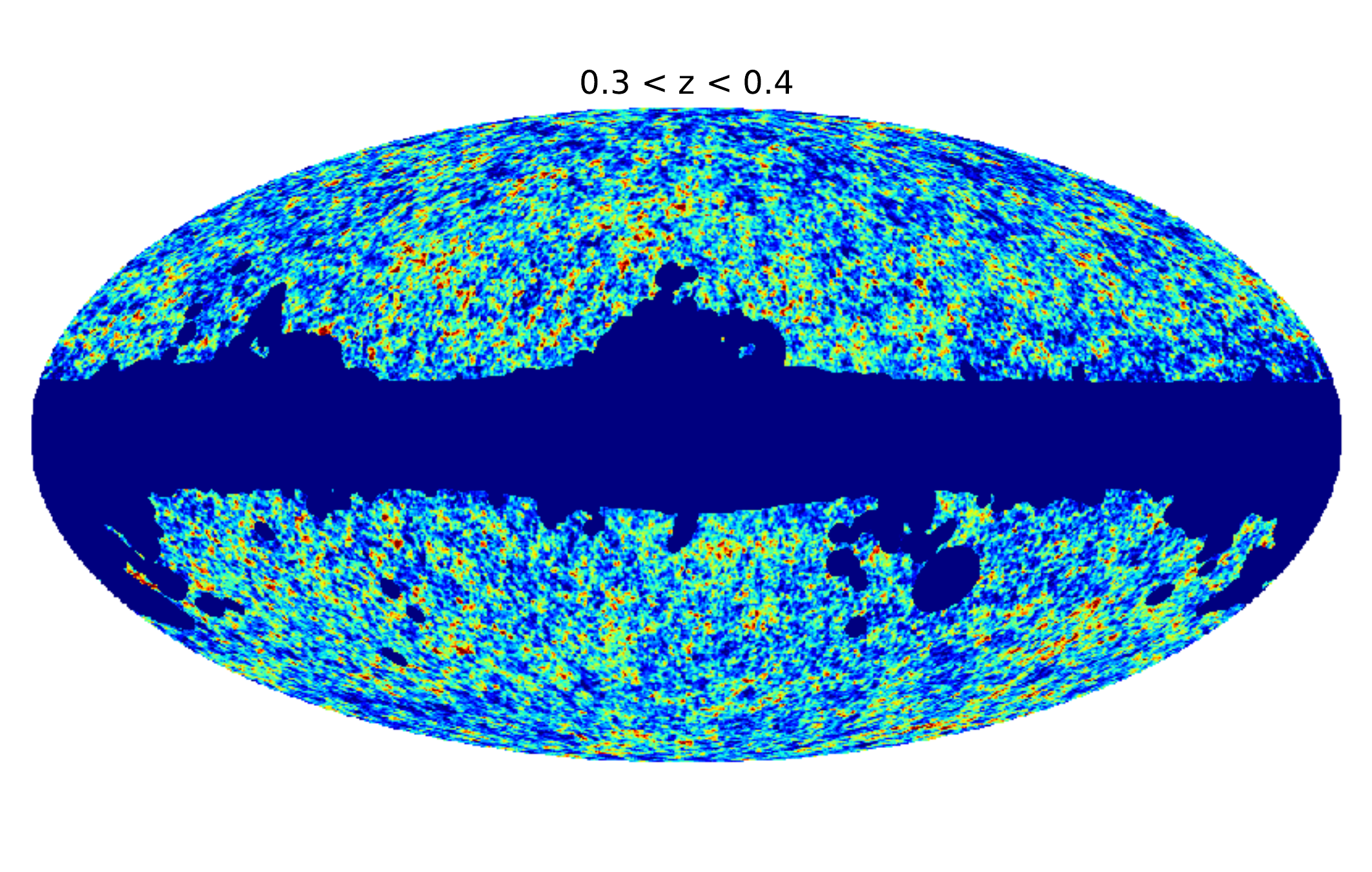}
\caption{\label{Fig: LSSslices}The large-scale structure at $z=0.1-0.4$: all-sky 
projections of $\Delta z = 0.1$ \phz\ slices from the WISE\ti{}SuperCOSMOS galaxy catalog, illustrating the power of this dataset in mapping the cosmic web at these redshifts.}
\end{figure*}

With these reservations, Fig.\ \ref{Fig: LSSslices} shows examples of three \phz\ shells of $\Delta z = 0.1$ extracted from the full \WISC\ sample, centered respectively on $z=0.15$, $z=0.25$ and $z=0.35$, and including respectively 7.3, 7.4 and 1.7 million sources. In order to produce these images, we have made one final correction for the redshift-dependent stellar contamination discussed above, which serves to mitigate large-scale non-uniformities, especially in the lowest redshift shell. We correlated the surface density in the slice with the total WISE sky density (treated as a proxy for stellar density) and removed the appropriate scaled fraction in order to remove the stellar gradients. This process is successful in yielding redshift slices with no apparent large-scale artefacts. These are the most comprehensive illustrations of all-sky galaxy distribution at these redshifts available so far, revealing new large-scale structures. In particular, neither 2MASS, nor especially 2MRS or PSC$z$, could reach to depths of $z>0.15$ in a comprehensive manner. Three-dimensional sampling of the cosmic web at these scales has been so far possible only with SDSS, covering 3 times less sky than our catalog, and being less complete beyond $r>17.77$ as far as the spectroscopic data are concerned.

\subsection{ {Comparison with external redshift samples}}
\label{Sec: External redshifts}

As a final `blind' test of the photometric redshifts in our catalog, and to verify whether they exhibit noticeable variations in performance over the sky, we cross-matched the sample with a number of external redshift catalogs. Such an exercise is only meaningful for auxiliary datasets which are complete to a depth at least similar to the present sample. For that reason the cross-matched catalogs mostly cover small fields: outside of GAMA, at present there are no other wide-angle spectroscopic datasets complete to its depth (otherwise we would of course have used them for the \phz\ training). In fact, however, part of the GAMA data themselves is included in this test, as some of them were not used in the \phz\ derivation and tests described above. These include the equatorial data from the G09 and G15 fields above $\delta_{1950}>2.5\degree$ (cf.\ \S\ref{Sec: tests on GAMA}), as well as the Southern G02 and G23 fields (less complete than the equatorial ones, \citealt{Liske15}), and the much deeper G10/COSMOS field \citep{Davies15} . Except for the very deep but small ($\sim2$ \dsqu) G10 field, all of these have high matching rates with \WISC\ as they have the same (G09 and G15) or very similar (G02 and G23) preselections as the `fiducial' GAMA data used for the \phz\ training. 

We have paired up several other publicly available datasets\footnote{A comprehensive list of of galaxy (redshift) surveys is provided at \url{http://www.astro.ljmu.ac.uk/\~ikb/research/galaxy-redshift-surveys.html}.} with \WISC; only a fraction of them had however a significant matching rate, and we will discuss here only those which had at least 500 common sources with our catalog. In all cases we used a $2"$ matching radius, which is a compromise between minimizing spurious cross-matches which could result from imprecise astrometry, and maximizing the matching rate for cases of not well defined centroids. 

Details regarding the fields' central coordinates, areas, as well as the \WISC\ \phz\ statistics (calculated with respect to the redshifts provided in the external datasets) are provided below and listed in Table \ref{Table external}. The samples included are:

{
\def\fix{\!\!\!\!\!\!\!\!\!\!\!\!\!\!\!\!\!\!\!\!\!\!\!\!\!\!\!\!\!\!\!\!\!\!\!}
\setlength\tabcolsep{3pt}
\begin{deluxetable*}{l c c c c c c c c c c c c}
\tabletypesize{\footnotesize}
\tablecolumns{13} 
\tablecaption{\label{Table external}\small {Statistics for the photometric redshift estimation, calculated for WISE\ti{}SuperCOSMOS sources cross-matched with external redshift catalogs. See text for details on the auxiliary datasets.}}

\tablehead{
external &
\colhead{\fix area} &
\colhead{coords.} &
\colhead{\# of} & 
 \multicolumn{2}{c}  {mean $\langle z \rangle$} & 
 \multicolumn{2}{c} { median $\overline{z}$} &
 \colhead{ 1$\sigma$ sc.\tablenotemark{e}} &
 \colhead{ scaled } &
  \colhead{ norm. } &
 \colhead{ bias\tablenotemark{h} } & 
 \colhead{ median } 
 \\
dataset &
\colhead{\fix [\dsqu]} &
\colhead{($\alpha, \delta$)\tablenotemark{a}} &
\colhead{sources\tablenotemark{b}} &
\colhead{spec\tablenotemark{c}} & \colhead{phot\tablenotemark{d}} &
\colhead{spec\tablenotemark{c}} & \colhead{phot\tablenotemark{d}} &
\colhead{ $\sigma_{\delta z/(1+z)}$} &
\colhead{ MAD\tablenotemark{f} } &
\colhead{ SMAD\tablenotemark{g} } &
\colhead{$\langle \delta z \rangle$} &
\colhead{ error\tablenotemark{i} } 
}

\startdata

\sidehead{GAMA}

~~equatorial & \fix & $(135, 2.65)$ \& \\
~~$\delta_{1950} >2.5$ & \fix 16.8 & $(217.5, 2.65)$ & 9,404 & 0.213 & 0.203 & 0.202 & 0.205 & 0.041 & 0.041 & 0.034 & $-$9.9e$-$3  & 14.8\% \\

 \\

~~G02 & \fix 55.9 & $(34.5, -6.95)$ & 17,812 & 0.211 & 0.205 & 0.207 & 0.207 & 0.039 & 0.038 & 0.031 & $-$6.2e$-$3 & 13.4\% \\

 \\

~~G23 & \fix 87.2 & $(345, -32.5)$ & 29,270 & 0.200 & 0.202 & 0.201 & 0.205 & 0.039 & 0.039 & 0.033 & $+$1.6e$-$3 &14.4\% \\

\\

AGES &  \fix 10.4 & $(217.9, 35.9)$ & 4,770 & 0.209 & 0.199 & 0.195 & 0.198 & 0.042 & 0.038 & 0.032 & $-$9.7e$-$3 & 13.8\% \\

\\

SHELS & \fix 4.8 &  $(139.9,  30)$ &  2,272 &  0.222 & 0.208 & 0.216 & 0.209 & 0.043 & 0.043 & 0.035 & $-$9.8e$-$3 & 14.6\% \\

\\

\rlap{G10/COSMOS} & \fix 2 & $(150.1, 2.2)$ & 603 &  0.226 & 0.202 & 0.218 & 0.201 & 0.068 & 0.039 & 0.033 & $-$2.4e$-$2 &  15.6\%  \\

\\

PRIMUS\tablenotemark{j} 
& \fix $\sim$\,10 & $-44<\delta<3$\tablenotemark{k}  &  3,222 & 0.227 & 0.215 & 0.215 & 0.217 & 0.059 & 0.046 & 0.038 & $-$1.2e$-$2 & 15.3\% \\

\\

COSMOS\tablenotemark{l} 
& \fix 2 &  $(150.1, 2.2)$ & 918 & 0.214 & 0.207 & 0.220 & 0.209 & 0.045 & 0.045 & 0.037 & $-$7.5e$-$3 & 15.8\% \\

\enddata
\tablenotetext{a}{Central coordinates of the field(s), in degrees.}
\tablenotetext{b}{In the cross-match with the \WISC\ fiducial sample.}
\tablenotetext{c}{External redshifts for a sample cross-matched with \WISC.}
\tablenotetext{d}{\WISC\ photometric redshifts for a sample cross-matched with the external dataset.}
\tablenotetext{e}{Normalized 1$\sigma$ scatter between the spectroscopic and photometric redshifts, $\sigma_{\delta z/(1+z)}$; unclipped.}
\tablenotetext{f}{Scaled median absolute deviation, $\mathrm{SMAD}(\delta z) = 1.48 \times \mathrm{med}(|\delta z-\mathrm{med}(\delta z)|)$.}
\tablenotetext{g}{Scaled median absolute deviation of the normalized bias, $\mathrm{SMAD}(\delta z/(1+z_\mrm{spec}))$.}
\tablenotetext{h}{Mean bias of $z_{\rm phot}$: $\langle \delta z \rangle = \langle z_{\rm phot}-z_{\rm spec} \rangle$; unclipped.}
\tablenotetext{i}{Median of the relative  error, $\mathrm{med}(|\delta z|/z_\mrm{spec})$; unclipped.}
\tablenotetext{j}{Low-resolution spectroscopy, $\sigma_{\delta z/(1+z)}=0.005$.}
\tablenotetext{k}{Nine fields over this declination range.}
\tablenotetext{l}{Accurate photometric redshifts, $\sigma_{\delta z/(1+z)}=0.007$.}
\end{deluxetable*}
}

\begin{itemize}

\item GAMA data in the equatorial G09 and G15 fields located at $\delta_{1950}>2.5\degree$; this is part of the sample comprehensively described in \S\ref{Sec: x-matches with GAMA}, removed from the \phz\ training and test phase due to the SuperCOSMOS North-South band difference (cf.\ \S\ref{Sec: SuperCOSMOS});  this sample includes 17,523 galaxies with $\zmed=0.217$ after cuts of  $z>0.002$ and redshift quality $\mathtt{NQ} \geq 3$.

\item GAMA G02 field (\ttt{TilingCat v04}): a spectroscopic redshift survey of a $\sim56$ \dsqu\ field centered at $\alpha=34.5\degree,\delta=-7\degree$, currently not publicly available (part of GAMA-II, \citealt{Liske15}),  with targets preselected from SDSS DR8 \citep{SDSS.DR8} and CFHTLenS \citep{CFHTLenS} photometric datasets to a limiting magnitude of $r<19.8$, although not fully complete to this limit over the whole field \citep{Liske15}; this sample includes 33,677 galaxies with $\zmed=0.226$ after cuts of  $z>0.002$ and redshift quality $\mathtt{NQ} \geq 3$.

\item GAMA G23 field (\ttt{TilingCat v11}): a spectroscopic redshift survey of a $\sim87$ \dsqu\ field centered at $\alpha=345\degree,\delta=-32.5\degree$, currently not publicly available (part of GAMA-II, \citealt{Liske15}), with targets preselected from KiDS \citep{KiDS} to a limiting magnitude of $i<19.2$, although not fully complete to this limit over the whole field \citep{Liske15}; this sample includes 47,489 galaxies with $\zmed=0.208$ after cuts of $z>0.002$ and redshift quality $\mathtt{NQ} \geq 3$.

\item AGES (AGN and Galaxy Evolution Survey; \citealt{AGES}): a spectroscopic redshift survey covering $\sim10$ \dsqu, centered at $\alpha=217.8\degree, \delta=34.3\degree$, with targets preselected to a limiting magnitude of $I<20$ from several datasets at various wavelengths; this sample includes 21,805 galaxies with $\zmed=0.342$ after a cut of $z_\mrm{spec}>0$ (measured spectro-$z$).

\item SHELS (Smithsonian Hectospec Lensing Survey; \citealt{SHELS}): a spectroscopic redshift survey of a $\sim5$ \dsqu\ field centered at $\alpha=140\degree,\delta=30\degree$, with targets preselected from the Deep Lens Survey \citep{DLS}, complete to a limiting magnitude of $R\leq20.6$ with part of the sources beyond this limit; this sample includes 15,591 galaxies with $\zmed=0.317$ after a cut on the redshift error $e_z<0.001$ and using only unmasked sources.

\item G10/COSMOS dataset: a publicly available redshift catalog \citep{Davies15}, part of GAMA, covering 2 \dsqu\ in the COSMOS field ($\alpha=150.1\degree,\delta=2.2\degree$), obtained by re-reducing archival spectroscopic zCOSMOS-bright data \citep{zCOSMOS07,zCOSMOS09} together with input from other sources (PRIMUS, SDSS, VVDS); the sample includes 16,128 galaxies with $\zmed=0.533$ after applying the flag $\mathtt{Z\_USE}=1$ (reliable high resolution spectroscopic redshift) and a cut on redshift $z>0$.

 \item PRIMUS (PRIsm MUlti-object Survey; \citealt{PRIMUS1,PRIMUS2}): a low-resolution spectroscopic redshift survey, covering a total of $\sim10$ \dsqu\  in nine fields to a depth of $i_\mrm{AB}\sim23.5$, preselected from several imaging datasets; here we used only the $\mathtt{ZQUALITY}=4$ sources (highest quality redshifts) which gave 87,742 objects in total with $\zmed=0.476$; one should bear in mind however that even for these sources, the PRIMUS \textit{spectroscopic} redshift precision is $\sigma_{\delta z/(1+z)}=0.005$ \citep{PRIMUS2}.

\item COSMOS photometric redshift catalog \citep{COSMOS}, providing very accurate \phzs\ based on 30-band photometry from UV through mid-IR; this is the same field as in the case of G10/COSMOS GAMA release, but the number of sources is much larger as there is no requirement of spec-$z$ availability; we used version 1.5 of the catalog, magnitude-limited to $I<25$, which includes 304,999 objects with \phzs\ measured ($0<\mathtt{zp\_best}<9.99$ in the catalog), with $\zmed=0.888$; at the bright end of interest for this exercise, the COSMOS \phzs\ have a scatter of $\sigma_{\delta z/(1+z)}=0.007$ \citep{COSMOS}, which is comparable to the accuracy of PRIMUS low-res spectro-$z$.

\end{itemize}

Several other datasets were tested, but they either gave below 500 cross-matches with \WISC\ each, or were too shallow for meaningful \phz\ statistics. The latter case includes the SDSS spectroscopic data: while it does provide a very high matching rate with our catalog (over 1/3 of SDSS DR12 spectroscopic galaxies are also in the \WISC\ fiducial sample), nevertheless the specific preselections of SDSS targets lead to biased \phz\ statistics. In order to circumvent this, one could try applying some weighting procedures such as discussed in \cite{Lima08}, which is beyond the scope of the present work.

The mean and median redshifts quoted in Table \ref{Table external} differ from those in Table \ref{Table GAMA fields} (flux-limited case) either because of different preselections of the external fields and/or, as in the case of GAMA equatorial data at $\delta_{1950}>2.5$, due to the bright-end flux cut of $W1=13.8$ applied to the fiducial \WISC\ sample, but not to the one referred to in Table \ref{Table GAMA fields}. This is a minor detail not influencing the conclusions.

Overall, the statistics provided in Table \ref{Table external} do not indicate significant variations of the \WISC\ \phz\ quality among the tested samples, and hence over the sky. In particular, in most cases the scatter in $\delta z/(1+z)$ (as measured through the SMAD) is within $0.035$, the mean bias of $\delta z$ is $<0.01$ and the median error in $|\delta z|/z$ is within 15\%. These numbers are very similar to those obtained in the GAMA equatorial fields in the test phase summarized in Table \ref{Table GAMA fields}.
On the other hand, the apparently worse results for the G10/COSMOS, PRIMUS and COSMOS catalogs should be taken with a grain of salt. These datasets have either a very low matching rate with our catalog (being very deep but covering small areas), and/or have low-resolution spectroscopy, or provide only photometric redshifts, even if of very high accuracy. In particular, the PRIMUS scatter in spectroscopic redshifts of $\sim0.005$ and the COSMOS scatter in \phz\ of $\sim0.007$ are comparable to the mean bias of \WISC\ \phzs\ and are less than an order of magnitude smaller than the SMAD of the latter. This means that for these samples, the comparison of the `true' redshift with the \WISC\ photometric one brings in uncertainty in the former parameter, which will lead to apparent deterioration of the \phz\ statistics.

\section{Summary, conclusions, future prospects}
\label{Sec: Summary}

In this paper we presented a novel photometric redshift galaxy catalog based on two the largest existing all-sky photometric surveys, WISE and SuperCOSMOS. A union of these two samples, once cleaned of stellar contamination, provides access to redshifts of $z<0.4$ on unmatched angular scales. Its angular coverage ($\simeq 3\pi$~sr) is a major advance with respect to existing surveys covering these redshifts \citep[e.g.][]{DAbrusco07,Oyaizu08,Brescia14,Beck16}.

We envisage manifold possible applications of our catalog, most simply from being able to improve the statistics on analyses of shallower datasets, such as 2MPZ \citep{AS14,XWH14,Alonso15}, 2MRS and 2MASS \citep{GH12} or the 2MASS PSC -- WISE combination \citep{Yoon14}. Our catalog can also be regarded as a testbed for currently being compiled or forthcoming more precise and deeper wide-angle samples, such as from DES, SKA or Euclid. A particularly interesting class of application involves `tomography':  slicing the dataset into redshift bins.  Cross-correlations with other wide-angle astrophysical probes at various wavelengths should be especially fruitful, owing to their insensitivity to any remaining small systematics. Such analyses include, for instance, CMB temperature maps for the integrated Sachs-Wolfe effect searches \citep[e.g.][]{G08,FP10}; CMB lensing measurements to constrain non-Gaussianity \citep{GP14} or neutrino mass \citep{PZ14}; or the gamma-ray background provided by the Fermi satellite to constrain the sources of this emission \citep{Xia15} or to search for dark matter \citep{Cuoco15}. In addition, we expect the  \WISC\ sample to be useful for studies on Faraday rotation of extragalactic sources \citep{Vacca15}, identification of galaxies in the SKA pathfinder WALLABY \citep{Popping12} or in planned CMB missions such as CoRE+ \citep{deZotti15}.  {It should be also appropriate in searches for electromagnetic counterparts of extragalactic gravitational wave sources, as -- together with 2MPZ (cf.\ \citealt{AH16}) -- it extends well beyond the catalogs currently used for that purpose \citep[e.g.][]{GWGC}. In addition, both 2MPZ and the present catalog provide two crucial parameters for such studies: the $B$-band magnitude (a proxy for black hole and neutron star merger rate), and the $W1$ magnitude, directly related to the galaxy's stellar mass.} In the nearer future, its bright end ($R\lesssim18$) may be employed as one of the input catalogs for the forthcoming TAIPAN survey \citep{TAIPAN}.

The fact that the median redshift of WISE galaxies is much higher than that of SCOS (\S\ref{Sec: x-matches with GAMA}; \citealt{Jarrett16}) makes it desirable to extend the present analysis beyond the latter sample. However, WISE on its own will not allow for precise photometric redshifts, as at its full depth it provides only two mid-IR bands. To obtain \phz\ coverage beyond the SDSS area, it will be necessary to combine WISE with forthcoming catalogs, such as Pan-STARRS \citep{Pan-STARRS} or the VISTA Hemisphere Survey \citep{VHS}. A supplementary approach to derive redshift estimates for WISE can be the one of \cite{Menard13}, which is indeed already being undertaken (A.\ Mendez, priv.\ comm.). One of the requirements for such studies to succeed will be the ability to reliably separate galaxies from stars and quasars in WISE; a report on ongoing machine-learning efforts towards this goal is presented in \cite{Kurcz16}.

 {The WISE~$\times$~SuperCOSMOS photometric redshift catalog is made publicly available through the Wide Field Astronomy Unit at the Institute for Astronomy, Edinburgh at \url{http://ssa.roe.ac.uk/WISExSCOS}.}

\section*{Acknowledgements}
\begin{small}

We thank the Wide Field Astronomy Unit at the Institute for Astronomy, Edinburgh, for archiving the 
WISE~$\times$~SuperCOSMOS catalog.

We acknowledge assistance from Eddie Schlafly with extinction corrections. We also benefited from discussions with Malcom Fairbairn, Robert Hogan, and Mike Read.

Special thanks to Mark Taylor for the TOPCAT \citep{TOPCAT} and STILTS \citep{STILTS} software, and for his assistance. Some of the results in this paper have been derived using the HEALPix package \citep{HEALPIX}.

The financial assistance of the South African National Research Foundation (NRF) towards this research is hereby acknowledged. M.\ Bilicki was supported by the Netherlands Organization for Scientific Research, NWO, through grant number 614.001.451, and through FP7 grant number 279396 from the  European Research Council. M.\ Bilicki and A.\ Solarz were partially supported by the Polish National Science Center under contract \#UMO-2012/07/D/ST9/02785. A.\ Solarz was supported by the Polish National Science Center under contract \#UMO-2015/16/S/ST9/00438 .

This publication makes use of data products from the Wide-field Infrared Survey Explorer, which is a joint project of the University of California, Los Angeles, and the Jet Propulsion Laboratory/California Institute of Technology, and NEOWISE, which is a project of the Jet Propulsion Laboratory/California Institute of Technology. WISE and NEOWISE are funded by the National Aeronautics and Space Administration.

This research has made use of data obtained from the SuperCOSMOS Science Archive, prepared and hosted by the Wide Field Astronomy Unit, Institute for Astronomy, University of Edinburgh, which is funded by the UK Science and Technology Facilities Council. 

GAMA is a joint European-Australasian project based around a spectroscopic campaign using the Anglo-Australian Telescope. The GAMA input catalog is based on data taken from the Sloan Digital Sky Survey and the UKIRT Infrared Deep Sky Survey. Complementary imaging of the GAMA regions is being obtained by a number of independent survey programs including GALEX MIS, VST KiDS, VISTA VIKING, WISE, Herschel-ATLAS, GMRT and ASKAP providing UV to radio coverage. GAMA is funded by the STFC (UK), the ARC (Australia), the AAO, and the participating institutions. The GAMA website is \url{http://www.gama-survey.org/}. 

Funding for SDSS-III has been provided by the Alfred P. Sloan Foundation, the Participating Institutions, the National Science Foundation, and the U.S. Department of Energy Office of Science. The SDSS-III web site is \url{http://www.sdss3.org/}. SDSS-III is managed by the Astrophysical Research Consortium for the Participating Institutions of the SDSS-III Collaboration including the University of Arizona, the Brazilian Participation Group, Brookhaven National Laboratory, Carnegie Mellon University, University of Florida, the French Participation Group, the German Participation Group, Harvard University, the Instituto de Astrofisica de Canarias, the Michigan State/Notre Dame/JINA Participation Group, Johns Hopkins University, Lawrence Berkeley National Laboratory, Max Planck Institute for Astrophysics, Max Planck Institute for Extraterrestrial Physics, New Mexico State University, New York University, Ohio State University, Pennsylvania State University, University of Portsmouth, Princeton University, the Spanish Participation Group, University of Tokyo, University of Utah, Vanderbilt University, University of Virginia, University of Washington, and Yale University. 


\end{small}

\bibliography{refs.bib}

\appendix
\section{Position-dependent cuts to remove stars from the cross-matched sample}
\label{App: position-dependent cuts}
Here we provide the details on the cuts applied to our dataset in order to clean it from stellar contamination (\S\ref{Sec: Star removal}). This procedure was then followed by more sophisticated masking of problematic areas that persisted after the cleanup procedure (\S\ref{Sec: Mask}).

We start with the modified longitude given by
\begin{equation}
\ell_\mrm{mod} = \left\{ \begin{array}{ll}
\ell & \mrm{for} \quad 0\degree \leq  \ell \leq 180\degree\;, \\
180\degree - \ell & \mrm{for} \quad \ell > 180\degree\;.
\end{array} \right.
\end{equation}
Next we define the limiting latitude for the Bulge cutout (in degrees) as
\begin{equation}
b_\mrm{Bulge} = 6 + \frac{11}{1+\left( \ell_\mrm{mod} / 60 \right)^2}\;.
\end{equation}
Anything with $|b|<b_\mrm{Bulge}$ is removed from the sample. This limiting latitude would go from $|b|=17\degree$ at the Galactic Center to $|b|=7\degree$ at the Anticenter; however, in our sample we already have only  $|b|>10\degree$ sources, so this cut is effective up to $\sim80\degree$ in longitude from the Center.

In addition, we masked out by hand the three most prominent nearby galaxies: the LMC, SMC and M31, applying circular cuts of radius respectively 8, 2 \& 2 degrees.

The next step is to define a position dependent color-cut for star removal, such that it would be equal to $W1-W2=0$ at high Galactic latitudes and be gradually increased to $W1-W2=0.12$ near the Galactic Plane. Using
\begin{equation}
b_\mrm{lim} = 5 + \frac{10}{1+\left( \ell_\mrm{mod} / 60 \right)^2}\;.
\end{equation}
and taking $\Delta b = |b|-b_\mrm{lim}$, we define the following threshold:
\begin{equation}
W12_\mrm{lim} = 0.12 \exp\left[-\left(\frac{\Delta b}{15}\right)^2\right]\;.
\end{equation}
Anything with $W1-W2<W12_\mrm{lim}$ is removed from the sample. The results of this procedure are illustrated in Fig.\ \ref{Fig: histograms cleanup} (\S\ref{Sec: Catalog cleanup}).

\begin{deluxetable*}{cccccc}
\tabletypesize{\footnotesize}
\tablewidth{0pt}
 \tablecolumns{6} 
\tablecaption{\label{Table extracts 3D}\small Extract from tables showing photometric redshift statistics in bins of redshift (separately spectroscopic and photometric), observed $B-R$ color and apparent $W1$ magnitude.}

\tablehead{
\colhead{ redshift bin } &
\colhead{ $B-R$ color bin } &
\colhead{ $W1$ mag bin } &
\colhead { \% of  sources\tablenotemark{a} } &
\colhead{ mean bias\tablenotemark{b} }  &
\colhead{ 1$\sigma$ scatter\tablenotemark{c} } 
}
\startdata
\sidehead{\bf Trained and tested on GAMA}
\sidehead{\underline{binned in spectro-$z$}}
$0.2 < z_\mrm{spec} < 0.3$	& $1.5 < B-R < 2.0$	& $ W1 < 14$			& 113		& $-0.0135$ & $0.0251$ \\
														&										&	$14 < W1 < 15$	& 2667	& $\,\;0.0111$	& $0.0245$ \\
														&										& $15 < W1 < 16 $	& 4001	& $\,\;0.0152$	&	$0.0236$ \\
														&										&	$16 < W1< 17$		& 92		&	$-0.0131$	&	$0.0284$ \\

\sidehead{\underline{binned in photo-$z$}}

$0.2 < z_\mrm{phot} < 0.3$	& $1.5 < B-R < 2.0$	& $ W1 < 14$			& 112		& $-0.0135$	& $0.0287$ \\
														&										& $ 14 < W1 < 15$ 	& 2591	& $-0.0002$	& $0.0261$ \\
														&										& $15 < W1 < 16$	& 4179	& $-0.0015$	& $0.0336$ \\
														&										& $16 < W1 <17$		& 144		& $-0.0025$	&	$0.0437$														
\enddata
\tablenotetext{a}{In the test set.}
\tablenotetext{b}{Mean bias of $z_{\rm phot}$: $\langle \delta z \rangle = \langle z_{\rm phot}-z_{\rm spec} \rangle$; unclipped.}
\tablenotetext{c}{Normalized 1$\sigma$ scatter between the spectroscopic and photometric redshifts, $\sigma_{\delta z/(1+z)}$; unclipped.}
\end{deluxetable*}


\end{document}